\documentclass[twocolumn,hyperpdf,
amsmath,amssymb,
aps,prd,10pt,
superscriptaddress,nofootinbib,noeprint,preprintnumbers]{revtex4-1}

\usepackage{graphicx}
\usepackage{dcolumn}
\usepackage{bm}
\usepackage{amsmath}
\usepackage{amsfonts}
\usepackage{slashed}
\usepackage{hyperref}
\usepackage{multirow}
\usepackage{soul}
\usepackage{mathtools}
\usepackage{float}
\usepackage{afterpage}
\usepackage{color}

\newcommand{\cm}{\ensuremath{\mathsf{cm}}}
\newcommand{\sm}{\ensuremath{\text{-}}}

\hypersetup{
pdftitle = {Resonances in coupled $\pi K, \eta K$ scattering from lattice QCD},
pdfsubject = {QCD},
pdfkeywords = {QCD, Lattice, scattering, hadron, spectrum, Kaon, pion, eta},
pdfauthor = {David J. Wilson, Jozef J. Dudek, Robert G. Edwards, Christopher E. Thomas},
colorlinks = {false},
}{}

\begin{document}

\preprint{JLAB-THY-14-1892}
\preprint{DAMTP-2014-81}

\title{Resonances in coupled $\pi K, \eta K$ scattering from lattice QCD}

\author{David~J.~Wilson}
 \affiliation{Department of Physics, Old Dominion University, Norfolk, VA 23529, USA}

\author{Jozef~J.~Dudek}
\email{dudek@jlab.org}
\affiliation{Theory Center, Jefferson Lab, 12000 Jefferson Avenue, Newport News, VA 23606, USA}
\affiliation{Department of Physics, Old Dominion University, Norfolk, VA 23529, USA}

\author{Robert~G.~Edwards}
\affiliation{Theory Center, Jefferson Lab, 12000 Jefferson Avenue, Newport News, VA 23606, USA}

\author{Christopher~E.~Thomas}
\affiliation{Department of Applied Mathematics and Theoretical Physics, Centre for Mathematical Sciences, University of Cambridge, Wilberforce Road, Cambridge, CB3 0WA, UK}

\collaboration{for the Hadron Spectrum Collaboration}
\date{\today}

\begin{abstract}
Coupled-channel $\pi K$ and $\eta K$ scattering amplitudes are determined by studying the finite-volume energy spectra obtained from dynamical lattice QCD calculations. Using a large basis of interpolating operators, including both those resembling a $q\bar{q}$ construction and those resembling a pair of mesons with relative momentum, a reliable excited-state spectrum can be obtained. Working at ${m_\pi=391\,\mathrm{MeV}}$, we find a gradual increase in the $J^P=0^+$ $\pi K$ phase-shift which may be identified with a broad scalar resonance that couples strongly to $\pi K$ and weakly to $\eta K$. The low-energy behavior of this amplitude suggests a virtual bound-state that may be related to the $\kappa$ resonance. A bound state with $J^P=1^-$ is found very close to the $\pi K$ threshold energy, whose coupling to the $\pi K$ channel is compatible with that of the experimental $K^\star(892)$. Evidence is found for a narrow resonance in $J^P=2^+$. Isospin--3/2 $\pi K$ scattering is also studied and non-resonant phase-shifts spanning the whole elastic scattering region are obtained.
\end{abstract}

\maketitle


Understanding the spectrum and properties of excited hadron states directly from the underlying theory of quarks and gluons, Quantum Chromodynamics (QCD), remains an unsolved problem. One challenge lies in the fact that excited hadrons are not asymptotically observable states, but rather appear as resonant enhancements in the scattering of lighter stable hadrons. Another challenge is the difficulty of computation within QCD which, at the energy scales of relevance, is a non-perturbative, relativistic theory. One technique which has shown significant progress when applied to hadron spectroscopy is lattice QCD. Lattice QCD is a systematically improvable calculational scheme in which the quark and gluon fields are discretized on a finite cubic grid, rendering the theory amenable to numerical computation. Monte-Carlo sampling of possible field configurations leads to estimates for hadronic correlation functions whose spectral content can then be explored.

The interactions of the lightest octet of pseudoscalar mesons are important since they are the stable particles to which excited hadrons decay. In this manuscript we will explore $\pi K$ scattering using lattice QCD techniques. This channel, having net strangeness, cannot proceed through intermediate quarkless states, which simplifies the phenomenology with respect to isospin--0 channels in which glueball states may appear.

The bulk of our knowledge of kaon scattering amplitudes comes from kaon beam experiments at SLAC in the 1970s and 80s. $\pi K$ scattering amplitudes were extracted from reactions using a proton target by extrapolating to small momentum transfer, $t$, where nearly-on-shell pion exchange dominates. Phase-shift analysis of the flavor exotic isospin--3/2 amplitudes as extracted from ${K^+ p \to K^+ \pi^+ n}$ and ${K^- p \to K^- \pi^- \Delta^{++}}$ by Estabrooks et al \cite{Estabrooks:1977xe} indicates a weak repulsive interaction in $S$-wave and very weak interactions in $P$-wave and higher. 

In isospin--1/2, as well as the phase-shift analysis of Estabrooks et al, there is a considerable set of $\pi K$ scattering results provided by the LASS experiment -- of particular relevance here are the final states $\pi K$ \cite{Aston:1987ir}, $\eta K$ \cite{Aston:1987ey} and $\pi\pi K$ \cite{Aston:1986jb}. In the partial-wave analysis of $\pi K \to \pi K$, a peaking amplitude in $S$-wave is interpreted as a broad $K^\star_0(1430)$ resonance which appears to saturate unitarity. The narrow elastic vector resonance, $K^\star(892)$, presents itself as a rapid rise in the $P$-wave phase-shift. The $D$-wave amplitude has a peak, well below the unitarity limit, that can be interpreted as an inelastic $K^\star_2(1430)$ resonance. Further resonances in the ``natural parity" series ($J^P=3^-,\, 4^+,\, 5^-$) are observed at higher energies.

$\eta K$ is the first inelastic channel to open, but LASS reports no significant amplitude into $\eta K$ for ${E_\mathsf{cm} < 2 \,\mathrm{GeV}}$ in $S,P$ and $D$ waves. Indeed the inelasticity in $P,D$-waves and higher appears to come first from the $\pi\pi K$ final state, where a significant amplitude is seen in $1^-$ above $1.3\,\mathrm{GeV}$ and a peak in $2^+$ at the $K^\star_2(1430)$. $\pi\pi K$ also couples to the ``unnatural parity" series, notably to $J^P=1^+$, where peaking behavior is observed that is commonly described in terms of two axial resonances, $K_1(1270),\, K_1(1400)$.

Resonances may or may not appear as bumps in hadron scattering amplitudes, and the least model-dependent way to describe them is to consider them as pole singularities in the analytic continuation of a scattering amplitude to complex values of energy. Narrow resonances, corresponding to sharp peaks in amplitudes, or rapid phase motion, appear as poles that lie close to the real energy axis where scattering amplitudes are determined experimentally. Poles that lie further away can lead to less rapid variation in the physical amplitudes -- a relevant example is the $\kappa$ resonance (the strange analogue of the $\sigma$ in $\pi\pi$), located at $\sqrt{s} \approx (650 - i\,  280) \,\mathrm{MeV}$. Strongly constrained analysis, using both experimental data and theoretical constraints, is required to determine the presence and location of such poles~\cite{DescotesGenon:2006uk}.

Our task here is to compute hadron scattering amplitudes within lattice QCD and to explore their singularity content. The explicit relationship between elastic scattering amplitudes and the discrete rest-frame spectrum in a finite periodic volume has been known for some time \cite{Luscher:1990ux,Luscher:1991cf}, along with later extensions considering the case of moving frames \cite{Rummukainen:1995vs, Kim:2005gf, Fu:2011xz, Leskovec:2012gb}. We have previously utilized these relations to determine, from first-principles lattice QCD computation, the detailed energy dependence of the scattering amplitudes for non-resonant $\pi\pi$ isospin--2 elastic scattering \cite{Dudek:2010ew,Dudek:2012gj} as well as the resonant isospin--1 case in which the $\rho$ appears \cite{Dudek:2012xn}. Recently we have seen the extension of the finite-volume formalism to the case of coupled-channel scattering \cite{Hansen:2012tf,Briceno:2012yi,Guo:2012hv}.

In order to extract the discrete spectrum of eigenstates of QCD in a finite volume, we will compute a matrix of two-point correlation functions, $\big\langle 0 \big| \mathcal{O}^{}_i(t) \mathcal{O}^\dag_j(0) \big| 0 \big\rangle$, using a large basis of operators, $\{ \mathcal{O}_i \}$, constructed from quark and gluon fields. The basis will include constructions resembling a single $q\bar{q}$-like meson, $\bar{\psi}\mathbf{\Gamma}\psi$, as well as others which resemble a pair of mesons having definite relative momentum, $\big(\bar{\psi} \mathbf{\Gamma}_1\psi\big)_{\vec{p}_1}  \big(\bar{\psi} \mathbf{\Gamma}_2\psi\big)_{\vec{p}_2}$. The matrix of correlation functions, which can be efficiently computed using the \emph{distillation} framework \cite{Peardon:2009gh}, is analyzed variationally to obtain a reliable extraction of many excited energy levels. 

With the finite-volume spectrum from a range of volumes and frames in hand, we can attempt to extract scattering amplitudes as a function of energy. This is a challenge since the energy of any single eigenstate of finite-volume QCD is a function of the scattering amplitudes at that energy for \emph{all} kinematically open scattering channels. The approach we will follow in this paper (following \cite{Guo:2012hv}) is to parameterise the energy-dependence of scattering amplitudes and attempt to describe the energy spectrum of many states at once by varying the parameters. The analytic forms for the scattering amplitudes can then be examined for their resonant pole content and relative couplings to scattering channels.

Our calculations herein will use an artificially heavy light quark mass, such that the pion has a mass of $391\,\mathrm{MeV}$ and the kaon has a mass of $549\,\mathrm{MeV}$. As such these first results can only be compared qualitatively to the experimental situation. We will present a determination of the scattering amplitudes for the lowest few natural-parity partial waves, $J^P = 0^+,\, 1^-,\, 2^+$, with $I=\tfrac{1}{2}$ and $\tfrac{3}{2}$.

$\pi K$ scattering has been studied previously using lattice QCD methods. Beane \emph{et al}~\cite{Beane:2006gj} studied $I=\frac{3}{2}$ scattering at threshold with 2+1 flavors of dynamical quarks and extracted the $S$-wave scattering length at four different pion masses by obtaining the energy levels corresponding to the mesons at rest on the lattice.  Sasaki \emph{et al} also performed a calculation to extract the $I=\frac{3}{2}$ threshold behavior and futhermore obtained the $I=\frac{1}{2}$ scattering length~\cite{Sasaki:2013vxa} by including a $q\bar{q}$--like operator and allowing for quark line annihilation. An earlier study of the scattering lengths in the quenched approximation appears in \cite{Nagata:2008wk}. Fu obtained scattering lengths~\cite{Fu:2011wc} in $S$-wave at six quark masses, and Fu~\emph{et al} have studied the $I=\frac{1}{2}$, $J^P=1^-$ interactions~\cite{Fu:2012tj}, although their determination neglects $S$-wave interactions which are known to be sizable. Lang, Prelovsek~\emph{et al}~\cite{Lang:2012sv,Prelovsek:2013ela} have studied both isospin combinations and have extracted scattering lengths and resonance parameters in small volumes and without dynamical strange quarks. Alternative strategies to extract $\pi K$ scattering information from finite-volume lattice QCD computations, based upon unitarization of a chiral lagrangian, are presented in \cite{Doring:2011nd, Doring:2012eu}.

Some results described in this paper previously appeared in \cite{PhysRevLett.113.182001} -- herein we expand considerably upon the details of the calculation. The remainder of the manuscript is structured as follows: In Section~\ref{sec_calc_details} we summarize the details of the lattices used and present relevant parameters, masses and thresholds. In Section~\ref{sec_symmetries} we discuss consequences of the reduced symmetry of a finite cubic lattice and the partial-wave mixing that occurs. In Section~\ref{sec_spec_determinations} we introduce our methods for constructing operators and obtaining correlation functions, and describe some typical spectra. In Section~\ref{sec_fvs} we describe the methods used to obtain infinite-volume scattering amplitudes from finite-volume energy spectra. Section~\ref{sec_one_half} contains our analysis of isospin--1/2 scattering; beginning with a limited set of data obtained at rest we obtain the coupled-channel $S$-wave amplitudes in isolation, then adding information from in-flight spectra we simultaneously describe $S$- and $P$-waves in the elastic scattering region below $\eta K$ threshold. We then present our main result using a large set of data to constrain the coupled-channel $S$- and $P$-waves, before presenting an extraction of the $D$-wave amplitude. Section~\ref{sec_one_half} concludes with a discussion of the resonant state content of the determined amplitudes. In Section~\ref{sec_three_half} we present the case of $\pi K$ $I=\frac{3}{2}$ scattering, before we summarize our findings in Section~\ref{sec_summary}. Appendices follow discussing $SU(3)$ flavor relations, the Chew-Mandelstam phase space and presenting the operator basis used to determine finite-volume spectra.

\section{Calculation details}
\label{sec_calc_details}

The discrete energy spectrum of a quantum field theory in a finite volume can be
determined from the exponential time dependence of Euclidean correlation
functions. These functions are averaged over a finite ensemble of gauge
fields, and as such, there is some statistical uncertainty in their
determination which complicates a reliable extraction of the energy.
To ameliorate this issue, we have employed the use of an
anisotropic lattice formulation with a temporal lattice spacing that
is smaller than the spatial lattice spacing. This fine
temporal resolution allows for a more precise determination of the energy while
reducing the computational cost relative to a fully isotropic lattice
calculation with equivalently fine resolution.

We have chosen to use an anisotropic Symanzik improved gauge action, and a dynamical Clover fermion action with
two flavors of light quarks and one heavier strange quark. The
boundary conditions are periodic in space and anti-periodic in
time. Details of the formulation are presented in
Refs.~\cite{Edwards:2008ja,Lin:2008pr}.  

We work in the isospin limit
where the $u,d$ quark masses are set equal, with a strange quark somewhat heavier. The dynamical quark
mass parameters are set to ${a_t m_l = -0.0840}$ and ${a_t m_s = -0.0743}$
which gives a pion mass of around $391$ MeV and a kaon around ${549\,\mathrm{MeV}}$.
The anisotropy of the lattice, determined from the dispersion relation
of the pion, is ${\xi \equiv a_s/a_t =3.444(6)}$ \cite{Dudek:2012gj}; the spatial lattice spacing is $\sim 0.12\,\mathrm{fm}$ and the
temporal lattice spacing is about $0.035\, \mathrm{fm}$.
The lattice volumes used in this work, ${16^3\!\times\! 128, \, 20^3\!\times\!
128}$  and ${24^3\!\times\! 128}$, correspond to spatial extents $L\sim
2\,\mathrm{fm},\, 2.5\,\mathrm{fm},\, 3\,\mathrm{fm}$. 
Some details of
the lattices and quark propagators used in the correlation function construction
are provided in Table~\ref{tab:lattices}.

\begin{table}
  \begin{tabular}{c|ccc}
  $(L/a_s)^3 \times (T/a_t)$	& $N_\mathrm{cfgs}$ 	& $N_{t_\mathrm{srcs}}$	& $N_\mathrm{vecs}$ \\
  \hline
  $16^3 \times 128$		& 479				& 4-8				& 64 \\
  $20^3 \times 128$		& 603				& 2-6				& 128 \\  
  $24^3 \times 128$		& 553				& 2-6				& 162
  \end{tabular}
  \caption{The lattice ensembles and propagators used in this
    paper. Shown are the lattice sizes, the number of configurations, the number of time-sources (which varies somewhat according to the correlator momentum and irrep) and the number of distillation vectors $N_{\mathrm{vecs}}$  featuring in the correlator construction \cite{Peardon:2009gh}.}
  \label{tab:lattices}
\end{table}

This anisotropic lattice formulation has been used successfully in previous
calculations of the light meson
spectrum~\cite{Peardon:2009gh,Dudek:2009qf,Dudek:2010wm,Dudek:2011tt,Dudek:2012xn,Dudek:2013yja}, baryon spectrum ~\cite{Edwards:2011jj, Dudek:2012ag, Edwards:2012fx}, $\pi\pi$ scattering~\cite{Dudek:2010ew,Dudek:2012gj} and observables involving charm quarks ~\cite{Liu:2012ze, Padmanath:2013zfa, Moir:2013ub}.

\begin{table}
  \begin{tabular}{r|l}
  	& $a_t m$\\
	\hline
    $\pi$  & 0.06906(13) \\
    $K$    & 0.09698(9) \\  
    $\eta$ & 0.10406(56) \\
    $\omega$ & 0.15678(41)\\	
    $\eta^\prime$ & 0.1750(54)\\
   \end{tabular}
   \hspace{0.5cm}
  \begin{tabular}{r|l}
    	& $a_t E_\mathsf{thr}$\\
		\hline
     $\pi K$ & 0.16604(15) \\
    $\eta K$ & 0.20104(57) \\
    $\pi \pi K$ & 0.23510(28)\\
    $\omega K$ & 0.25376(42) \\
    $\pi \eta K$ & 0.27010(58)\\
    $\eta^\prime K$ & 0.2764(54) \\
    $\pi\pi\pi K$ & 0.30416(40) \\
  \end{tabular}
  \caption{Stable meson masses, $a_t m$, determined on the lattice
    ensembles in Table~\ref{tab:lattices}. Pion and kaon masses are from
    an infinite-volume extrapolation, while the $\eta$, $\omega$ and $\eta'$
    are those evaluated on the $24^3$ lattices. 
    Also shown are the threshold energies, $a_t E_\mathsf{thr}$.
    }   
  \label{tab_masses}
\end{table}

Some computed masses and thresholds on these lattices are listed in
Table~\ref{tab_masses}. The kaon mass, computed on the three volumes, is extrapolated to
infinite volume to give the value in Table \ref{tab_masses} (using the
same method presented in \cite{Dudek:2012gj} for the pion mass). From the
kaon dispersion relation we determine an anisotropy $\xi_K =
3.449(4)$ which is compatible with the value determined from the pion
dispersion quoted above.

In this work we will make use of the volume dependence of the spectrum which arises from hadronic interactions.  There can also be exponentially-suppressed volume corrections to hadron energies that are not related to interactions -- the largest of these typically fall off exponentially with $m_\pi L$ and so, with $m_\pi L$ ranging from 3.8 to 5.7 in this work, we expect these effects to be small; previous investigations~\cite{Dudek:2012gj,Beane:2011sc} have not found a large variation of the pion mass with $L$ on these lattices.  In addition, here and in other studies~\cite{Dudek:2012gj,Dudek:2012xn} we obtain a good fit when data from the three volumes is fit simultaneously, and when scattering phase shifts are extracted on each volume independently these are generally consistent between volumes.

We will primarily present dimensionful results in units of the inverse temporal lattice
spacing to avoid unnecessary ambiguity with how one sets the lattice scale. When
required to quote a value in physical units, we will choose our usual
scale setting procedure where ${a_t = \frac{a_t m_\Omega}{m_\Omega^\mathrm{phys}}}$, using the $\Omega$ baryon mass
determined on these lattices, \mbox{$a_t m_\Omega = 0.2951$}, and the
physical $\Omega$ baryon mass \mbox{$m_\Omega^\mathrm{phys} = 1672
  \,\mathrm{MeV}$}.

\section{Reduced symmetry of a finite cubic lattice}
\label{sec_symmetries}

The symmetry of a lattice in a finite volume is reduced compared to that of continuous space in an infinite volume.  Our implementation, a spatially cubic lattice discretization in a cubic box with periodic boundary conditions, has the symmetry of a cube.  The relevant symmetry group for a system of hadrons overall at rest is therefore the double cover of the octahedral (or cubic) group with parity, $O_h^D$.  For a system ``in-flight'' with overall non-zero momentum, $\vec{P} \neq \vec{0}$, the appropriate symmetry is reduced further to that of the \emph{little group}~\cite{Moore:2005dw}, $\textrm{LG}(\vec{P})$, the subgroup of $O_h^D$ which leaves $\vec{P}$ invariant\footnote{for notational convenience we define $\textrm{LG}(\vec{0}) = O_h^D$}.  The spatially periodic boundary conditions quantize the allowed momenta, $\vec{p} = \frac{2\pi}{L}(n,m,p)$, where $L$ is the spatial extent of the lattice in physical units and $n,m,p$ are integers; we write this in a compact notation as $\vec{p} = [n,m,p]$ or $[n m p]$.

The consequences of this reduced symmetry for scattering have been discussed in detail in Refs.~\cite{Dudek:2012gj,Dudek:2012xn}.  In brief, at zero momentum the continuum spin, $J$, is not a good quantum number and states are instead labelled by irreducible representations, \emph{irreps}, of $O_h^D$.  Parity, $P$, and any relevant flavor quantum numbers are still good.  For $\vec{P} \neq \vec{0}$, $J$ and the helicity, $\lambda$, are not good quantum numbers and states are classified by the irreps of $\textrm{LG}(\vec{P})$.  Any relevant flavor quantum numbers are still good but, in general, parity is \emph{not} apart from for the $\lambda = 0$ components where $\tilde{\eta} \equiv P (-1)^J$ is a good quantum number~\cite{Thomas:2011rh}.  In this study we consider scattering of two unequal-mass hadrons and so there is no `extra' symmetry that arises with two identical hadrons, or two hadrons degenerate in mass related by some flavor symmetry~\cite{Dudek:2012xn}, which forbids odd and even partial waves (of opposite parities) from mixing when $\vec{P} \neq \vec{0}$. 

The manner in which the various components of a spin $J$ state (for $\vec{P}=\vec{0}$) or various helicities (for $\vec{P}\neq\vec{0}$) are distributed, or \emph{subduced}, across the relevant lattice irreps, $\Lambda$, is presented in Table II of Ref.~\cite{Dudek:2012gj}.  The subduction of $\pi K$ partial waves with $\ell \leq 4$ into lattice irreps is shown in Table~\ref{tab_pwa_irrep} -- the pattern is the same for both isospins and for $\eta K$ scattering.  It is apparent from the table that odd and even partial waves (with $P=-$ and $+$ respectively) appear in the same irreps when $\vec{P} \neq \vec{0}$ -- this features arises when the scattering particles are of unequal mass.

\begin{table}
  \begin{center}
    \begin{tabular}{cc|c|l|l}
      \hline
      \hline
           &&&&\\[-1.7ex]
      \multirow{2}{*}{$\vec{P}$} & \multirow{2}{*}{LG$(\vec{P})  $ }  & \;  \multirow{2}{*}{$\Lambda$} \;& $\,\,J^P (\vec{P}=\vec{0})$    & \; \multirow{2}{*}{$\pi K$ $\ell^N$} \\  
      &               &       & $\left|\lambda\right|^{({\tilde{\eta}})}(\vec{P}\neq\vec{0})$ & \\[0.5ex]
      \hline \hline
           &&&&\\[-1.5ex]
      \multirow{7}{*}{$\left[0,0,0\right]$}&     \multirow{7}{*}{$\textrm{O}_h^\textrm{D}$ ($\textrm{O}_h$)} & $A_1^+$   	  
      & $0^+,\, 4^+$          &\; $0^1,\, 4^1$\\
      && $T_1^-$    & $1^-,\, 3^-,\, \mathit{(4^-)}$ &\; $1^1,\, 3^1$\\
      && $E^+$      & $2^+,\, 4^+$          &\; $2^1,\, 4^1$\\
      && $T_2^+$    & $2^+,\, 4^+,\,  \mathit{(3^+)}\,$  &\; $2^1,\, 4^1$\\
      && $T_1^+$    & $4^+, \, \mathit{(1^+,3^+)}$ &\; $4^1$ \\
      && $T_2^-$    & $3^-,\, \mathit{(2^-,4^-)}$ &\; $3^1$ \\
      && $A_2^-$    & $3^-$               &\; $3^1$\\[0.5ex]
      \hline
      \hline 
      &&&&\\[-1.5ex]
      
      \multirow{5}{*}{$\left[0,0,n\right]$} & \multirow{5}{*}{Dic$_4$ ($\textrm{C}_{4\textrm{v}}$)}
      & $A_1$      & $0^+,\, 4$        &\; $0^1,\, 1^1,\, 2^1,\, 3^1,\, 4^2$ \\
      && $E_2$      & $1,\, 3$          &\; $1^1,\, 2^1,\, 3^2,\, 4^2$\\
      && $B_1$      & $2$               &\; $2^1,\, 3^1,\, 4^1$ \\
      && $B_2$      & $2$               &\; $2^1,\, 3^1,\, 4^1$ \\ 
      && $A_2$      & $4,\, \mathit{(0^-)}$      &\; $4^1$ \\[0.5ex]	
      \hline
      &&&&\\[-1.5ex]
      
      \multirow{4}{*}{$\left[0,n,n\right]$} & \multirow{4}{*}{Dic$_2$ ($\textrm{C}_{2\textrm{v}}$)} 
      & $A_1$     & $0^+,\, 2,\, 4$   &\; $0^1,\, 1^1,\, 2^2,\, 3^2,\, 4^3$ \\
      && $B_1$     & $1,\, 3$          &\; $1^1,\, 2^1,\, 3^2,\, 4^2$ \\
      && $B_2$     & $1,\, 3$          &\; $1^1,\, 2^1,\, 3^2,\, 4^2$ \\
      && $A_2$     & $2,\, 4, \, \mathit{(0^-)}$ &\; $2^1,\, 3^1,\, 4^2$ \\[0.5ex]	
      \hline
      &&&&\\[-1.5ex]
   
      \multirow{3}{*}{$\left[n,n,n\right]$} &       \multirow{3}{*}{Dic$_3$ ($\textrm{C}_{3\textrm{v}}$)}
      & $A_1$     & $0^+,\, 3$        &\;  $0^1,\, 1^1,\, 2^1,\, 3^2,\,  4^2$\\
      && $E_2$     & $1,\, 2,\, 4$     &\;  $1^1,\, 2^2,\, 3^2,\,  4^3$ \\
      && $A_2$     & $3, \, \mathit{(0^-)}$      &\;  $3^1,\, 4^1$\\[0.5ex]
      \hline
      \hline
    \end{tabular}
  \end{center}
  \caption{The pattern of subductions of $\pi K$ (or equivalently $\eta K$) partial-waves, $\ell \leq 4$, into lattice irreps, $\Lambda$, where $N$ is the number of embeddings of this $\ell$ in the irrep and $n$ is a non-zero integer.  This is derived from Table II of Ref.~\cite{Dudek:2012gj} by considering the subductions of $\ell$ when $\vec{P} = \vec{0}$ or the various helicity components for each $\ell$ when $\vec{P} \neq \vec{0}$, effectively Table VII of Ref.\cite{Dudek:2012gj} and Table III of Ref.\cite{Dudek:2012xn} combined.  The LG$(\vec{P})$ column shows the double-cover little group (the corresponding single-cover little group relevant for only integer spin is given in parentheses).  Also shown are the various $J \leq 4$ or $|\lambda| \leq 4$ that appear in each of the relevant irreps.  The $J^P$ values and $|\lambda|^{\tilde{\eta}}=0^-$ in italics are in the ``unnatural parity'' [$P = (-1)^{J+1}$] series and do not contribute to pseudoscalar-pseudoscalar scattering.}
  \label{tab_pwa_irrep}
\end{table}

\section{Spectrum determination}
\label{sec_spec_determinations}

We obtain the finite-volume spectrum in a given irrep by analyzing a matrix of Euclidean time correlation functions,
\begin{equation}
C_{ij}(t) = \big\langle 0 \big| \mathcal{O}^{}_i(t) \, \mathcal{O}_j^\dag(0) \big| 0 \big\rangle, \label{corr}
\end{equation}
where a basis of hadronic creation operators, $\{\mathcal{O}^\dag_i\}$, transforming with the desired quantum numbers, has been constructed from quark and gluon fields. Each correlation function in this matrix has a spectral decomposition featuring a common discrete spectrum of finite-volume eigenstates $\big| \mathfrak{n}\big\rangle$,
\begin{equation}
C_{ij}(t) = \sum_{\mathfrak{n}} e^{-E_\mathfrak{n} t}  \frac{1}{2E_\mathfrak{n}}
	\big\langle 0 \big| \mathcal{O}_i(0) \big| \mathfrak{n} \big\rangle  
	\big\langle \mathfrak{n} \big|	\mathcal{O}_j^\dag(0) \big| 0 \big\rangle. 
	\label{decomp}
\end{equation}
Within the chosen operator basis we seek the optimal linear combination for interpolation of each possible low-lying finite-volume eigenstate from the vacuum. This can be achieved in a variational manner \cite{Michael:1985ne,Luscher:1990ck} by solving a generalized eigenvalue problem~\cite{Blossier:2009kd},
\begin{equation}
C(t) v^\mathfrak{n}(t) = \lambda_\mathfrak{n}(t) C(t_0) v^\mathfrak{n}(t), \label{GEVP}
\end{equation}
where the eigenvalues $\lambda_\mathfrak{n} \sim e^{- E_\mathfrak{n}(t-t_0)}$ are fitted to determine the state energy, $E_\mathfrak{n}$, and where the eigenvectors provide the weights in construction of the optimal operators, $\Omega_\mathfrak{n}^\dag \sim \sum_i v^\mathfrak{n}_i \mathcal{O}^\dag_i$. The eigenvectors, which we extract independently on each timeslice, should be constant in time for $t > t_0$. They are related to the matrix elements, $\big\langle \mathfrak{n} \big| \mathcal{O}^\dag_i(0) \big| 0 \big\rangle$, whose relative values can provide some information on the internal structure of each eigenstate.

The spectral decomposition in Eq.~\ref{decomp} is strictly complete only in the limit that the time-extent of the lattice is infinite, $T\to \infty$. For finite values of $T$, there are small additional contributions which enter with amplitudes suppressed by a factor which is at worst $e^{- m_\pi T}$. As discussed in \cite{Dudek:2012gj} we can remove these, without invalidating any of the requirements for a variational solution, using a procedure of weighting the correlators with an appropriate exponential time dependence before forming the shifted correlator, $C(t) - C(t+\delta t)$.

The Hadron Spectrum Collaboration has utilized these techniques previously to extract a large number of excited states. See Refs.~\cite{Dudek:2007wv, Dudek:2010wm} where further implementation details can be found.

\subsection{Operator construction}
\label{sec_operators}

In Ref.~\cite{Dudek:2012xn}, which considered the $\rho$ resonance in $\pi\pi$ elastic scattering, it was found that in order to reliably extract the complete low-energy spectrum of finite-volume eigenstates, operators resembling both single-hadrons and multi-hadrons must be included in the basis. In a series of papers we have developed general methods for constructing such operators~\cite{Dudek:2009qf,Dudek:2010wm,Thomas:2011rh,Dudek:2012gj} having a range of different spin and spatial structures, respecting the symmetries of a finite-volume cubic lattice.  This technology has proven effective in applications to $\pi\pi$ scattering~\cite{Dudek:2010ew,Dudek:2012gj,Dudek:2012xn} -- we use analogous constructions in this work and so refer to Ref.~\cite{Dudek:2012xn} for a more extensive summary and to the aforementioned references for details of the constructions.  The only differences in the current work are in the flavor structure of the operators and in the particular combinations of momenta used to construct ``meson-meson"-like operators.

Our ``single-meson'' operators, projected onto definite quantised momentum, $\vec{k}$, are fermion bilinears 
$\sum_{\vec{x}}  e^{i\vec{k}\cdot\vec{x}} \sum_{ij}w_{ij} \, \bar{q}_i(\vec{x};t){\bf \Gamma}_t q_j(\vec{x};t)$, where the ${\bf \Gamma}_t$ are operators acting in space, color and Dirac spin-space on a time-slice, $t$, containing a Dirac gamma matrix structure combined with some number of gauge-covariant derivatives. The quark fields $q_i$ include the up, down and strange quarks, $\left[u, d, s\right]$. The sum over the quark field labels and weights $w_{ij}$ project the bilinear into a $SU(3)_F$ flavor representation with strangeness and total isospin, $(S,I)$, and $z$-component of the isospin, $I_z$. Examples of the isospin construction include strangeness--0, isospin--0 states with ${\bf 8}_F$ or ${\bf 1}_F$ in flavor. In this case, the isospin weights are diagonal in flavor with  $w={\rm diag}(\tfrac{1}{\sqrt{6}},\tfrac{1}{\sqrt{6}},\tfrac{-2}{\sqrt{6}})$ and ${\rm diag}(\tfrac{1}{\sqrt{3}},\tfrac{1}{\sqrt{3}},\tfrac{1}{\sqrt{3}})$, respectively. Since we consider $u$ and $d$ quarks which are lighter than the strange quark, $SU(3)_F$ is not exact, and the optimal operator to interpolate the lightest $S=0, I=0$ state, the $\eta$ meson, may be a linear superposition of these two flavor constructions. 

Calculations using these operators for light isovectors and kaons are discussed in Refs.~\cite{Dudek:2009qf,Dudek:2010wm} and for light isoscalars in Refs.~\cite{Dudek:2011tt,Dudek:2013yja}.  These operators are constructed to have definite continuum $J^P$ and then, for $\vec{k} = \vec{0}$, their various $J_z$ components are subduced into the relevant irreps of the octahedral group.  For $\vec{k} \neq \vec{0}$, from the $J^P$ operators we first form operators with definite helicity, $\lambda$, and then subduce the components into irreps of the little group, $\textrm{LG}(\vec{k})$~\cite{Thomas:2011rh}.  At both zero and non-zero momentum, the result is an operator labelled by the lattice irrep, $\Lambda$, and irrep row, $\mu$. The octahedral group construction of the fermion bilinears is done independently of the flavor representation.

In this study we construct multi-meson operators from products of operators for pseudoscalar $\pi$, $K$ and $\eta$ mesons.  The ``single-meson'' operators are in the one-dimensional $\Lambda^P = A_1^-$ irrep at rest and the one-dimensional $A_2$ irrep for all the non-zero momenta we consider; in addition, the $\pi$ and $\eta$ operators have negative and positive $G$-parity respectively.

We use ``optimized'' operators, constructed as the optimal linear combination of ``single-meson'' operators to interpolate each ground-state pseudoscalar, allowing us to perform analyses at smaller Euclidean times. For operators at rest, up to three spatial derivatives are used, while in-flight, up to two derivatives are used. The optimal linear combinations of operators to interpolate the $\eta$, containing both octet and singlet components, are obtained from a variational analysis~\cite{Dudek:2013yja} and the weights of the dominant octet constructions are used to form our projected $\eta$ operators -- this is done independently in each moving frame. The efficacy of the optimized operator procedure was demonstrated in Ref.~\cite{Dudek:2012gj} -- as a short hand we represent them by $\pi(\vec{k})$, $K(\vec{k})$ and $\eta(\vec{k})$. 

Following Ref.~\cite{Dudek:2012gj} we construct a general $\pi K$ creation operator as,
\begin{equation}
\label{eq_op_meson_pair}
\left(\pi K\right)^{\left[\vec{k}_1,\vec{k}_2\right]\dagger}_{\vec{P},\Lambda,\mu} = \sum_{\substack{\vec{k}_1 \in \{\vec{k}_1\}^{\star} \\ \vec{k}_2 \in \{\vec{k}_2\}^{\star} \\ \vec{k}_1 + \vec{k}_2 = \vec{P}}}
\!\!\!\mathcal{C}(\vec{P},\Lambda,\mu;\; \vec{k}_1;\,\vec{k}_2)\; \pi^{\dagger}(\vec{k}_1)\; K^{\dagger}(\vec{k}_2) ~,
\end{equation}
where the operator has overall momentum $\vec{P}$ and is in irrep $\Lambda$ (row $\mu$) of $\textrm{LG}(\vec{P})$.  For clarity we have suppressed the sum over isospin components to give total $I=\frac{1}{2}$ or $\frac{3}{2}$.  An exactly analogous construction is used to build a $\eta K$ operator: $\pi^\dagger(\vec{k})$ is replaced with $\eta^\dagger(\vec{k})$.  In this equation $\mathcal{C}$ is a generalised Clebsch-Gordan coefficient for $\Lambda_1 \otimes \Lambda_2 \rightarrow \Lambda$ with $\Lambda_{i} = A_1^-$ of $O_h^D$ if $\vec{k}_{i} = \vec{0}$ and $\Lambda_{i} = A_2$ of $\textrm{LG}(\vec{k}_{i})$ if $\vec{k}_{i} \neq \vec{0}$.  Subject to the constraint that $\vec{k}_1 + \vec{k}_2 = \vec{P}$, the sum over $\vec{k}_{i}$ is over all momenta in the \emph{star} of $\vec{k}_{i}$, denoted $\{\vec{k}_{i}\}^\star$, i.e. all momenta related to $\vec{k}_{i}$ by an allowed lattice rotation (in the cases we consider this is all momenta of magnitude $|\vec{k}_i|$).  Ref.~\cite{Dudek:2012gj} gives further details and explicit values of $\mathcal{C}$. In some cases we will use a shorthand notation where we label operators by $|\vec{k}|^2$; for example $\pi_1 K_2$ indicates $\vec{k}_1 = [001],\, \vec{k}_2 = [011]$.

In this study we extract spectra for the hadronic system with overall momentum $\vec{P} = [0,0,0]$, $[0,0,1]$, $[0,1,1]$, $[1,1,1]$ and $[0,0,2]$.  The combinations of ${\vec{k}_1,\,\vec{k}_2}$ used to construct $\pi K$ and $\eta K$ operators for $I=\frac{1}{2}$ and $\pi K$ operators for $I=\frac{3}{2}$ are given in 
Tables~\ref{table_ops}, \ref{ops_24}, \ref{ops_20}, \ref{ops_16} and \ref{ops_I3o2} of Appendix~\ref{app_op_tabs}.

In this calculation we do not include three-meson (or higher) operator constructions, nor do we include \emph{local} $qq\bar{q}\bar{q}$ (or higher) constructions. While all the operator constructions we have chosen should have some overlap with all states of a given quantum number, the overlap may be too small for adequate resolution via the variational method, in which case the obtained energy spectrum may not be precisely determined. This situation can happen in the energy region above a multi-meson threshold, and was observed and discussed in the study of isospin--1 $\pi\pi$ in Ref.~\cite{Dudek:2012xn}. The analogous situation in this work is the opening of three-meson thresholds, the lowest of which is $\pi\pi K$. We will comment more on the implications of such thresholds in later sections of the paper.

\subsection{Correlator construction}
\label{sec_correlators}

We make use of the \emph{distillation} framework \cite{Peardon:2009gh}
to evaluate the two-point correlation functions constructed from the
operators defined in the previous section. Distillation is a
quark-field smearing method that is designed to increase overlap onto
the low modes relevant in low-lying hadronic states. We define a smearing
operator on a time-slice, $t$, which acts in 3-space, $\vec{x}$, and color
space, $a$,
\begin{equation}
\Box(\vec{x}a,\vec{y}b;t) =
\sum_{n=1}^N\xi_n(\vec{x}\,a;t)\;\xi^\dag_n(\vec{y}\,b;t), \nonumber
\end{equation}
where we choose the fields $\{\xi_n\}$ to be the lowest $N$ eigenvectors
of the gauge-covariant Laplacian on time-slice $t$. The smearing of
the quark fields in a correlation function can be factorized allowing
the ``perambulators'', 
the combination of eigenvectors and the inverse of the lattice representation of the Dirac matrix, ${\cal M}_q^{-1}$, 
$\xi^\dag_n(t'){\cal M}_q^{-1}(t',t)\xi_m(t)\equiv \tau^{[q]}_{nm}(t',t)$,
to be constructed as matrices in distillation space for each quark, $q$. Similarly, the
quark smeared 
operators presented in Sec.~\ref{sec_operators}
can be factorized into a matrix representation in distillation space,
$(\xi^\dag_n(t) {\bf \Gamma}_t\xi_m(t))_{\vec{k}}\equiv\Phi_{nm}(\vec{k};t)$.
The resulting
correlation function traces are over the set of eigenvectors, which is much smaller than the full
lattice space. The perambulators used in this work,
corresponding to the light and strange quark Dirac inversions, 
have been previously computed and reused in several other computations
which also spell out the advantages of the method \cite{Dudek:2009qf,Dudek:2010wm,Dudek:2012gj,Dudek:2012xn}.
Some details are provided in Table~\ref{tab:lattices}.

\begin{figure}
\includegraphics[width=0.35\textwidth]{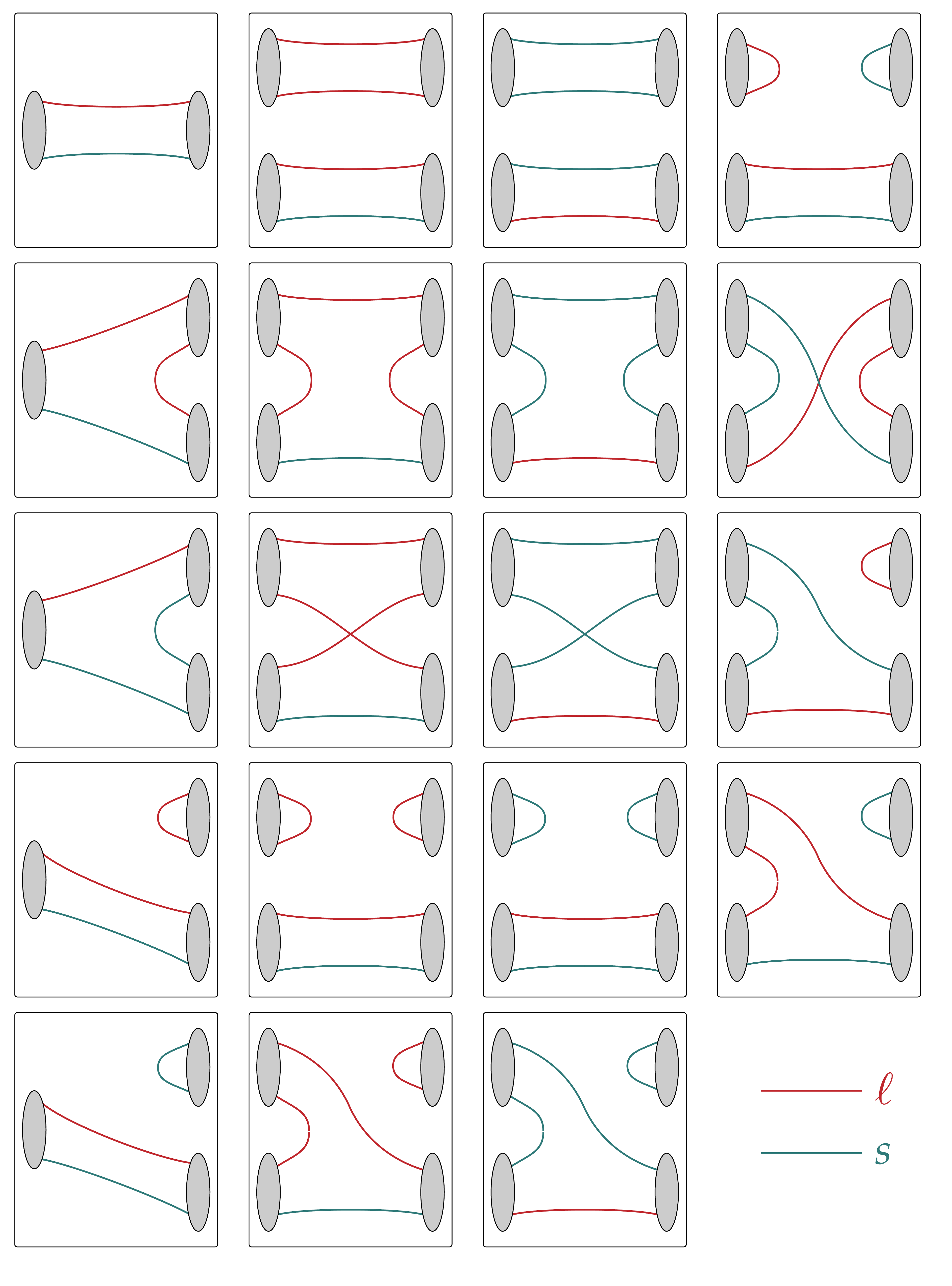}
\caption{Schematic Wick contractions required to compute correlation functions with the $\pi K$, $\eta K$ and kaon ``single-meson" operators described in the text. Also required is each of these contractions with source $\leftrightarrow$ sink.}
\label{wicks}
\end{figure}

Construction of correlators in isospin--3/2 necessarily involves only $\pi K$ operator constructions, while in isospin--1/2, we include ``single-meson'' operators as well as $\pi K$ and $\eta K$ multi-meson operator constructions as in Eq.~\ref{eq_op_meson_pair}. In the case of ``meson-meson" operators at both source and sink, the correlation function takes the form
%
\begin{align}
\sum_{\vec{k}_1,\vec{k}_2}&\mathcal{C}^\ast(\vec{P},\Lambda,\mu;\; \vec{k}_1;\,\vec{k}_2)\;
\sum_{\vec{k}_3,\vec{k}_4}\mathcal{C}(\vec{P},\Lambda,\mu;\;
\vec{k}_3;\,\vec{k}_4)\; \nonumber\\
&\times \Big\langle \left(\bar{q} \Box w^A\,\mathbf{\Gamma}_t^A\Box q\right)_{\vec{k}_1}
\left(\bar{q} \Box w^B\,\mathbf{\Gamma}_t^B\Box q\right)_{\vec{k}_2}
\nonumber\\
&\quad\quad \times\!  \left(\bar{q} \Box w^C\,\mathbf{\Gamma}_0^C \Box q\right)_{\vec{k}_3}^\dag\; 
\left(\bar{q} \Box w^D\,\mathbf{\Gamma}_0^D \Box q\right)_{\vec{k}_4}^\dag \Big\rangle ,\label{eq:projcorr}
\end{align}
%
where the time source is at time-slice 0, and the annihilation operator is at time-slice, $t$. The bilinears are projected into an appropriate flavor representation with weights $w$. As indicated in Table~\ref{tab:lattices}, we will average over several time-sources. The sums over Clebsch-Gordan coefficients project the creation operator, featuring the single particle constructions of operators $C$ and $D$, and the annihilation operators $A$ and $B$, onto total momentum $\vec{P}$. Integration of the quark fields, replacing the up and down quark labels with a light quark $\ell$, leads to terms featuring Wick contractions which include those of form
\begin{align}
{\rm Tr}\Big[\tau^{[q_1]}(0,t)&\Phi^A(\vec{k}_1; t)\tau^{[q_2]}(t,t)\Phi^B(\vec{k}_2;t)\nonumber\\
&\times \ \tau^{[q_3]}(t,0)\Phi^{C\dag }(\vec{k}_3;0)\tau^{[q_4]}(0,0)\Phi^{D\dag }(\vec{k}_4;0)\Big],
\nonumber
\end{align}
where the trace is over the distillation and Dirac spin indices with perambulators of some quark flavor $q_i$, either $\ell$ or $s$. A schematic representation of the required ``single-meson" and ``meson-meson" operator contractions is shown in Figure~\ref{wicks}. For correlators with $\eta K$ constructions at source and sink, there are 20 such diagrams. The Clebsch-Gordan projection of the creation and annihilation operators onto definite momentum imply there are many such sums for each set of momentum $\vec{k}_i$. The largest number of such pairs is 12 for momentum type  $\vec{k}=[011]$ projected onto total momentum $\vec{P}=\vec{0}$. In this case, there are 2880 diagrams after the Wick contractions of Eq.~\ref{eq:projcorr}. When evaluating correlation functions, we include all required Wick contractions.

\subsection{Typical determined spectra}

\begin{figure*}
\includegraphics[width=0.95\textwidth]{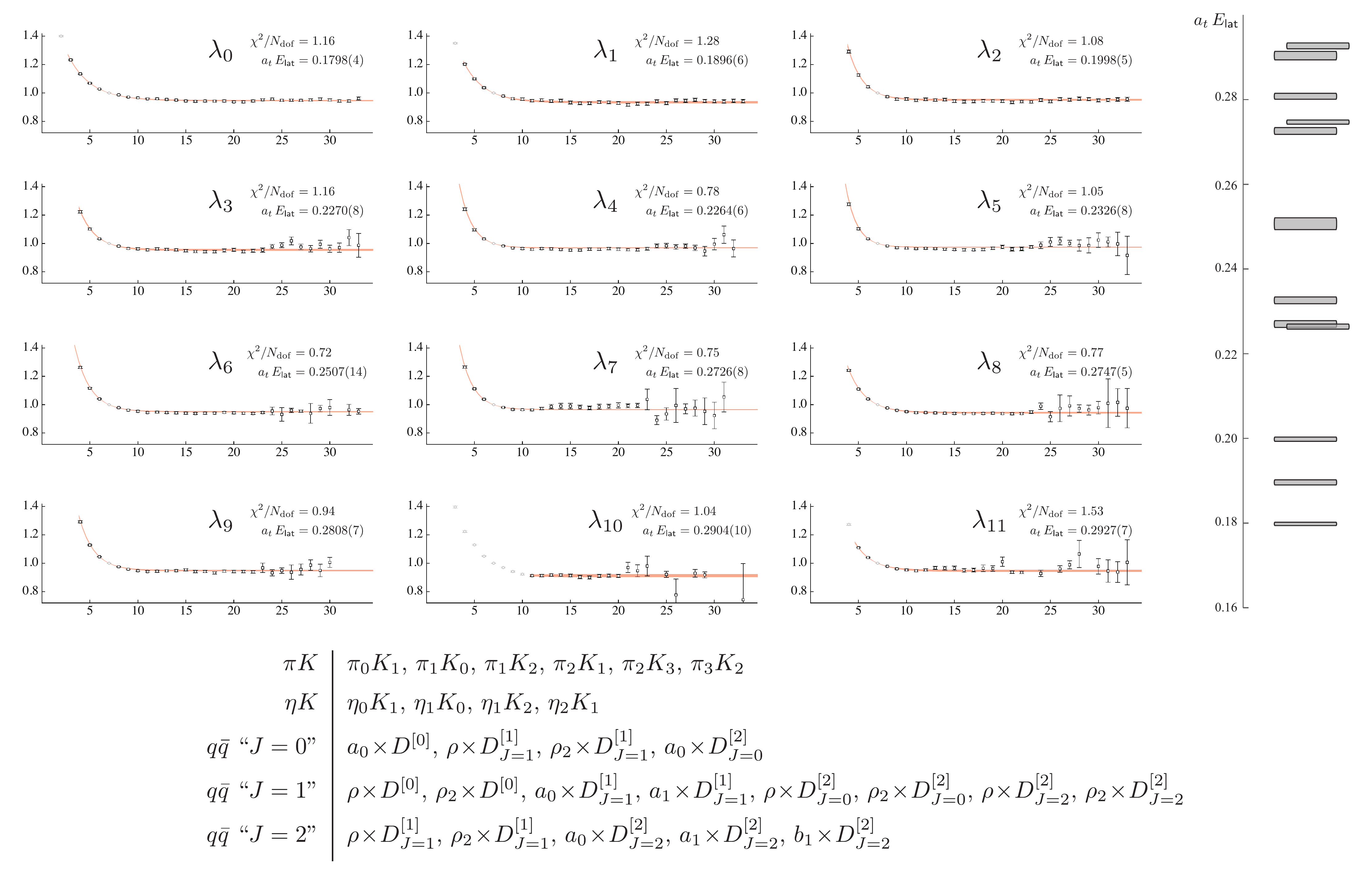}
\caption{Eigenvalues of Eq.~\ref{GEVP}, $\lambda_\mathfrak{n}(t)$, in the isospin--1/2 case $\vec{P}=[001]$, $\Lambda = A_1$ on the $24^3$ lattice. Plotted as $e^{E_\mathfrak{n}(t-t_0)} \lambda_\mathfrak{n}(t)$ are the data points and a timeslice-correlated fit of the form $\lambda_\mathfrak{n}(t) = (1-A_\mathfrak{n}) e^{-E_\mathfrak{n}(t-t_0)} + A_\mathfrak{n} e^{-E_\mathfrak{n}'(t-t_0)}$. The set of 27 operators used in the variational basis is listed beneath the plot.}
\label{D4A1_prin_corrs}
\end{figure*}

As an example of the quality of determined spectra we present in Figure \ref{D4A1_prin_corrs} the eigenvalues of Eq.~\ref{GEVP}, $\lambda_\mathfrak{n}(t)$, for the lowest 12 states in the isospin--1/2 $[001]\,A_1\, (24^3)$ channel extracted from the 27--dimensional correlation matrix built using the operator basis listed in the figure. We are clearly able to obtain a detailed spectrum, including near-degenerate states, with statistical precision on the energy values at or below 1\%. 

\begin{figure}
\includegraphics[width=0.5\textwidth]{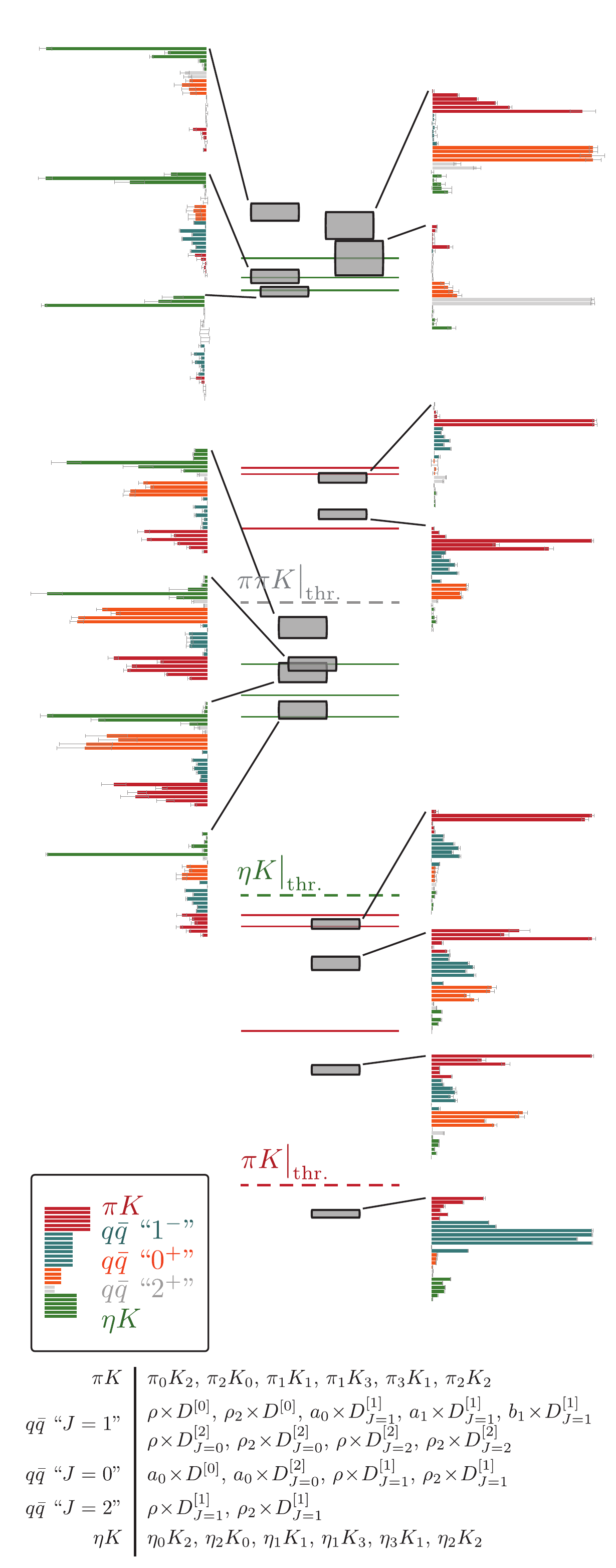}
\caption{The spectrum and relative operator overlaps with overall momentum $[011]$ in the $A_1$ irrep on the $24^3$ lattice for isospin--1/2. The grey boxes represent finite volume energy levels. The thin solid lines in the center indicate non-interacting energy levels while the dashed lines show kinematic thresholds. }
\label{D2A1_histo}
\end{figure}

The matrix elements $\big\langle \mathfrak{n} \big| \mathcal{O}^\dag_i(0) \big| 0 \big\rangle$ are also well determined in the solution of Eq. \ref{GEVP}, and their relative size can give us some insight into the make-up of the states in our excited spectrum. As an example, we show in Figure \ref{D2A1_histo}, the spectrum and relative overlap matrix-elements (normalized as in \cite{Dudek:2009qf}) of the lowest 15 states in the $[011]\, A_1\, (24^3)$ channel extracted from a 27--dimensional correlation matrix. In the main we observe a separation between states with significant overlap onto $\pi K$ operators from those with overlap onto $\eta K$ operators. This likely reflects the relatively small breaking of $SU(3)$ flavor symmetry in our calculation with $m_\pi = 391\,\mathrm{MeV},\, m_K = 549\,\mathrm{MeV}, \, m_\eta = 589 \,\mathrm{MeV}$. With $SU(3)$ flavor symmetry, the $J=0,2 \ldots$ channels have much reduced coupling to $\eta K$ compared to $\pi K$ \cite{Lipkin:1980tk, Aston:1987ey}. On the other hand, the $J=1,3\ldots$ channels have equal coupling to $\pi K, \, \eta K$, but since the first vector resonances above $\eta K$ threshold likely lie off the top of the scale we have presented, we are unlikely to see this coupling manifested. The origin of these $SU(3)$ flavor arguments is presented in Appendix \ref{app_su3f}.

If QCD were such that hadrons had no residual interactions, our ``meson-meson" operator basis would be diagonal, with for example an operator $\pi_{n_\pi^2} K_{n_K^2}$ producing an eigenstate of energy ${E^{\mathrm{n.i.}}_{\mathsf{cm}} = \sqrt{\big(E^{\mathrm{n.i.}}_{\mathsf{lat.}}\big)^2 - n^2_{\vec{P}}\!\left(\tfrac{2\pi}{L} \right)^2  }}$ where ${E^{\mathrm{n.i.}}_{\mathsf{lat.}} = \sqrt{m_\pi^2 + n_\pi^2 \!\left(\tfrac{2\pi}{L} \right)^2 } + \sqrt{m_K^2 + n_K^2 \!\left(\tfrac{2\pi}{L} \right)^2 }}$. If states appear in the spectrum that differ from these energies, there is some indication of interactions, including the possibility of resonances.

The presence in Fig.~\ref{D2A1_histo} of a state below $\pi K$ threshold, significantly below the first non-interacting $\pi K$ level, and which has strong overlap onto ``single-meson" operators, likely suggests a $J=1$ $K^\star$ state that is either bound, or barely above threshold. Above $\pi K$ threshold we observe several states displaced somewhat from non-interacting $\pi K$ positions, which show some degree of overlap onto both $\pi K$-like constructions and ``single-meson" operators. 
Above the $\eta K$ threshold, we observe states with strong overlap onto $\eta K$-like constructions lying quite close to non-interacting $\eta K$ positions. High in the spectrum, above $\pi\pi K$ threshold\footnote{but recall that we are not including $\pi\pi K$-like operators in the basis} a state is observed having strong overlap onto ``single-meson" operators which in the rest frame would overlap with $J=2$. These operators can overlap onto other $J$ at non-zero momentum -- a detailed discussion of this point can be found in \cite{Thomas:2011rh}.

While these qualitative features can guide us toward the resonant content of the theory, ultimately, rigorously correct determinations will come from a quantitative description of the scattering amplitudes which we can extract from the volume and frame dependence of the discrete spectra. In the next section we will describe how this can be achieved.

\section{Scattering amplitudes from finite-volume spectra}
\label{sec_fvs}

In order to connect the discrete finite-volume spectra obtained in our lattice calculation to infinite-volume scattering amplitudes, we make use of the formalism originally proposed by L\"uscher \cite{Luscher:1990ux} for elastic scattering in the rest-frame, and subsequently extended to in-flight systems~\cite{Rummukainen:1995vs,Kim:2005gf}, scattering of particles of unequal mass~\cite{Fu:2011xz,Leskovec:2012gb}, and multiple coupled-channels~\cite{He:2005ey,Hansen:2012tf,Briceno:2012yi,Guo:2012hv}.
For an $L \times L \times L$ box with periodic boundary conditions, the condition determining the spectrum in the irrep $\Lambda$, for a moving frame $\vec{P} =\tfrac{2\pi}{L}\vec{d}$, relevant to the case of any number of pseudoscalar-pseudoscalar scattering channels, can be expressed as
\begin{align}
\det\Big[&
\delta_{ij}\delta_{\ell\ell^\prime}\delta_{nn^\prime}  \nonumber \\
&\;+ i\rho^{}_i\, t^{(\ell)}_{i j}
\left(\delta_{\ell\ell^\prime} \delta_{nn^\prime}+
i  \mathcal{M}^{\vec{d},\Lambda}_{\ell n;\ell^\prime n^\prime}(q^2_i) \right)
\Big] = 0.
\label{eq_luescher_t}
\end{align}
In this expression the channels are labelled by an index $i$, with $\rho_i(E_\mathsf{cm}) = \frac{2 k_i}{E_\mathsf{cm}}$ the phase-space for that channel where $k_i$ is the momentum in the $\mathsf{cm}$ frame, ${k = \tfrac{1}{2 E_\mathsf{cm}}\big(E_\mathsf{cm}^2 - (m_1+m_2)^2 \big)^{1/2} \big(E_\mathsf{cm}^2 - (m_2- m_1)^2 \big)^{1/2} }$. The scattering amplitudes for partial wave $\ell$ appear in the $t$-matrix, $t_{ij}^{(\ell)}(E_\mathsf{cm})$. The matrix 
\begin{align}
\mathcal{M}^{\vec{d},\Lambda}_{\ell n;\ell^\prime n^\prime}\delta_{\Lambda \Lambda^\prime}\delta_{\mu\mu^\prime} 
= \mathcal{S}^{\vec{d}\Lambda\mu n}_{\ell m} \, \mathcal{M}^{\vec{d}}_{\ell m;\ell^\prime m^\prime}\,  \mathcal{S}^{\vec{d}\Lambda^\prime\mu^\prime n^\prime}_{\ell^\prime m^\prime}
\label{eq_msub}
\end{align}
is a known function of the dimensionless variable ${q^2_i = \left(\tfrac{k_i L}{2\pi}\right)^2}$. The angular-momentum basis $\mathcal{M}_{\ell m;\ell^\prime m^\prime}$ is projected into the appropriate little-group irreps, $\Lambda$, using the subduction matrices, $\mathcal{S}$, presented in \cite{Thomas:2011rh}. The index $n$ indicates the $n^\mathrm{th}$ subduction of partial-wave $\ell$ into irrep $\Lambda$ -- Table \ref{tab_pwa_irrep} presents the subduced angular-momentum content of each irrep. $\mathcal{M}^{\vec{d}}_{\ell m;\ell^\prime m^\prime}$ is as given in \cite{Leskovec:2012gb}, as the extension to unequal scattering masses of Eq.~89 of Ref.~\cite{Rummukainen:1995vs}.


In the case of elastic scattering, where only a single channel is open, scattering in partial-wave $\ell$ can be described by a single real energy-dependent parameter, the phase-shift, $\delta_\ell(E_\mathsf{cm})$, which appears in the scattering amplitude as $t^{(\ell)} = \frac{1}{\rho} e^{i\delta_\ell} \sin \delta_\ell$. If only a single partial-wave appears in the quantization condition, Eq. \ref{eq_luescher_t}, then for each finite-volume energy eigenvalue, $E_\mathfrak{n}$, a value of $\delta(E_\mathfrak{n})$ can be extracted by solving Eq. \ref{eq_luescher_t}. Unfortunately such a situation is never realized exactly -- the matrix in Eq. \ref{eq_msub} is formally a matrix of infinite dimension in $\ell$ and thus Eq. \ref{eq_luescher_t} is simultaneously a function of many $\delta_\ell$.

The difficulty is illustrated in Table~\ref{tab_pwa_irrep} which shows the lowest few $\ell$ values appearing in each irrep. For example, the at-rest $A_1^+$ irrep, which we might expect to be the cubic analogue of $\ell=0$, contains also $\ell=4$ and higher partial-waves. In-flight irreps are seen to be even more dense in the low-lying $\ell$-space. In practice, close to threshold, the angular-momentum barrier ensures that phase-shifts have the behavior, $\delta_\ell \sim k_\mathsf{cm}^{2\ell+1}$, which typically suppresses higher partial-waves relative to lower $\ell$ such that we are justified in truncating the number of partial-waves included\footnote{see for example \cite{Dudek:2012gj} where the role of higher partial waves was explored in $\pi\pi$ isospin--2 scattering.}.

At higher energies, as new two-body channels open up, the full form of Eq.~\ref{eq_luescher_t}, as the determinant of a matrix in both angular-momentum and channel space, becomes the relevant quantization condition determining the spectrum in a finite volume\footnote{a kinematically closed channel can have an effect on the quantization condition in a limited energy region below its threshold as the elements of $\mathcal{M}$ are not exactly zero below threshold, but rather decay exponentially to the constant required to decouple the channel.}. Given knowledge of the energy dependence of a scattering matrix, $t^{(\ell)}_{ij}(E_\mathsf{cm})$, one can solve this condition for a discrete spectrum, $\{E^{\vec{P},\Lambda}_\mathfrak{n}\}(L)$, in volume $L\times L\times L$. Of course the practical problem at hand  is the reverse of this, to find the $t$-matrix given a lattice QCD calculation of the spectrum $\{E^{\vec{P},\Lambda}_\mathfrak{n}\}(L)$. The challenge is that, even in the case of dominance of a single partial-wave, $\ell$, for each level $E_\mathfrak{n}$ the quantization condition contains multiple unknowns, namely the elements of the $t$-matrix. Even accounting for the constraints from \mbox{$S$-matrix} unitarity and time-reversal invariance, this is an under-constrained problem once more than one channel is open.

One approach to solving this problem is to parameterize the energy-dependence of $t^{(\ell)}_{ij}(E_\mathsf{cm})$ in a manner satisfying $S$-matrix unitarity and time-reversal invariance and to then attempt to describe the entire spectrum $\{ E_\mathfrak{n}\}$ simultaneously by varying free parameters in the parameterization. Describing the spectrum for a range of volumes and in several moving frames with a relatively small number of parameters allows us to build an over-constrained system. The effectiveness of the procedure was tested in a toy model in \cite{Guo:2012hv}. Including multiple partial-waves is straightforward: an independent parameterization is constructed for each $\ell$ and included in Eq.~\ref{eq_luescher_t}.

Explicitly we minimize a $\chi^2$ function describing the difference between the lattice QCD obtained spectra, $E_\mathsf{cm}(L;\vec{P}\Lambda \mathfrak{n})$, and the spectra corresponding to a particular scattering parameterization, 
\begin{widetext}
\begin{equation}
	\chi^2\big( \{ a_j \} \big) = \sum_L 	
					\sum_{\substack{\vec{P}\Lambda \mathfrak{n} \\ \vec{P}'\Lambda' \mathfrak{n}'}}
					\Big[ E_\mathsf{cm}(L;\vec{P}\Lambda \mathfrak{n}) - E^{\mathrm{par.}}_\mathsf{cm}(L;\vec{P}\Lambda \mathfrak{n}; \{a_j\}) \Big]	
		\mathbb{C}^{-1}\big(L; 	\vec{P}\Lambda \mathfrak{n}; \vec{P}' \Lambda' \mathfrak{n}'	\big)		
		\Big[ E_\mathsf{cm}(L;\vec{P}'\Lambda' \mathfrak{n}') - E^{\mathrm{par.}}_\mathsf{cm}(L;\vec{P}'\Lambda' \mathfrak{n}'; \{a_j\}) \Big]				
\end{equation}
where $E^{\mathrm{par.}}_\mathsf{cm}(L;\vec{P}\Lambda \mathfrak{n}; \{a_j\})$ is the $\mathfrak{n}^\mathrm{th}$ solution of Eq.~\ref{eq_luescher_t} with a parameterized $t$-matrix depending upon parameters $\{ a_j\}$. Data covariance, $\mathbb{C}$, whose off-diagonal elements between energies evaluated on the same ensemble can be non-zero, can be estimated using jackknife.
\end{widetext}

\subsection{t-matrix parameterizations}

In parameterizing scattering amplitudes, as well as ensuring that $S$-matrix unitarity is respected, we should aim to use forms which can be analytically continued in the complex $s=E_\mathsf{cm}^2$ plane. This will allow us to examine the resulting amplitudes for poles, argued to be the least model-dependent way to describe bound-states and resonances.

In the case of elastic scattering, two convenient parameterizations are the effective range expansion and the relativistic Breit-Wigner. The effective-range expansion,
\begin{align}
k_\mathsf{cm}^{2\ell+1}\cot\delta_\ell = \frac{1}{a_\ell} + \frac{1}{2}r_\ell k^2_\mathsf{cm} +\mathcal{O}\!\left(k_\mathsf{cm}^4\right),
\label{eq_ere}
\end{align}
builds in the correct threshold behavior imposed by the angular-momentum barrier and characterizes the scattering by a series of constants, the first two of which, $a_\ell, r_\ell$, are known as the scattering length and the effective range. This parameterization is quite flexible, being capable of describing repulsive scattering, the presence of a bound-state or even a resonance. 

A common procedure to describe an elastic resonance is to use the relativistic Breit-Wigner form,
\begin{align}
t^{(\ell)}(s) =\frac{1}{\rho(s)} \frac{ \sqrt{s}\, \Gamma_\ell(s)}{m_R^2-s-i \sqrt{s}\,  \Gamma_\ell(s)},
\label{eq_bw}
\end{align}
where $m_R$ is the ``Breit-Wigner mass", and $\Gamma_\ell(s)$ is the energy-dependent width which may be parameterized in a form that ensures the correct behavior near threshold, $\Gamma_\ell(s)=\frac{g_R^2}{6\pi}\frac{k_\cm^{2\ell+1}}{s\, m_R^{2\left(\ell-1\right)}}$, with $g_R$ being a coupling. More sophisticated forms for the width capable of damping out the $k_\cm^{2\ell+1}$ behavior well above the threshold were discussed in \cite{Dudek:2012xn}.


In order for Eq.~\ref{eq_luescher_t} to have solutions we require our parameterizations satisfy $S$-matrix unitarity -- this is somewhat harder to ensure in the coupled-channel case than in the elastic case. One very convenient method is to use the $K$-matrix formalism in which we express the elements of the inverse of the $t$-matrix for partial-wave $\ell$ as
\begin{equation}
	t^{-1}_{ij}(s) = \frac{1}{(2k_i)^\ell} K^{-1}_{ij}(s) \frac{1}{(2k_j)^\ell} + I_{ij}(s).
\label{eq_t_matrix_k}
\end{equation} 

The factors $(2k_i)^{-\ell}$ ensure the correct behavior at kinematic thresholds \cite{Guo:2010gx}, while $K(s)$ is a real symmetric matrix to be parameterized. $S$-matrix unitarity is ensured if $\mathrm{Im}\, {I}_{ij}(s) = -\delta_{ij}\, \rho_i(s)$ for energies above the kinematic threshold in channel $i$, and $\mathrm{Im}\, {I}_{ij}(s) = 0$ below the threshold. There is however some flexibility in the choice of the real part of $I(s)$, with the simplest option being to set it equal to zero above threshold. A choice which captures more of the correct analytic properties of scattering amplitudes, known as the Chew-Mandelstam prescription~\cite{Chew:1960iv}, relates the real part to the imaginary part using a dispersion relation -- our implementation is described in Appendix~\ref{app_chew_man}.

The main freedom in this method lies in the parameterization of the $K$-matrix -- a simple choice, which can accommodate a wide range of scattering behaviors, is to construct it from a sum of poles plus a polynomial in $s$,
\begin{align}
K_{ij}(s) = \sum_p\frac{g^{(p)}_{i} g^{(p)}_{j}}{m_p^2-s} + \sum_n\gamma_{ij}^{(n)}s^n ,
\label{k-matrix}
\end{align}
where $g^{(p)}_{i}$ are real ``couplings" for pole $p$ in channel $i$, the $m_p$ are real, and $\gamma_{ij}^{(n)}$ form constant real symmetric matrices. The presence of poles in the $K$-matrix does not guarantee that the $t$-matrix will have poles close to the real-$s$ axis, but including a $K$-matrix pole is often an efficient way to describe a $t$-matrix pole if one needs to be present. 

Another alternative is to parameterize the inverse of the $K$-matrix as a symmetric matrix of polynomials,
\begin{align}
K^{-1}_{ij}(s) = \sum_{n=0}^{N_{ij}} c_{ij}^{(n)} s^n ,
\label{k-poly-inv}
\end{align}
with $c_{ij}^{(n)}$ being real parameters.

\section{$\pi K$, $\eta K$ coupled-channel scattering in isospin--1/2}
\label{sec_one_half}

\begin{figure}
\includegraphics[width=0.48\textwidth]{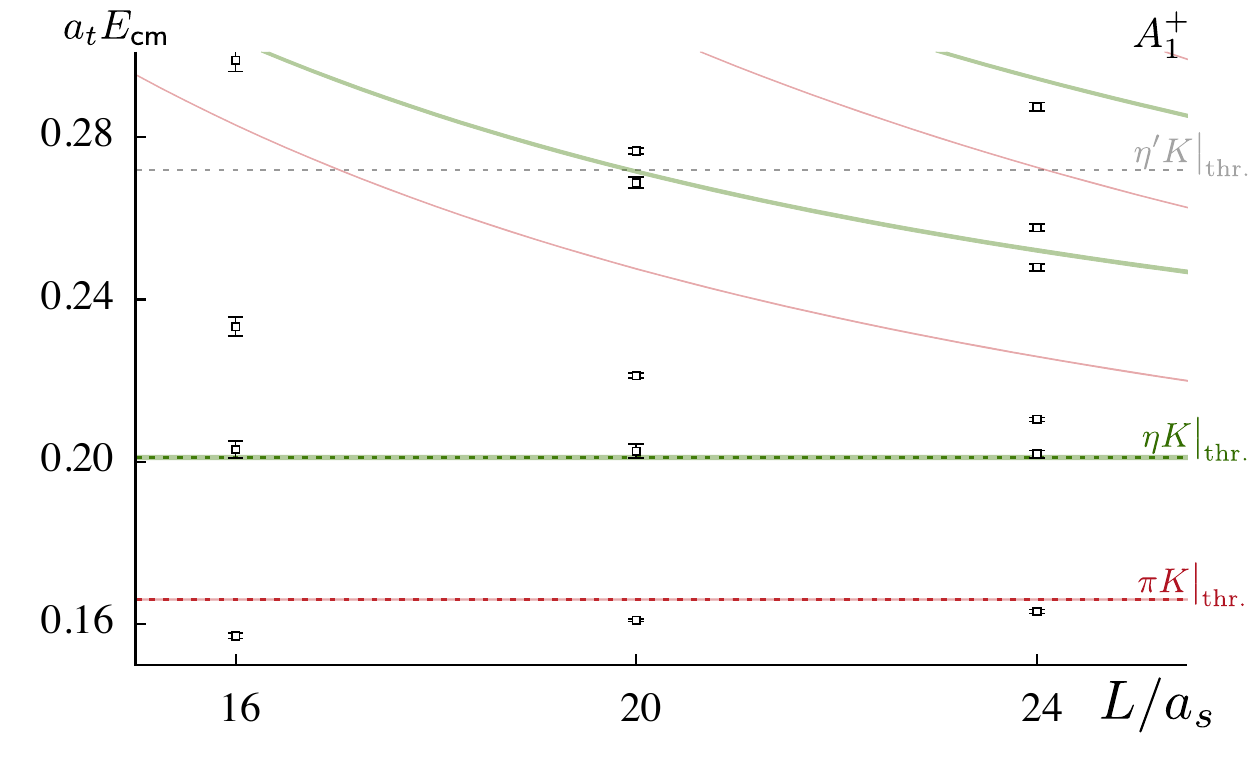}
\caption{$\vec{P}=[000]$ $A_1^+$ spectrum. The data points are the energies obtained from variational analysis of a correlation matrix featuring up to 8 ``single-meson" and up to 6 ``meson-meson" operators at $L/a_s = 16,\, 20,\, 24$. The red bands are the $\pi K$ non--interacting level positions, whilst the green bands represent the $\eta K$ non--interacting level positions (the width of the bands follows from the uncertainty on the meson masses). The dashed grey line shows the $\eta' K$ threshold.}
\label{spectrum_I1o2_P000_A1}
\end{figure}

Utilizing the methods described in Section \ref{sec_spec_determinations} we obtain matrices of correlation functions in a large number of irreps with $|\vec{P}|^2 \le 4$. Each of these are analyzed independently using the variational method and the energy levels obtained potentially provide information on the partial-waves subduced into that irrep. We begin by considering the $A_1^+$ irrep at rest, which is likely to be dominated by $\ell = 0$ at low energies, with the next lowest partial-wave, $\ell=4$, being heavily suppressed by the angular-momentum barrier.

\begin{figure*}
\includegraphics[width=0.78\textwidth]{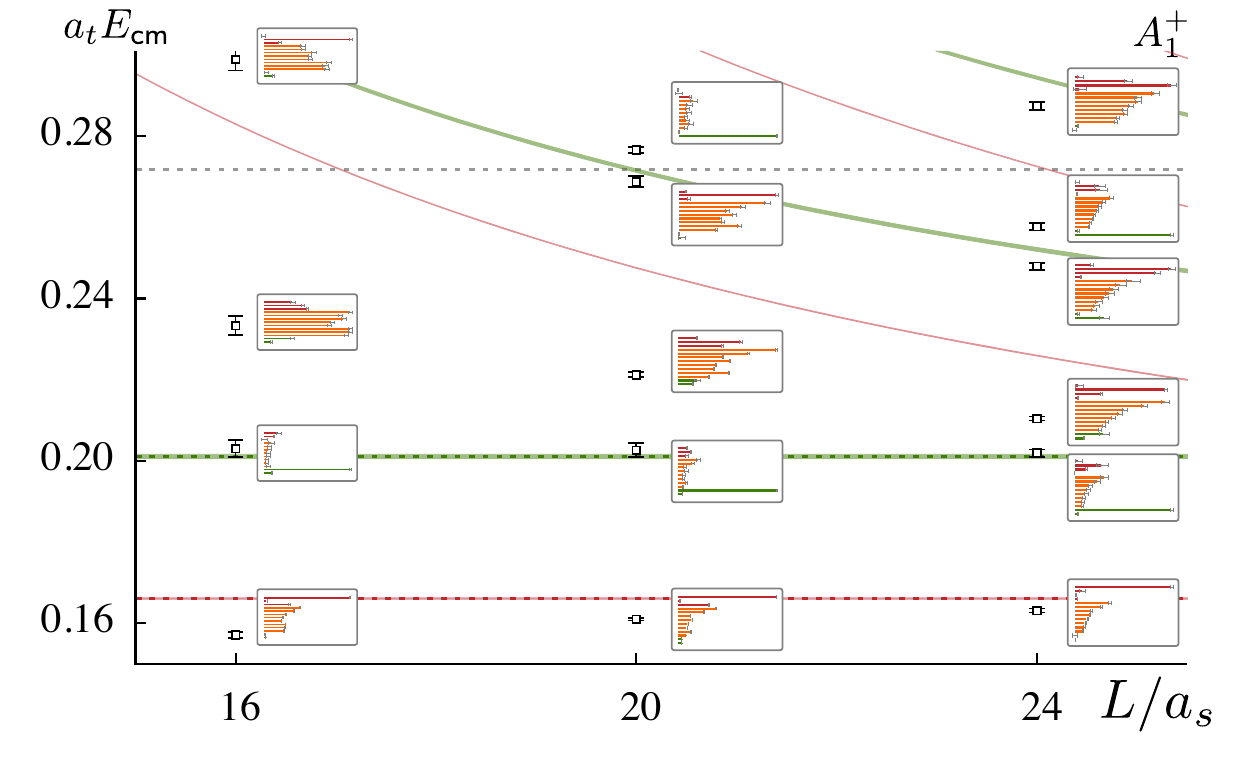}
\caption{$\vec{P}=[000]$ $A_1^+$ spectrum. For each state we show a histogram indicating the relative value of overlap $\langle \mathfrak{n} | \mathcal{O}_i | 0 \rangle$ for each operator in the basis: $\pi K$(red), ``single-hadron"(orange) and $\eta K$(green). }
\label{spectrum_I1o2_P000_A1_histo}
\end{figure*}

\subsection{$S$-wave at rest}

In Fig.~\ref{spectrum_I1o2_P000_A1} we show the spectrum of finite-volume eigenstates in the $\vec{P} = [000]$, $A_1^+$ irrep. Our use of three volumes provides 15 energy levels in the region of interest, between $\pi K$ and $\pi\pi\pi K$ thresholds. 

Before attempting a description in terms of coupled $\pi K, \eta K$ scattering amplitudes, we may examine the qualitative features of the spectrum in Fig.~\ref{spectrum_I1o2_P000_A1}. We note that there is always a state below $\pi K$ threshold -- these overlap strongly onto the operator $\pi_0 K_0$ (see Fig.~\ref{spectrum_I1o2_P000_A1_histo}) and likely indicate that $\pi K$ in $S$-wave is attractive at low energy. The presence of levels close to, but slightly above, each $\eta K$ non-interacting level may be interpretable as a weak, repulsive interaction in $\eta K$ $S$-wave scattering. At each volume there is clearly an ``additional" state beyond the number expected on the basis of non-interacting meson pairs that appears between $a_t E_\mathsf{cm} = 0.20$ and $0.24$. The position of this level, which has significant overlap onto the ``single-meson" operators in our basis, as well as to \mbox{$\pi K$-like} operators (see Fig.~\ref{spectrum_I1o2_P000_A1_histo}), is strongly volume dependent. This may be an indication of a broad scalar resonance coupling to $\pi K$. This qualitative description does not suggest strong coupling between the $\pi K$ and $\eta K$ channels. We include three points which lie slightly above the $\eta' K$ threshold without having included $\eta' K$-like operators in the variational basis. We will proceed assuming that these levels are reliable and we will not consider $\eta' K$ to be an open channel in the $t$-matrix.

We will explore a parameterization for the coupled $\pi K, \eta K$ scattering matrix in $S$-wave which has sufficient freedom to describe the presence of resonances which may couple to one or both channels, as well as non-resonant features including repulsion. We consider a simple \mbox{$K$-matrix} representation (c.f. Eq.~\ref{k-matrix}), including a single pole plus a constant term:
\begin{align}
K=\frac{1}{m^2-s}&\begin{bmatrix}
g_{\pi K}^2 & g_{\pi K}\;g_{\eta K} \\
g_{\pi K}\;g_{\eta K} & g_{\eta K}^2
\end{bmatrix}
+\begin{bmatrix}
\gamma_{\pi  K, \pi  K} & \gamma_{\pi K,  \eta K} \\
\gamma_{\pi  K, \eta K} & \gamma_{\eta K, \eta K}
\end{bmatrix}.
\label{eq_kma_S_exact}
\end{align}
The resulting $t$-matrix is constructed using the Chew-Mandelstam phase-space with ${\mathrm{Re}\,  I_{ij}(s =m^2) = 0}$ (see Appendix \ref{app_chew_man} for more details). Were this parameterization, with its six free parameters, to prove incapable of describing the data, it could be augmented with additional poles or a higher-order polynomial. Should parameters be redundant, this should be visible in the parameter correlation matrix. Later in the manuscript we will consider a broader set of possible parameterizations. 

\begin{figure}
\includegraphics[width=1.0\columnwidth]{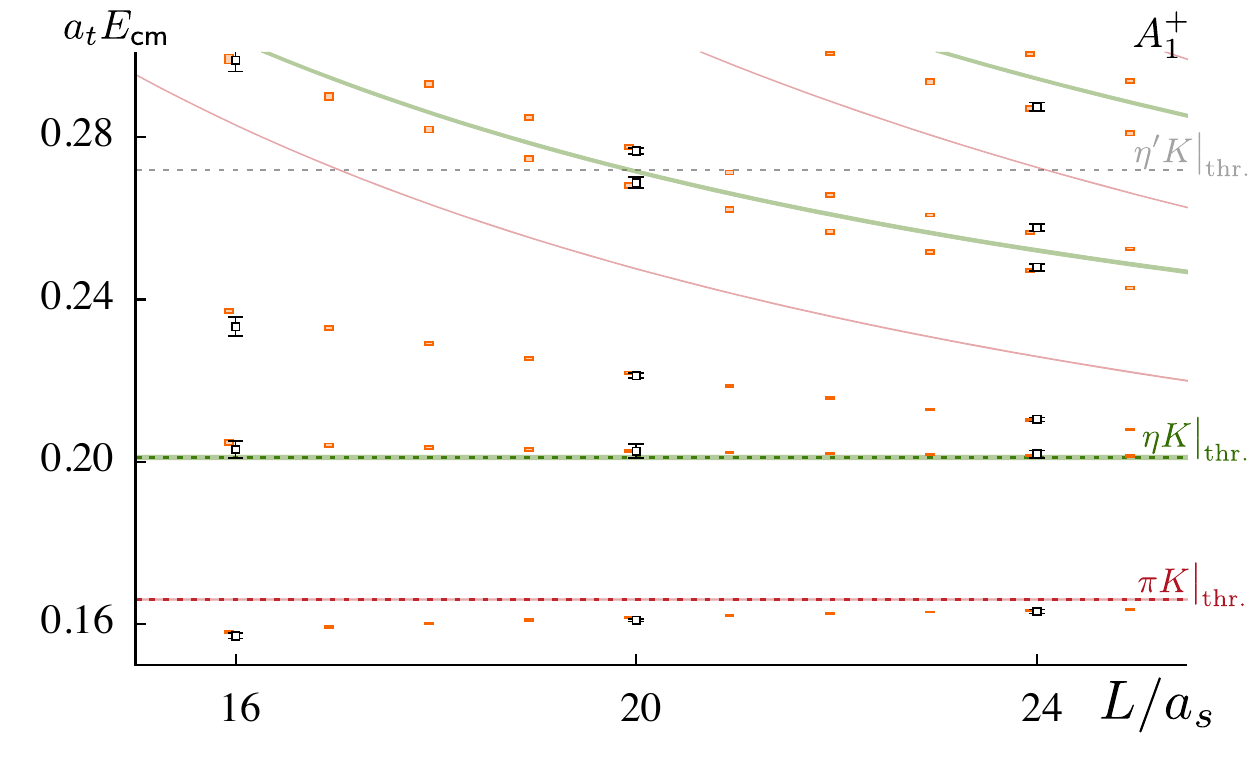}
\caption{$\vec{P}=[000]$ $A_1^+$ spectrum. 
Orange boxes: spectrum at each integer value of $L/a_s$ obtained by solving Eq. \ref{eq_luescher_t} for the parameterization in Eq. \ref{eq_kma_S_exact} with parameter values given by Eq. \ref{A1_fit_par_values}; the parameter errors and correlations are propagated through the calculation with the resulting uncertainty on the energy shown by the vertical size of the box. Original lattice QCD spectrum shown in black.}
\label{fig_swave_rest_model}
\end{figure}

As described in Section \ref{sec_fvs}, we minimize a $\chi^2$, varying the free parameters in the model, until the best agreement is obtained between the energy levels from the variational description of lattice QCD correlation functions, shown in Fig.~\ref{spectrum_I1o2_P000_A1}, and the discrete set of energies that satisfy Eq.~\ref{eq_luescher_t} for a given model $t$-matrix. The result of this fit is
\begin{widetext}
\begin{center}
\begin{tabular}{rll}
$m =$                         & $(0.2466 \pm 0.0020 \pm 0.0009) \cdot a_t^{-1}$   & 
\multirow{5}{*}{ $\begin{bmatrix*}[r] 1 &  0.35 &  -0.38 &  0.17 & 0.27 &  -0.19 \\ 
                                    	&  1    &  -0.05 & -0.16 & 0.85 &  0.08 \\
                                    	&       & 1     &  0.26 &  -0.11 &  0.64 \\
                                    	&       &       & 1     &  0.10 &  0.25 \\
                                    	&       &       &       & 1     &  0.05 \\
                                    	&       &       &       &       &  1    \end{bmatrix*}$ } \\
$g_{\pi K} =$                 & $(0.165 \pm 0.006 \pm 0.002) \cdot a_t^{-1}$   & \\
$g_{\eta K} =$                & $(0.033 \pm 0.010 \pm 0.003) \cdot a_t^{-1}$   & \\
$\gamma_{\pi K,\,\pi K} = $   & $\,\,\,\;0.184 \pm 0.054 \pm 0.030 $   & \\
$\gamma_{\pi K,\,\eta K} = $  & $-0.52 \pm 0.20 \pm 0.06 $   & \\
$\gamma_{\eta K,\,\eta K} = $ & $-0.37 \pm 0.07 \pm 0.05 $   & \\[1.3ex]
&\multicolumn{2}{l}{ $\chi^2/ N_\mathrm{dof} = \frac{6.40}{15-6} = 0.71 $\,.}
\end{tabular}
\end{center}
\vspace{-1cm}
\begin{equation} \label{A1_fit_par_values}\end{equation}
\end{widetext}
In these fit results, the first quoted error is statistical and corresponds in the usual way to an increase in $\chi^2$ by one unit, while the second reflects the uncertainty in the scattering meson masses, $a_t m_\pi$, $a_t m_K$, $a_t m_\eta$ and anisotropy, $\xi$. The parameter correlation matrix is also shown, indicating that in general there are not particularly large correlations between parameters. We can plot the finite-volume energy levels corresponding to this best-fit model $t$-matrix obtained by solving Eq.~\ref{eq_luescher_t} for the parameterization in Eq.~\ref{eq_kma_S_exact} with parameter values given by Eq.~\ref{A1_fit_par_values} alongside those obtained in the lattice QCD calculation -- this is shown in Fig.~\ref{fig_swave_rest_model}, where the agreement is clear, as one would expect from a fit with a $\chi^2/N_\mathrm{dof}$ close to unity.

\begin{figure}[htb]
\includegraphics[width=1.0\columnwidth]{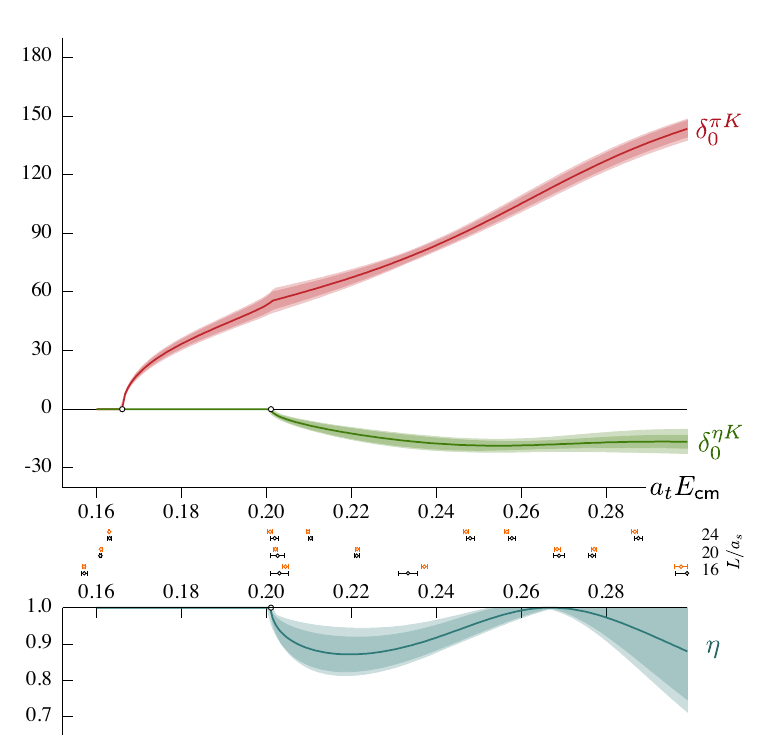}
\caption{
The curves show the phase-shifts and inelasticity as defined in Eq. \ref{del_del_eta} for the parameterization in Eq. \ref{eq_kma_S_exact} with parameter values given by Eq.~\ref{A1_fit_par_values}. The inner and outer error bands reflect the two sets of errors (statistical and variation in meson masses and anisotropy) quoted. Top: $\delta_{\ell=0}^{\pi K}$ and $\delta_{\ell=0}^{\eta K}$ in degrees. Middle: Minimisation result, model energies with uncertainties in orange, lattice QCD energies in black. Bottom: the inelasticity. Note the position of the three lowest points below $\pi K$ threshold that enter in the fit and tightly constrain the $t$-matrix near threshold.
}
\label{fig_swave_rest}
\end{figure}

In Fig.~\ref{fig_swave_rest} we take the $t$-matrix resulting from this minimization and plot the multichannel phase-shifts, $\delta_i(s)$, with ${i = \pi K, \eta K}$ and inelasticity, $\eta(s)$, defined in the usual manner,
\begin{equation}
t_{ij}= \left\{ \begin{matrix} \frac{\eta \, e^{2i\delta_i}-1}{2i \, \rho_i}  & (i=j)\\[1.1ex]
					\frac{   \sqrt{1-\eta^2} \,   e^{i(\delta_i+\delta_j)}}{2 \,   \sqrt{\rho_i \, \rho_j} } & (i\ne j)
					\end{matrix} \right. ,
\label{del_del_eta}
\end{equation}
where $\rho_i(s)=2k_i/\sqrt{s}$ is the phase space for channel $i$.

\begin{figure}[htb]
\includegraphics[width=1.0\columnwidth]{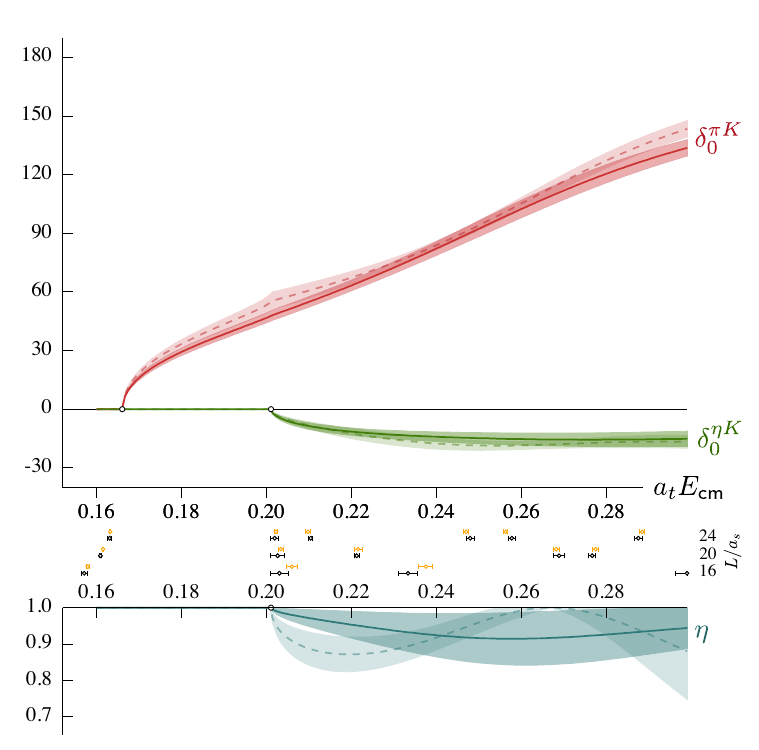}
\caption{Dashed curves as in Figure~\ref{fig_swave_rest}. Solid curves show the the phase-shifts and inelasticity for an alternative parameterization of the $K$-matrix given by Eq.~\ref{k-poly-inv} and described in the text. 
}
\label{fig_swave_rest_compare}
\end{figure}

To assess whether features present in Figure~\ref{fig_swave_rest} are truly required to describe the finite-volume spectra, or whether they are artifacts of the particular parameterization utilized, we also attempt a description using a different form for the $K$-matrix. This second fit uses Eq.~\ref{k-poly-inv} with $N_{\pi K , \pi K} = N_{\pi K , \eta K} = N_{\eta K , \eta K} = 1$ and is able to describe the spectra with $\chi^2/N_\mathrm{dof} = 12.2/(15-3) = 1.36$. The resulting phase-shifts and inelasticity are plotted in Figure~\ref{fig_swave_rest_compare} along with the previous fit. We see that the large-scale behavior is the same in both fits, although two detailed features prove to not be robust under change in paramaterization: the visible cusp in $\delta^{\pi K}$ at the opening of the $\eta K$ threshold and the degree of deviation from unity of the inelasticity  below $a_t E_\mathsf{cm} \sim 0.24$.

Note that our earlier suspicion that $\pi K$ and $\eta K$ are essentially decoupled is manifested in the fit results, Figure~\ref{fig_swave_rest_compare} shows the inelasticity which, while it has a large uncertainty, and does vary somewhat under change in parameterization, hardly deviates away from unity, indicating complete decoupling, over the entire constrained energy region. Arguments based upon $SU(3)_F$ flavor symmetry, outlined in Appendix~\ref{app_su3f}, suggest than in even-$\ell$ partial waves, the resonant octet coupling to $\pi K$ is strongly enhanced over coupling to $\eta K$, leading to an approximate decoupling. As mentioned in Section~\ref{intro}, such a decoupling is observed experimentally in the $J^P=0^+,2^+$ channels \cite{Aston:1987ir,Aston:1987ey}.

The $S$-wave amplitudes we have constrained using this limited set of data contain some suggestive properties. A phase-shift rising through $90^\circ$, as shown in Figs.~\ref{fig_swave_rest},\ref{fig_swave_rest_compare} is often indicative of a resonance. It appears from this fit that such a resonance may be coupled to $\pi K$ and not $\eta K$, but the uncertainty on the inelasticity is large. To obtain a more constrained description of the scattering we require more data -- we now proceed to investigate a much larger set of irrep spectra.

\subsection{Finite-volume spectra}

\begin{figure}[b]
\includegraphics[width=0.48\textwidth]{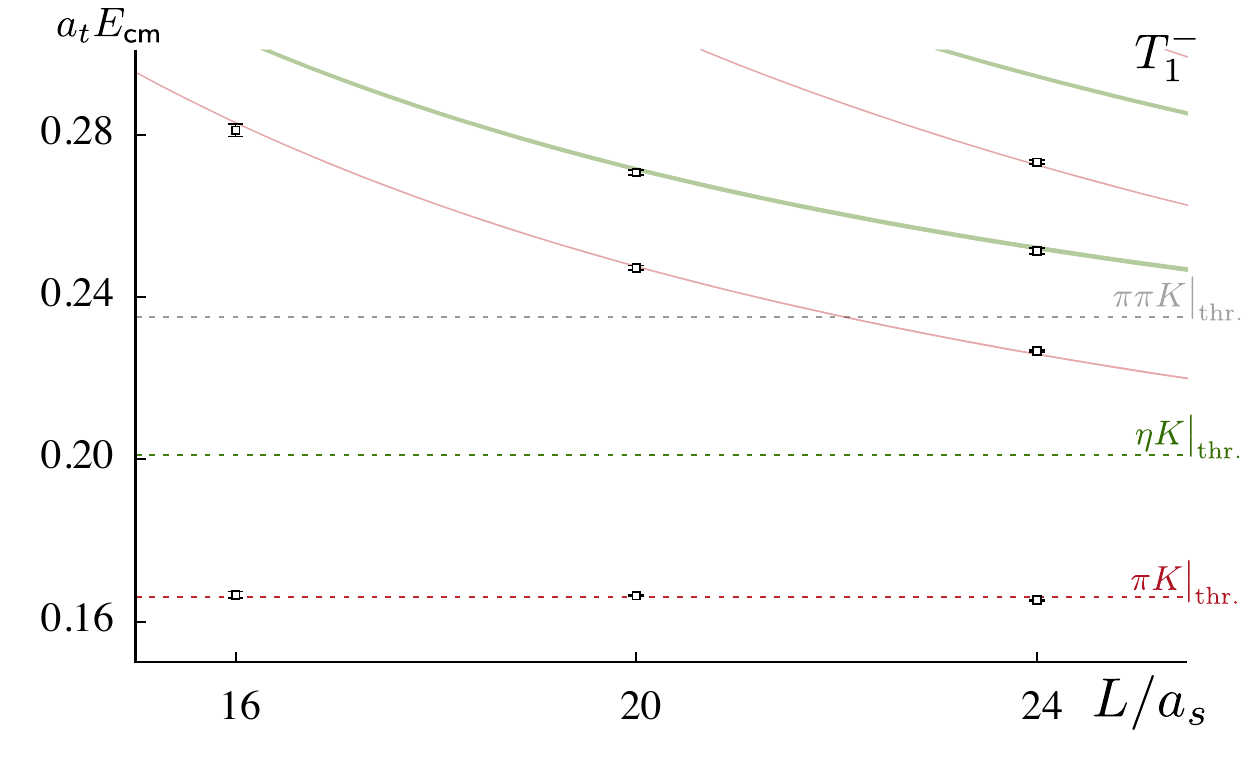}
\caption{As for Fig.~\ref{spectrum_I1o2_P000_A1} but for the $T_1^-$ irrep ($J^P = 1^-,3^-\ldots$). Note that there are no non-interacting energy levels at thresholds in this case. The $\pi\pi K$ threshold is indicated by a dashed gray horizontal line.  }
\label{spectrum_I1o2_P000_T1}
\end{figure}

We now begin the task of improving our description of the $S$-wave and determining the behavior of higher partial waves. 

In Fig.~\ref{spectrum_I1o2_P000_T1} we show the spectrum in the $T_1^-$ irrep on our three volumes, which we expect to be dominated by the $\ell=1$ partial wave. We have not included $\pi\pi K$-like operators in our basis, and as such we expect our spectrum near and above the $\pi\pi K$ threshold to be incomplete and/or inaccurate. Even if we had obtained the complete spectrum, the formalism for relating scattering amplitudes to finite-volume spectra when three-body channels are open is not yet completely mature~\cite{Polejaeva:2012ut,Briceno:2012rv,Hansen:2013dla}. As such we will largely limit our consideration to energies below the $\pi\pi K$ threshold at $a_t E_\mathsf{cm} = 0.235$.  We note that for each volume there is a state very close to the $\pi K$ threshold, which would not be expected in a non-interacting theory where the first level would appear much higher and correspond to $\pi_1 K_1$ (the lowest red curve in Fig~\ref{spectrum_I1o2_P000_T1}). The observed near-threshold level overlaps strongly with the ``single-meson" operators in the variational basis -- this, along with the lack of any significant volume dependence, is strongly suggestive of a low-lying vector meson; we will explore this further below.

\begin{figure*}[htb]
\includegraphics[width=.85\textwidth]{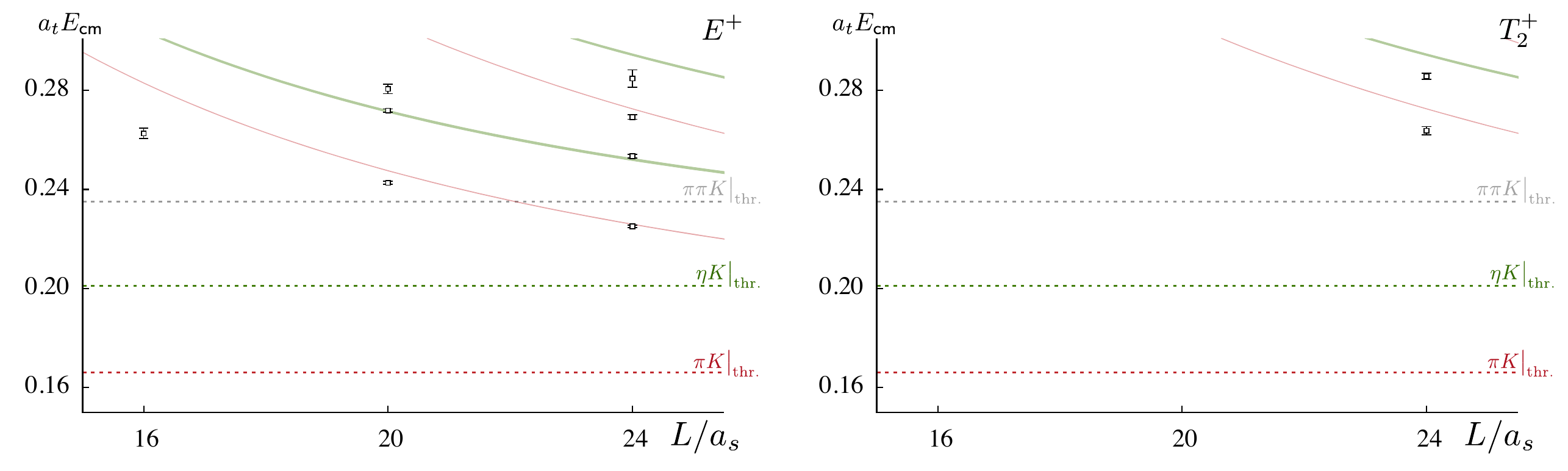}
\caption{As for Fig.~\ref{spectrum_I1o2_P000_T1} but for the $E^+$ ($J^P=2^+, 4^+\ldots$) and $T_2^+$ ($J^P=2^+,3^+, 4^+\ldots$) irreps. Note that in the $T_2^+$ case a spectrum was computed only on the $24^3$ volume.}
\label{spectrum_I1o2_P000_J2}
\end{figure*}

$\pi K$ and $\eta K$ scattering with $\ell=2$ are the lowest angular momentum contributions in the $E^+$ and $T_2^+$ irreps shown in Fig.~\ref{spectrum_I1o2_P000_J2}. We note that there may be an excess of states around $a_t E_\mathsf{cm} = 0.28$ compared to the non-interacting spectrum -- this may signal the presence of a narrow $J^P=2^+$ resonance.

\begin{figure*}
\includegraphics[width=.90\textwidth]{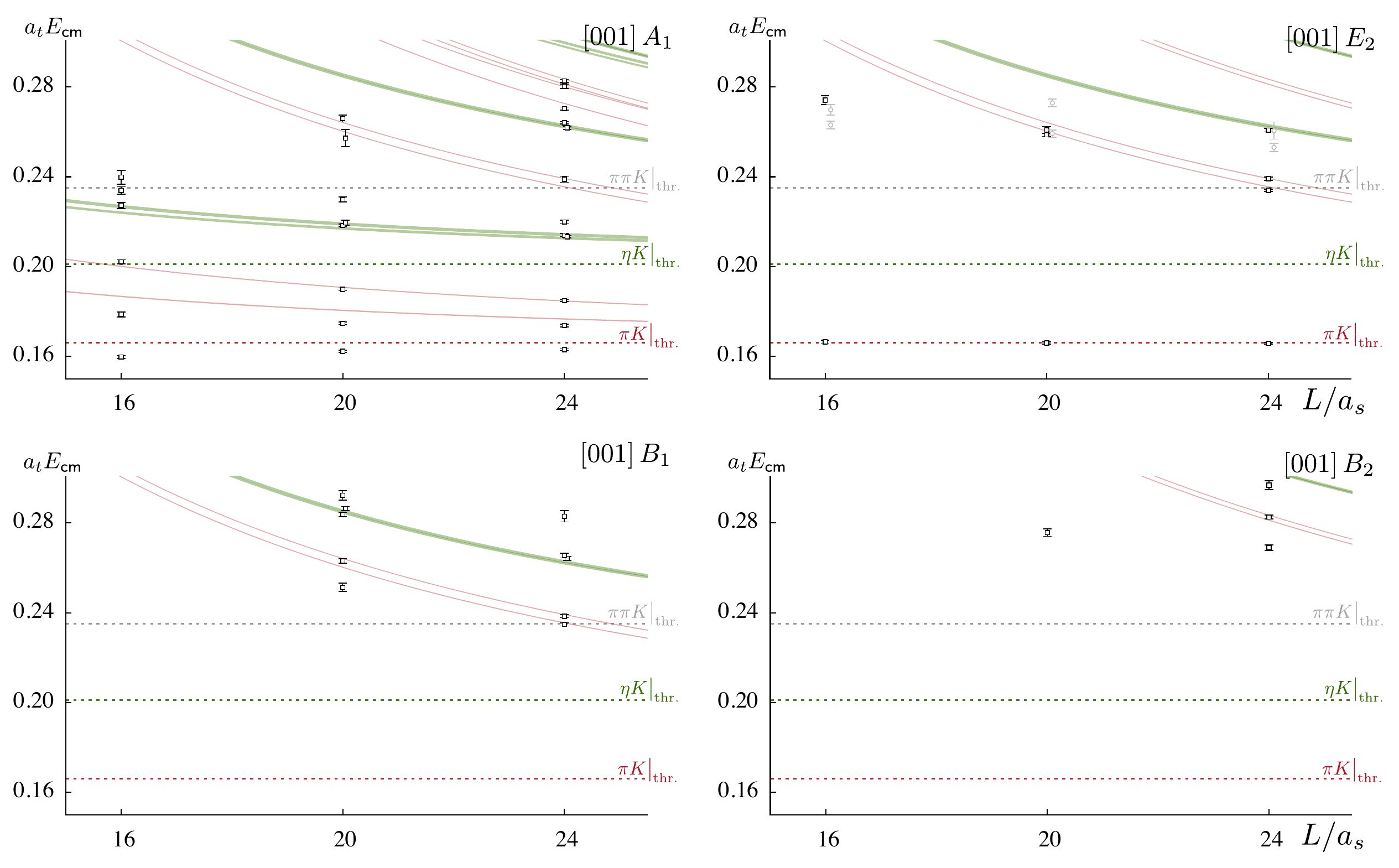}
\caption{
$\vec{P}=[001]$ finite-volume spectra for irreps $A_1,\, E_2, \, B_1$ and $B_2$. Light grey points in $A_1, E_2$ have large overlap onto ``single-meson" operators that we identify with unnatural parity $J^P=1^+$ states which cannot couple to $\pi K$ or $\eta K$.
}
\label{spectrum_I1o2_P100}
\end{figure*}

\begin{figure*}
\includegraphics[width=.92\textwidth]{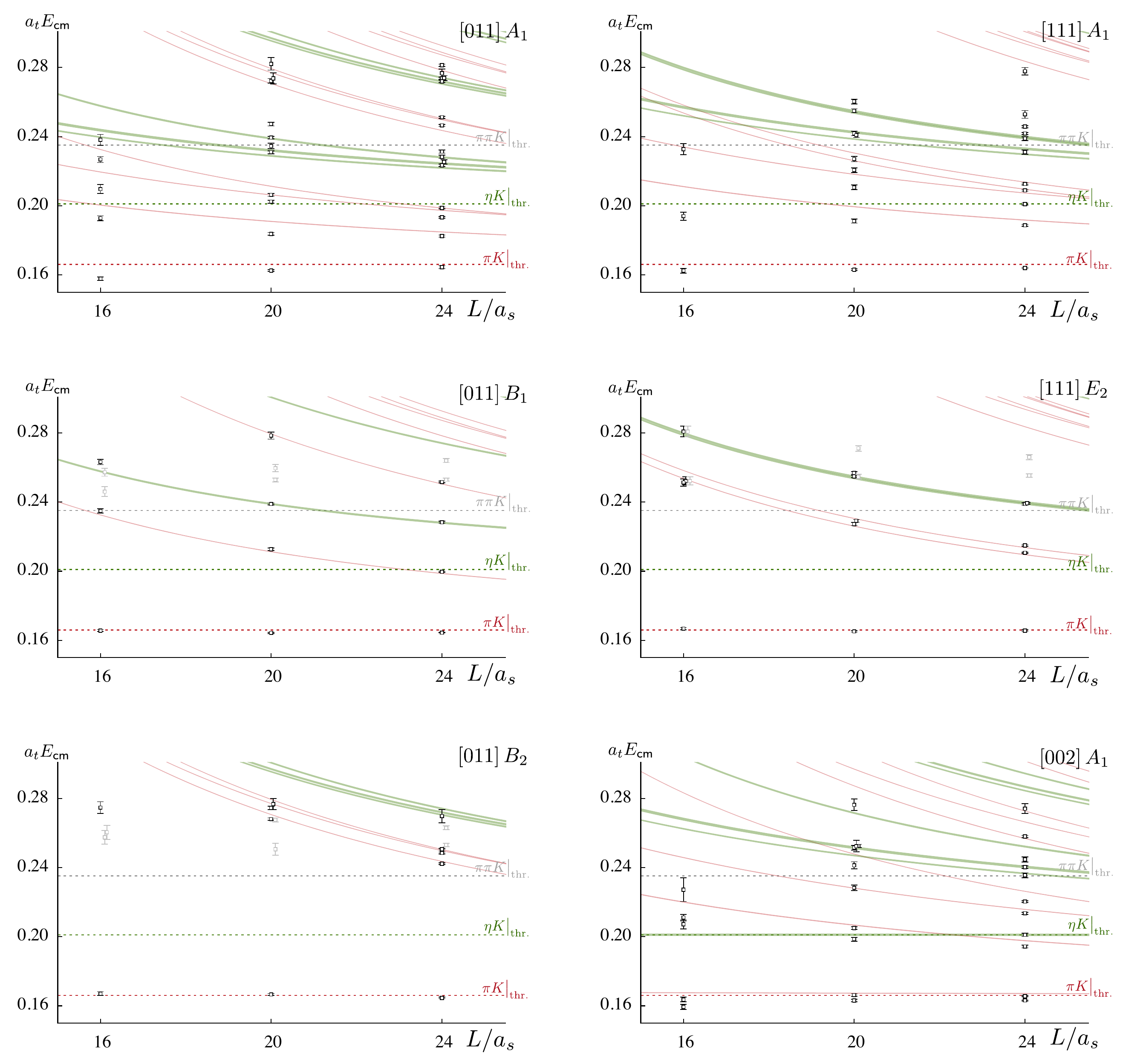}
\caption{As Figure~\ref{spectrum_I1o2_P100} for $\vec{P}=[011], [111], [002]$. Note the non-interacting levels very close to thresholds in the case $[002]\, A_1$.
}
\label{spectrum_I1o2_PBig}
\end{figure*}

In Figs.~\ref{spectrum_I1o2_P100}~and~\ref{spectrum_I1o2_PBig} we show the energy levels extracted when the scattering system is in-flight with respect to the lattice. There are typically more levels in the same energy region compared to the at-rest case since the allowed values of lattice momentum lead to many more non-interacting energy combinations. In the unequal mass case that we consider here, there is a ``duplication" of certain levels when compared to the equal mass case since ${\pi(\vec{k}_1)K(\vec{k}_2)\ne \pi(\vec{k}_2)K(\vec{k}_1)}$ when ${|\vec{k}_1|\ne |\vec{k}_2|}$.

In Table~\ref{tab_pwa_irrep}, considering in-flight irreps, we see that all partial-waves appear in $A_1$. In $[001]$ $B_1$ and $B_2$ the lowest allowed partial-wave is $\ell=2$, whilst the other irreps we consider have $\ell=1$ as their lowest partial-wave. In-flight, typically, unless there is some symmetry preventing it, there is a lowest allowed partial-wave and all higher partial-waves contribute. 

A near-threshold state, as noted earlier in the $[000]\, T_1^-$ case, appears in every irrep where the $\ell=1$ partial-wave features. In particular, since $\ell=1$ has a helicity zero component that is subduced into all in-flight $A_1$ irreps, it will always appear there, complicating the extraction of an $S$-wave amplitude near threshold.

As was hinted at in the at-rest $E^+, T_2^+$ discussion, there appears to be a $J^P=2^+$ resonance near ${a_t E_\cm=0.27}$ which can be seen most clearly in Fig.~\ref{spectrum_I1o2_P100} as the lowest level, overlapping strongly with ``single-meson" operators, in the $[001]\, B_2$ spectrum. 

We will begin analyzing these spectra by considering the $P$-wave at low energies.


\subsection{A near-threshold $J^P=1^-$ state.}

In every irrep which contains a subduction of $\ell=1$ we observe a finite-volume eigenstate very close to $\pi K$ threshold. We begin by considering irreps in which $\ell=1$ is the lowest allowed partial-wave, these being $T_1^-$ at rest, $[001]\, E_2$, $[011]\, B_{1,2}$ and $[111]\, E_2$. The $\ell=2$ amplitude is expected to be very small in this region as we will verify later.

We will explore single-channel elastic parameterizations to describe the spectrum in the energy region $0.16<a_t E_\cm<0.18$ which, on the basis of the qualitative observations made above, we expect to feature either a bound-state or a resonance only very slightly above threshold. A form capable of describing either of these possibilities is the relativistic Breit-Wigner of Eq.~\ref{eq_bw}; this describes a bound-state if the mass parameter takes a value below the threshold energy, since then the ``width" term becomes real and acts as an self-energy correction to the mass of the bound-state. We make use of the energy levels from all relevant irreps on the $20^3$ and $24^3$ volumes, and the at-rest $T_1^-$ energy level from the $16^3$ lattice, leading to 11 data points to constrain the fit. The best fit description is given by

\vspace{0.2cm}
\begin{tabular}{rll}
$m_R=$ & $(0.16488 \pm 0.00014 \pm 0.00012) \cdot a_t^{-1}$ & \multirow{2}{*}{ $\begin{bmatrix*}[r] 1 & -0.77 \\ & 1\end{bmatrix*}$ } \\
$g_R=$ & $5.72 \pm 0.45 \pm 0.27$   & \\[1.3ex]
&\multicolumn{2}{l}{ $\chi^2/ N_\mathrm{dof} = \frac{7.84}{11 - 2} = 0.87, $}  \\
\end{tabular}
\vspace{.1cm}

\noindent where we observe that the mass parameter is found to be below the $\pi K$ threshold (at $a_t E_\mathsf{cm} = 0.16604(15)$), and that we are describing a bound-state rather than a resonance. The zero of the denominator of $t(s)$ is shifted very slightly from $\sqrt{s} = m_R$ by the continuation of $i\sqrt{s}\,  \Gamma_1(s)$ which becomes real below threshold. That the width term is appearing as what amounts to a ``self-energy" correction likely explains the relatively large correlation between the $m_R$ and $g_R$ parameters.

There is additional information we can utilize to further constrain our description of this amplitude, which comes from the helicity zero components of the $P$-wave amplitude that are subduced into the in-flight $A_1$ irreps. There is a state near threshold in each of those irreps, as can be seen in Figs.~\ref{spectrum_I1o2_P100}~and~\ref{spectrum_I1o2_PBig}. The challenge presented in using these is that we require knowledge of the $\ell=0$ amplitude at the corresponding energy to reliably extract information for the $\ell=1$ amplitude from Eq.~\ref{eq_luescher_t}. 

We attack this by taking the coupled-channel $K$-matrix fit result described in Eqs.~\ref{eq_kma_S_exact} and \ref{A1_fit_par_values}, obtained from the $A_1^+$ spectrum at rest, shown in Fig.~\ref{fig_swave_rest}, to fix the value of the $\ell=0$ phase-shift at the appropriate level energies. Making the reasonable assumption that $\ell \ge 2$ amplitudes are negligible, with the known value of $\delta_{\ell=0}$ in hand the coupled Eq.~\ref{eq_luescher_t} for $\delta_{\ell=0}, \delta_{\ell=1}$ has only $\delta_{\ell=1}$ unknown which can be solved for. In this way we may re-fit including a further 8 points to constrain the amplitude:

\vspace{0.2cm}
\begin{tabular}{rll}
$m_R=$ & $(0.16482 \pm 0.00009 \pm 0.00009) \cdot a_t^{-1}$ & \multirow{2}{*}{ $\begin{bmatrix*}[r] 1 & -0.46 \\ & 1\end{bmatrix*}$ } \\
$g_R=$ & $5.93 \pm 0.26 \pm 0.14$   & \\[1.3ex]
&\multicolumn{2}{l}{ $\chi^2/ N_\mathrm{dof} = \frac{9.23}{19 - 2} = 0.54   \;. $}  \\
\end{tabular}
\vspace{-.8cm}
\begin{equation}  \label{bw} \end{equation}

We notice that the statistical uncertainties are reduced with the larger set of data, and we also observe a smaller degree of correlation between $m_R$ and $g_R$ which may be due to the fact that we are making use of data over a larger energy region such that the energy dependence in the $i\sqrt{s} \, \Gamma_1(s)$ term of the denominator is being sampled.

Since the energy levels span the scattering threshold, plotting the elastic phase-shift, which changes from real to imaginary as we cross the threshold from above, is not ideal. One convenient option is to plot $k^{2\ell+1}\cot \delta_\ell$ against energy -- this quantity is continuous and real through the threshold and for the relativistic Breit-Wigner, Eq.~\ref{eq_bw}, it vanishes at $E_\mathsf{cm} = m_R$,
\begin{align}
k^3 \cot \delta_1 = (m_{R}^2-s)  \frac{6\pi \sqrt{s} }{g_{{R}}^2}\,.
\label{kcotd}
\end{align} 

\begin{figure}
\includegraphics[width=1.0\columnwidth]{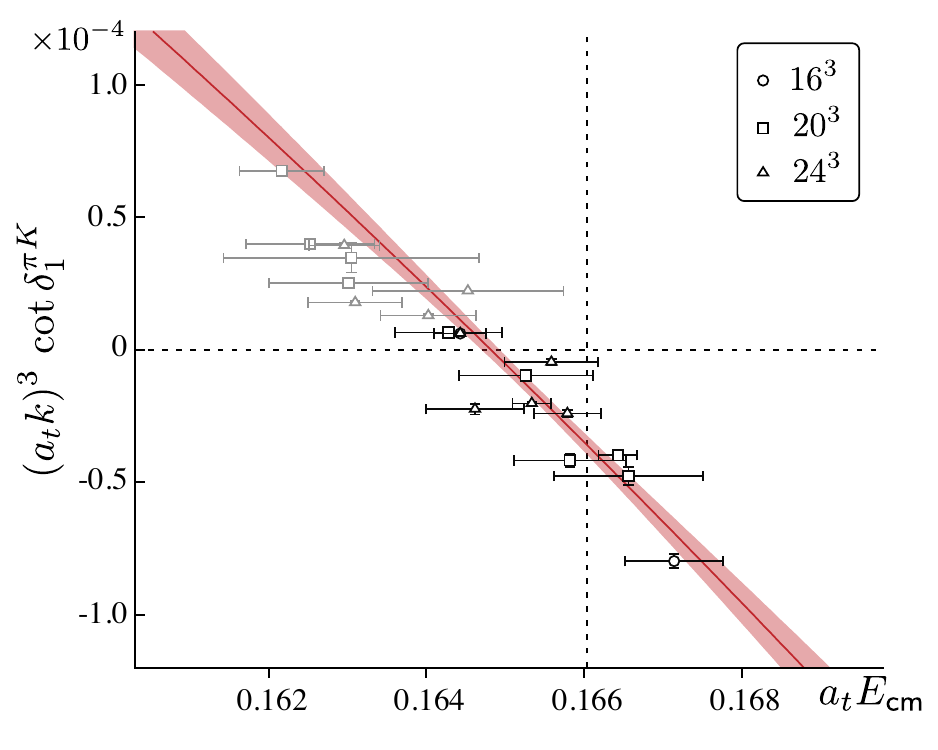}
\caption{The $\pi K$ threshold region in the $\ell=1$ partial wave plotted as $k^3\cot\delta$. 
The darker points are obtained from irreps where $\ell=1$ is the lowest allowed, while the lighter points are extracted from $A_1$ irreps where the $\ell=0$ contribution is accounted for as described in the text. The curve shows the fit to a relativistic Breit-Wigner with the parameters in Eq. \ref{bw}.
}
\label{fig_pwave_kcot_bw}
\end{figure}

We plot this quantity in Fig.~\ref{fig_pwave_kcot_bw}. The spread of points in energy is to be expected even for a bound-state, as the $\mathcal{M}$ function in the finite-volume quantization condition, Eq.~\ref{eq_luescher_t}, varies irrep-to-irrep. Note that $A_1$ points are systematically lower in energy than those from the $E$ and $B$ irreps which is a consequence of this (the effect of the attractive $S$-wave interaction is found to be small at these energies). The fit curve, corresponding to Eq.~\ref{bw}, is also shown, where it is clear that inclusion of the $A_1$ irrep levels better constrains the slope, which determines $g_R$.

We conclude that there is a vector meson bound-state in this calculation and we will return to the interpretation of this state later.

\subsection{$\pi K$ elastic scattering below $\eta K$ threshold}

We now briefly study the elastic $\pi K$ scattering region below the $\eta K$ threshold where single-channel parameterizations are justified. For the $S$-wave, an effective range expansion is adopted, while for the $P$-wave we first consider a Breit-Wigner as in the previous section. Since now we are considering a larger energy region (out to $a_t E_\mathsf{cm} = 0.201$) it is not guaranteed that the Breit-Wigner will still be capable of describing the amplitude.

We proceed with simultaneous inclusion of $\ell =0$ and $\ell=1$ waves in Eq.~\ref{eq_luescher_t} with the parameterizations described above, where we are assuming that $\ell=2$ and higher amplitudes play a negligible role at these low energies. Fitting to all energy levels below $a_t E_\mathsf{cm} = 0.201$ in irreps $A_1^+$, $T_1^-$ from all three volumes, and irreps $[001]\,A_1$, $[001]\,E_2$, $[011]\,A_1$, $[011]\,B_1$, $[011]\,B_2$, $[111]\,A_1$, $[111]\,E_2$ and $[002]\,A_1$ from the the $20^3$ and $24^3$ volumes we obtain, 
\begin{center}
\begin{tabular}{rll}
$a_{\ell=0} = $ & $(17.2 \pm 0.9 \pm 1.2) \cdot a_t$ & \multirow{3}{*}{ $\begin{bmatrix*}[r] 1 &  0.1 & 0.2 \\ & 1 & -0.2 \\ & & 1\end{bmatrix*}$ } \\
$m_{R} = $ & $(0.16498 \pm 0.00009 \pm 0.00024) \cdot a_t^{-1}$ & \\
$g_{R} = $ & $(4.72 \pm 0.17 \pm 0.28) $   & \\[1.3ex]
&\multicolumn{2}{l}{ $\chi^2/ N_\mathrm{dof} = \frac{42.8}{37 - 3} = 1.26 $\;,}  \\
\end{tabular}
\vspace{-.8cm}
\begin{equation} \label{elastic_SP_BW}\end{equation}
\end{center}

\noindent in the case that we restrict the $S$-wave effective range expansion to a scattering length. Adding an effective range term to the $S$-wave amplitude does not improve the fit, and thus we explore adjusting the $P$-wave parameterization. Replacing the Breit-Wigner with a single-channel version of a $P$-wave $K$-matrix featuring a single pole plus a constant, ${K(s) = g^2 / (m^2 -s) + \gamma}$, and using the Chew-Mandelstam phase-space subtracted at the pole, improves the $\chi^2/N_\mathrm{dof}$,
\begin{center}
\hspace*{-0.4cm}
\begin{tabular}{rll}
$a_{\ell =0} = $ & $(17.4 \pm 0.9 \pm 1.2) \cdot a_t$ & \!\!\!\!\!\!\multirow{4}{*}{ $\begin{bmatrix*}[r] 1 &  0.0 & 0.1 & 0.0 \\ & 1 & \sm0.6 & \sm0.5 \\ & & 1 & 0.9  \\ & & & 1\end{bmatrix*}$ } \\
$m = $ & $(0.16480 \pm 0.00014 \pm 0.00011) \!\cdot\! a_t^{-1}$ & \\
$g = $ & $0.480 \pm 0.023 \pm 0.027$   & \\
$\gamma = $ & $(10.5 \pm 2.3 \pm 2.4) \cdot a_t^2$   & \\[1.3ex]
&\multicolumn{2}{l}{ $\chi^2/ N_\mathrm{dof} = \frac{20.5}{37 - 4} = 0.62 $\,.}  \\
\end{tabular}
\vspace{-.8cm}
\begin{equation} \label{elastic_SP_K}\end{equation}
\end{center}

\begin{figure}
\vspace{-0.5cm}
\includegraphics[width=1.0\columnwidth]{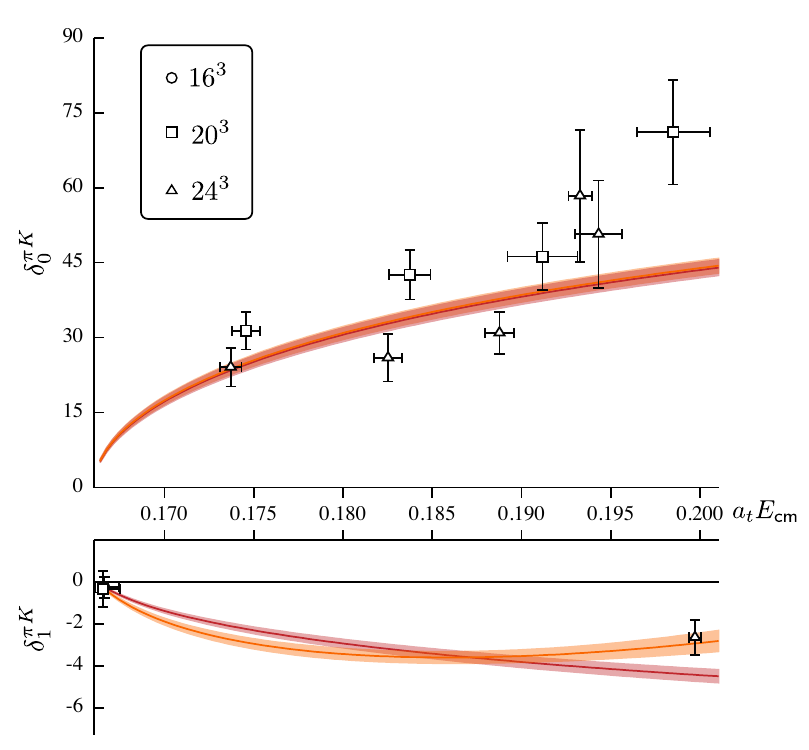}
\vspace{-0.5cm}
\caption{ $\pi K$ $S$-wave (upper) and $P$-wave (lower) elastic scattering phase-shifts. Red curve: scattering length in $S$-wave and Breit-Wigner in $P$-wave (Eq.~\ref{elastic_SP_BW}). Orange curve: scattering length in $S$-wave and $K$-matrix pole plus constant in $P$-wave (Eq.~\ref{elastic_SP_K}). The points were determined using Eq.~\ref{eq_luescher_t} as described in the text. In $P$-wave there are three overlapping points very slightly above threshold. }
\label{fig_SP_elastic_phases}
\end{figure}

In Fig.~\ref{fig_SP_elastic_phases} we show the phase-shifts corresponding to these two fits. We obtain the elastic phase-shift points for each energy level using Eq.~\ref{eq_luescher_t}. For the irreps where \mbox{$P$-wave} is the lowest, it is straightforward to neglect \mbox{$D$-wave} and higher. For irreps where the \mbox{$S$-wave} is lowest, we fix the \mbox{$P$-wave} using the fit result given in Eq.~\ref{bw} and use Eq.~\ref{eq_luescher_t} again assuming \mbox{$D$-wave} and higher may be neglected. The Breit-Wigner parameterization gives a good description in the energy region around the bound-state, however there is one \mbox{$P$-wave} point at higher energy that is poorly described. The added freedom in the \mbox{$P$-wave} \mbox{$K$-matrix} resolves this.

\subsection{$\pi K, \eta K$ amplitudes constrained by 80 energy levels}

We now embark upon a description of the bulk of the spectrum data presented in Figures \ref{spectrum_I1o2_P000_A1}, \ref{spectrum_I1o2_P000_T1}, \ref{spectrum_I1o2_P000_J2}, \ref{spectrum_I1o2_P100} and \ref{spectrum_I1o2_PBig}, in terms of a coupled $\pi K, \eta K$ scattering system. We will restrict ourselves initially to energies below $\pi\pi K$ threshold, except for the $A_1^+$ at-rest irrep which is dominated by $J^P=0^+$ which does not couple to $\pi\pi K$ -- in this case we consider energy levels up to $\pi\pi\pi K$ threshold at $a_t E_\mathsf{cm} = 0.304$. 

A description of the spectra is sought using $K$-matrix parameterizations in each partial-wave as defined in Eq.~\ref{eq_t_matrix_k} and Eq.~\ref{k-matrix}. We have explored many variations of this, including using a simple phase-space rather than the Chew-Mandelstam type, using powers of the phase-space instead of momenta to provide the threshold behavior, varying the subtraction point of the Chew-Mandelstam functions and using the $K$-matrix parameterization of Eq.~\ref{k-poly-inv} with a range of different polynomial orders. The resulting phase-shifts and inelasticities are found to be broadly the same in every fit with error bands that overlap for much of the region -- further discussion of these systematic variations will appear in Section~\ref{sec_fit_systematics}.

Our preferred choice is to parameterize the coupled $\pi K, \eta K$ $t$-matrix using a $K$-matrix featuring a single pole coupled to both channels plus a constant matrix (see Eq.~\ref{eq_kma_S_exact}). We opt to use the Chew-Mandelstam phase-space subtracted such that $\mathrm{Re}\, I_{ij}(s=m^2) = 0$ where $m^2$ is the $K$-matrix pole position. Such a parameterization can be used in both $S$ and $P$-waves according to Eq.~\ref{eq_t_matrix_k}. Initially we will assume that the $D$-wave makes no significant contribution -- we will explore the sensitivity to this assumption later in the manuscript.

We used levels from the following irreps: $[000]\, T_1$ on $L/a_s=16,20,24$, $[001]\, E_2$ on $L/a_s=20,24$, $[011]\, B_1, B_2$ on $L/a_s=20,24$ and $[111]\, E_2$ on $L/a_s=20,24$ -- in total 19 energy levels, to constrain a fit describing the $P$-wave amplitude. In this case we choose to use a constant term only in the $\gamma_{\eta K, \eta K}$ position, with $\gamma_{\pi K, \pi K} = \gamma_{\pi K, \eta K} = 0$. The result of the fit, which has $\chi^2/N_\mathrm{dof} = 15.0/(19-5) = 1.00$, is:
\begin{widetext}
\vspace{0.5cm}
\begin{center}
\begin{tabular}{rll}
$m = $ 						& $(0.16497 \pm 0.00012 \pm 0.00002) \cdot a_t^{-1}$ & 
          \multirow{4}{*}{ $\begin{bmatrix*}[r] 1   &  0.0 & -0.6 &  -0.5  \\ 
          											&    1 & -0.4 & -0.2  \\ 
													&      &    1 &  0.8  \\
													&	   &      &    1   \end{bmatrix*} .$ } \\
$g_{\pi K} = $ 				&  $0.321 \pm 0.022 \pm 0.032 $ & \\
$g_{\eta K} = $ 			&  $0.65 \pm 0.11 \pm 0.11 $   & \\
$\gamma_{\eta K, \eta K} = $ & $(17.3 \pm 7.8 \pm 6.1)  \cdot a_t^{-2}$    & 
\end{tabular}
\end{center}

Fixing the $P$-wave amplitude to that presented above, we vary $S$-wave parameters to describe 61 energy levels taken from $A_1$ irreps: $[000](16,20,24)$, $[001](20,24)$, $[011](20,24)$, $[111](20,24)$ and $[002](20,24)$. The result, with $\chi^2/N_\mathrm{dof} = 49.1/(61-6) = 0.89$, is:

\begin{center}
\begin{tabular}{rll}
$m = $ 						& $(0.2458 \pm 0.0014 \pm 0.0004) \cdot a_t^{-1}$ & 
          \multirow{6}{*}{ $\begin{bmatrix*}[r] 1   &  0.5 & -0.3 &  0.0 &  0.1 & -0.1 \\ 
          											&    1 & -0.4 & -0.7 &  0.5 & -0.1 \\ 
													&      &    1 &  0.3 & -0.6 &  0.3 \\
													&	   &      &    1 &  0.1 & -0.1 \\
													&      &      &      &    1 & -0.3 \\
													&      &      &      &      &    1  \end{bmatrix*} .$ } \\
$g_{\pi K} = $ 				&  $(0.156 \pm 0.004 \pm 0.001) \cdot a_t^{-1}$ & \\
$g_{\eta K} = $ 			& $( 0.027 \pm 0.008 \pm 0.008) \cdot a_t^{-1}$   & \\
$\gamma_{\pi K, \pi K} = $ 	& $0.082 \pm 0.046 \pm 0.022$   & \\
$\gamma_{\pi K, \eta K} = $ & $0.33 \pm 0.13 \pm 0.06$   & \\
$\gamma_{\eta K, \eta K} = $ & $-0.41 \pm 0.05 \pm 0.07$   & 
\end{tabular}
\begin{equation} \label{global}\end{equation}
\end{center}
\vspace{1cm}

\begin{figure*}
\includegraphics[width=0.57\textwidth]{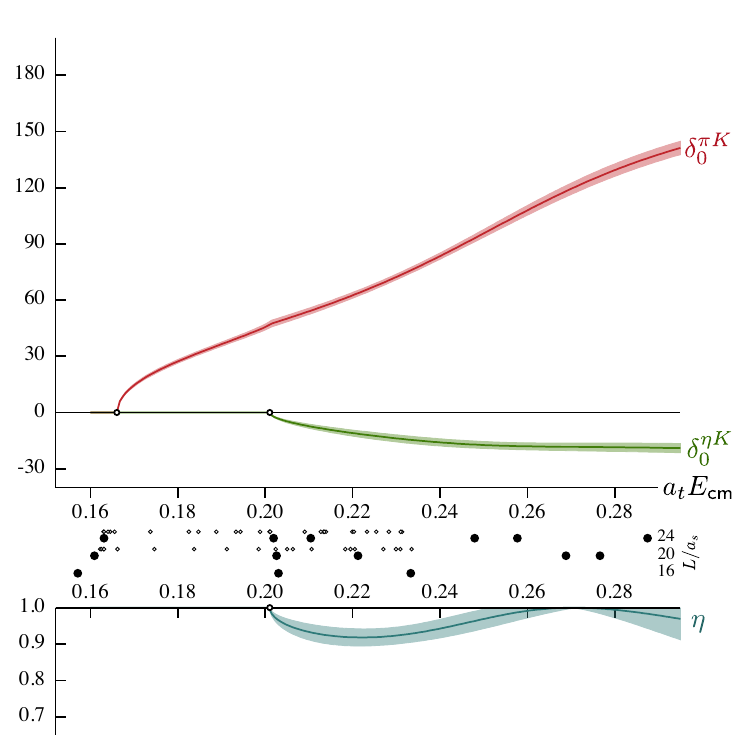}
\caption{
$S$-wave phase-shifts (in degrees) and inelasticity from the $K$-matrix description, with parameters given in Eq.~\ref{global}, of a large set of energy levels. Finite volume energy levels constraining the fit are shown as points in the middle with the at-rest data marked with filled circles and the in-flight data with hollow circles.}
\label{fig_S_wave_global}
\end{figure*}

\end{widetext}

\begin{figure}[H]
\includegraphics[width=1.0\columnwidth]{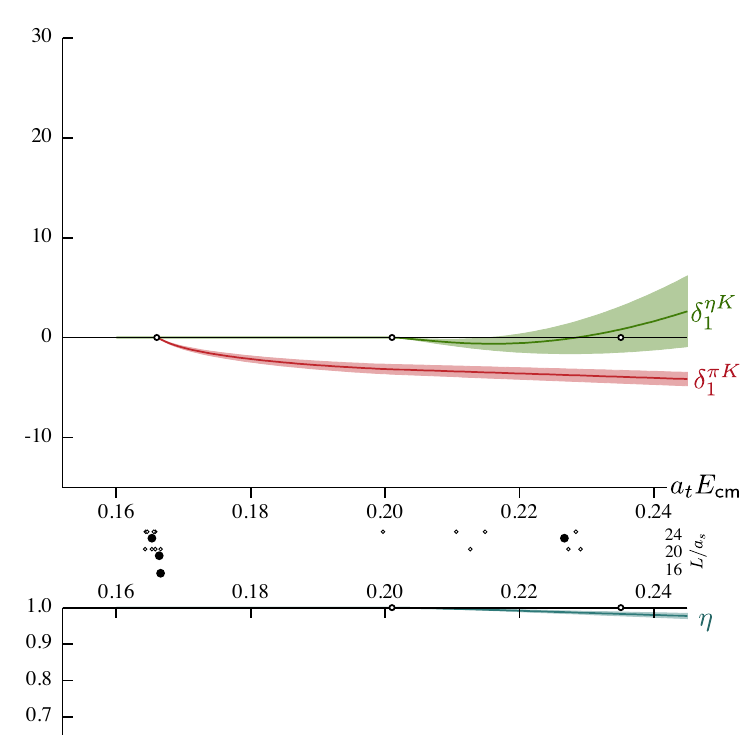}
\caption{As Fig.~\ref{fig_S_wave_global} for the $P$-wave.
}
\label{fig_P_wave_global}
\end{figure}

The phase-shifts and inelasticity corresponding to this fit are shown in Fig.~\ref{fig_S_wave_global} for the $S$-wave and in Fig.~\ref{fig_P_wave_global} for the $P$-wave. An alternative approach in which all 80 levels are considered together, varying the $S$ and $P$-wave parameters simultaneously, leads to a solution statistically compatible with the one presented above.

As with the $S$-wave fit using only at-rest points, we find only very weak coupling between the $\pi K$ and $\eta K$ channels, with an apparent weak repulsive interaction in the $\eta K$ channel and a gradual rise in the $\pi K$ phase-shift. As previously we note the rapid rise in the $\pi K$ phase-shift at threshold, followed by a slow increase through $90^\circ$ at higher energies. In Section~\ref{sec_poles_residues} we will analyze the resulting $t$-matrix for its singularity structure and corresponding resonance interpretation and consider a wider range of amplitude parameterization forms.

\begin{figure}[H]
\includegraphics[width=1.0\columnwidth]{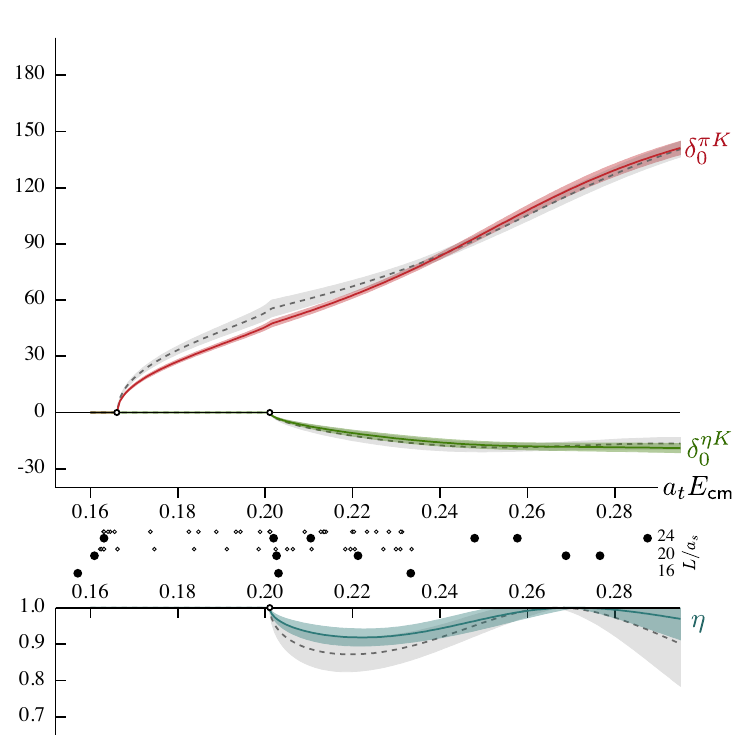}
\caption{
$S$-wave phase-shifts and inelasticity from the \mbox{$K$-matrix} description, with parameters given in Eq.~\ref{global}, of a large set of energy levels (colored curves), compared to the at-rest only fit, Eq.~\ref{A1_fit_par_values} (grey dashed curves).
}
\label{fig_S_wave_global_versus_rest}
\end{figure}

Comparing to the earlier description of the at-rest $A_1^+$ data alone, Eq.~\ref{A1_fit_par_values}, we observe in Fig.~\ref{fig_S_wave_global_versus_rest}, that the additional in-flight data has reduced the statistical uncertainties, weakened the prominent cusp in $\delta_0^{\pi K}$ and reduced the degree of inelasticity.

\newpage


\subsection{$D$-wave $\pi K, \eta K$ scattering}

In the energy region below $\pi\pi K$ threshold, in the volumes we have considered, there are insufficient energy levels in irreps which have $\ell =2$ as the lowest partial-wave to constrain the amplitude. As we have previously discussed, the spectrum we have computed without $\pi\pi K$-like operators should not necessarily be either complete or accurate above $\pi\pi K$ threshold, nor are we strictly justified in describing it solely using the $2 \to 2$ scattering formalism of Section \ref{sec_fvs}. Nevertheless we will proceed in a cavalier manner and attempt to describe the spectrum up to $\pi\pi\pi K$ threshold, assuming without justification that there is negligible coupling between $\pi\pi K$ and $\pi K, \eta K$ in $D$-wave.

We refer the reader to \cite{Giudice:2012tg}, in particular to their Figure 16, where the result of applying a $2 \to 2$ formalism in an energy region where higher-multiplicity scattering is occurring is shown. They observe that the resulting phase-shift points do not lie on a single curve in the inelastic region. Such an observation would be a signal that our assumption of a negligible role for $\pi \pi K$ is unjustified.

We proceed with an attempt to describe the spectra in irreps having $\ell=2$ as their lowest partial-wave -- there are 24 such levels which come from $E^+$, $T_2^+$, $[001]\,B_1, B_2$ irreps. Under the assumption that the $\ell \ge 3$ partial-waves are negligible in this energy region, we fit the energy levels using a coupled $\pi K, \eta K$ $K$-matrix model of the ``pole plus constant" form we have used previously, and find,

\begin{widetext}
\begin{center}
\begin{tabular}{rll}
$m =$                         & $(0.2789 \pm 0.0011 \pm 0.0002) \cdot a_t^{-1}$   & 
\multirow{5}{*}{ $\begin{bmatrix*}[r] 1    &  -0.03 & -0.34 &  0.50 &  0.04 &  0.45 \\ 
                                    & 1    & -0.41 & -0.34 &  -0.27 &  0.30 \\
                                    &       & 1     &  -0.26 &  0.64 & -0.67 \\
                                    &       &       & 1     & -0.03 & 0.35 \\
                                    &       &       &       & 1     &  0.10 \\
                                    &       &       &       &       &  1    \end{bmatrix*}$ } \\
$g_{\pi K} =$                 & $( 1.25 \pm 0.06 \pm 0.01 ) \cdot a_t$   & \\
$g_{\eta K} =$                & $( 0.29 \pm 0.64 \pm 0.03) \cdot a_t$   & \\
$\gamma_{\pi K,\,\pi K} = $   & $( 21 \pm 13 \pm 5) \cdot a_t^4$   & \\
$\gamma_{\pi K,\,\eta K} = $  & $( 34 \pm 55 \pm 7) \cdot a_t^4$   & \\
$\gamma_{\eta K,\,\eta K} = $ & $(-8 \pm 30 \pm 13) \cdot a_t^4$   & \\[1.3ex]
&\multicolumn{2}{l}{ $\chi^2/ N_\mathrm{dof} = \frac{16.0}{24-6} = 0.89 $\,.}  \\
\end{tabular}
\vspace{-.8cm}
\begin{equation} \label{Dfit}\end{equation}
\end{center}
\end{widetext}

\begin{figure}[H]
\includegraphics[width=1.0\columnwidth]{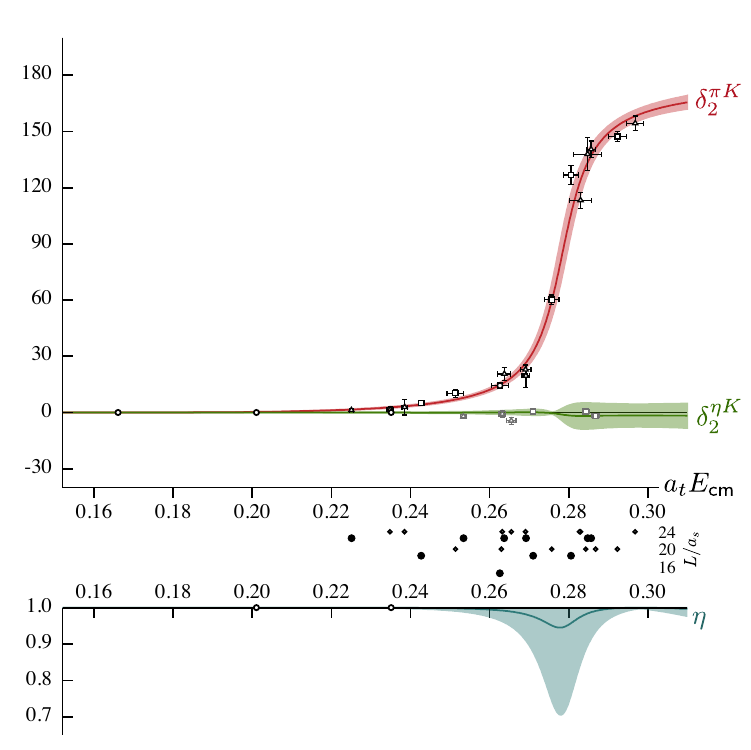}
\caption{
The $D$-wave phase-shifts (in degrees) and inelasticity as obtained from our lattice data from states with energies up to the $\pi\pi\pi K$ threshold.}
\label{fig_Dwave}
\end{figure}

The resulting phase-shifts and inelasticity are presented in Figure~\ref{fig_Dwave}. As with the $S$-wave, this description is entirely consistent with $\pi K$--$\eta K$ decoupling. The same $SU(3)_F$ logic, outlined in Appendix~\ref{app_su3f}, applies to the $D$-wave as applied to the $S$-wave. Under the assumption of complete decoupling, we can attempt to independently directly extract $\pi K$ and $\eta K$ phase-shifts using Eq.~\ref{eq_luescher_t} from levels identified as being ``$\pi K$" or ``$\eta K$" by their overlaps (states which overlap strongly with $q\bar{q}$-like operators typically also overlap with $\pi K$ and not $\eta K$ and are included in the $\pi K$ list). These points are included in Figure~\ref{fig_Dwave}, where we note immediately that the $\pi K$ phase-shift points are compatible with lying on a single curve. This, and the quite reasonable $\chi^2/N_\mathrm{dof}$ for the fit in Eq.~\ref{Dfit} may suggest that our neglect of $\pi\pi K$ scattering in $D$-wave is justified at these energies.

Figure~\ref{fig_Dwave} clearly shows a resonance-like behavior in $\pi K$ between $a_t E_\mathsf{cm} = 0.26$ and $0.29$. The rapid rise in the phase-shift suggests a narrow resonance and indeed an elastic relativistic Breit-Wigner description of just the levels with large overlap onto $\pi K$-like operators is very successful with $a_t m_R = 0.2785(8)$ and ${g_R = 9.26(36)}$, where the energy-dependent width is given by ${\Gamma_{\ell=2}(s) = \frac{g_R^2}{6\pi}\frac{k^5}{s \, m_R^2}   }$. 

\begin{figure}
\includegraphics[width=1.0\columnwidth]{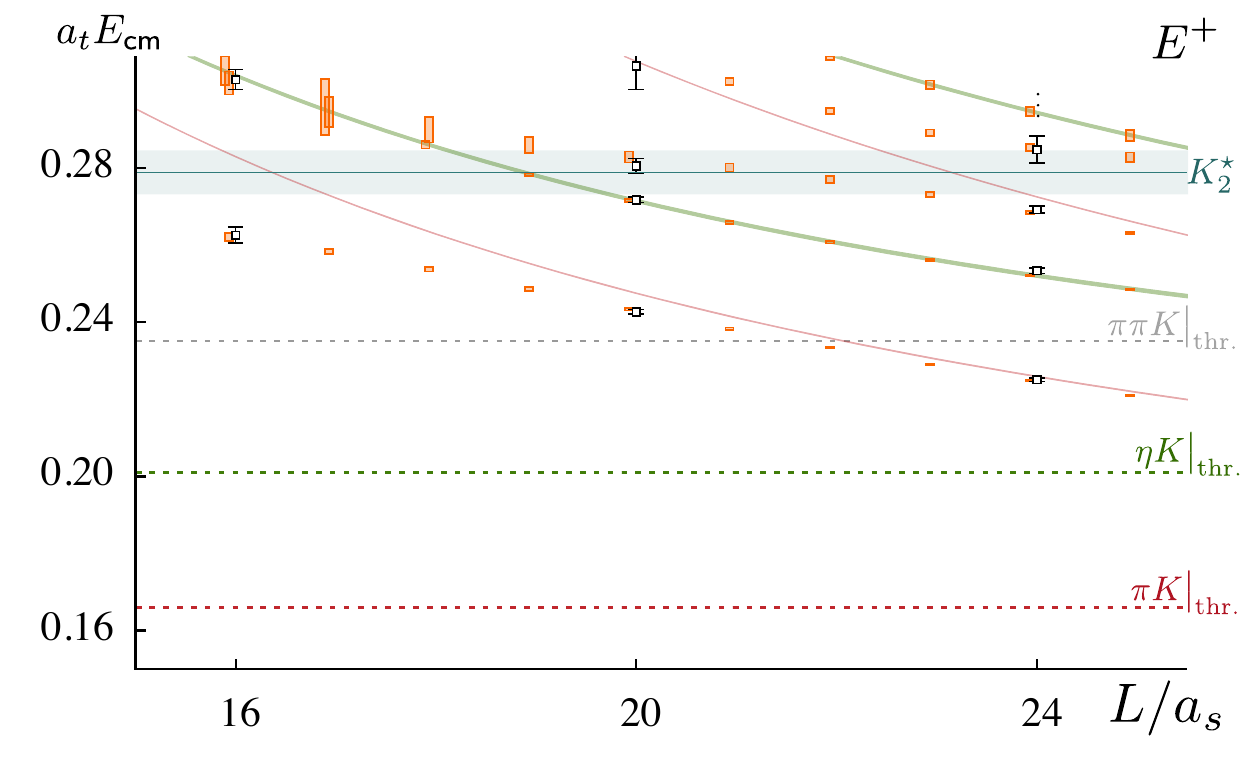}
\caption{The finite-volume spectrum in the $E^+$ irrep at integer values of $L/a_s$ determined by solving Eq.~\ref{eq_luescher_t} for the model in Eq.~\ref{Dfit} (orange), compared with the lattice QCD energies (black).}
\label{fig_Dwave_spectrum}
\end{figure}

The description of the spectrum in $E^+$ by the model of Eq.~\ref{Dfit} is shown in Figure~\ref{fig_Dwave_spectrum} where it is seen to be quite successful and where we explicitly see the expected avoided level crossings (for $L/a_s = 16,24$) as $\pi K$ non-interacting levels cross the energy region where the resonant behavior is present. Note that at $L/a_s = 20$, where an $\eta K$ non-interacting level is crossing the resonance, there is not an avoided level crossing, indicative that the resonance is not coupled to $\eta K$ as is born out in the fit, Eq.~\ref{Dfit}.

Importantly, in the energy region below $\pi\pi K$ threshold, where we obtained the $S$ and $P$ wave amplitudes above, the $D$-wave phase-shifts are tiny as can be seen in Fig.~\ref{fig_Dwave}. We implemented the fitted $D$-wave amplitude as a fixed entry into the $S,P$-wave fit to explore whether our earlier neglect of the $D$-wave introduced a significant error in the above $S$ and $P$-wave amplitudes and found that including it induced negligible changes in the determined amplitudes.

\subsection{Resonance poles}
\label{sec_poles_residues}

Resonances and bound-states may be identified with pole singularities of the $t$-matrix when it is analytically continued to complex values of $s$. $S$-matrix unitarity implies that $t(s)$ is a multi-sheeted function of $s$, with a square-root branch point at the opening of each kinematic threshold -- a common choice is to have the resulting branch cut run along the positive real $s$ axis and to consider physical scattering to occur just above that cut (at $s+i\epsilon$). Passing through the cut from above takes one from the ``first" or ``physical" sheet to the ``second" or ``unphysical" sheet. Poles off the real axis should not appear on the physical sheet but may be present on unphysical sheets where they appear in complex-conjugate pairs, $s_\mathrm{r} \pm i s_\mathrm{i}$. These poles correspond to resonances and a common convention is to express the pole position in the lower half-plane as $\sqrt{s} = m - i \Gamma/2$, calling $m$ and $\Gamma$, the pole mass and width of the resonance respectively. 

Close to a pole, the elements of the $t$-matrix can be expressed as
\begin{equation}
t_{ij}(s \sim s_0 ) \sim \frac{c_i \, c_j}{s_0 - s},
\end{equation}
where the residue of the pole has been factorized into couplings that can be interpreted as the coupling of the resonance to the channels $i,j$ (not to be confused with the couplings $g_i$ in the $K$-matrix which do not in general have a simple physical interpretation).

In single-channel scattering there are just two sheets and they may be differentiated by the sign of the imaginary part of $k$ -- on the physical sheet, we have $\mathrm{Im}\, k > 0$ while on the unphysical sheet we have $\mathrm{Im}\, k < 0$. If there are multiple scattering channels the number of sheets increases, but it remains possible to label them in a similar manner. For example, in our two channel case, $\pi K$, $\eta K$, the physical sheet, sheet $\mathsf{I}$, corresponds to $\mathrm{Im}\, k_{\pi K} > 0$, $\mathrm{Im}\, k_{\eta K} > 0$. The most relevant unphysical sheet, usually called sheet $\mathsf{II}$, reached by going through the $\pi K$ cut, but not the $\eta K$ cut, has $\mathrm{Im}\, k_{\pi K} < 0$, $\mathrm{Im}\, k_{\eta K} > 0$. Another unphysical sheet, sheet $\mathsf{III}$, has $\mathrm{Im}\, k_{\pi K} < 0$, $\mathrm{Im}\, k_{\eta K} < 0$, while sheet $\mathsf{IV}$ has $\mathrm{Im}\, k_{\pi K} > 0$, $\mathrm{Im}\, k_{\eta K} < 0$. In coupled-channel scattering, a pole corresponding to a single resonance can appear on more than one unphysical sheet and may not have precisely the same position or residue on different sheets.

It is also possible to have pole singularities of $t(s)$ on the real axis below threshold. If such a pole occurs on the physical sheet it corresponds to a bound-state, while if it appears on an unphysical sheet it is termed a ``virtual bound state". A familiar example is in nucleon-nucleon scattering where the attractive triplet channel contains a bound-state pole (the deuteron), while the singlet channel is not attractive enough to support a bound-state, but does feature a virtual bound-state pole.


In the previous section we obtained parameterized descriptions of scattering amplitudes. These were constrained by comparison between the finite-volume spectra such amplitudes imply (according to Eq.~\ref{eq_luescher_t}) and the finite-volume spectra obtained in explicit lattice QCD computation. Thus far we have presented only the behavior of these amplitudes for real values of $s$. We now turn to the structure of these amplitudes for complex values of $s$, and in particular the presence of any pole singularities. 

In the $\pi K$ $P$-wave, we described spectra over a limited energy region around threshold by a Breit-Wigner form, Eq.~\ref{bw}, and over a larger region up to $\eta K$ threshold using a single-channel $K$-matrix, Eq.~\ref{elastic_SP_K}. The Breit-Wigner amplitude has a pole on the physical sheet on the real axis at ${a_t \sqrt{s_0} = 0.16477(17)}$, and the $K$-matrix has a pole on the physical sheet on the real axis, $a_t \sqrt{s_0} = 0.16474(10)$, with couplings ${c_{\pi K} = i\, (9.8\pm 1.4) \!\times\! 10^{-3} \cdot a_t^{-1}}$ in the Breit-Wigner case and ${i\, (10.0\pm 0.7) \!\times\! 10^{-3} \cdot a_t^{-1}}$ in the $K$-matrix case. We thus interpret this as a vector bound-state.

In the coupled $\pi K, \eta K$ $S$-wave, we obtained a description of spectra using a two-channel $K$-matrix with parameters given in Eq.~\ref{global}. The corresponding $t$-matrix is found to have poles in the lower half-plane of the unphysical sheets $\mathsf{II}$ and $\mathsf{III}$ at positions
\begin{alignat}{2}
&a_t \sqrt{s_0}\big|_{\mathsf{II}}  &&= 0.2473(37) - \tfrac{i}{2} 0.099(14) \nonumber \\
&a_t \sqrt{s_0}\big|_{\mathsf{III}} &&= 0.2563(27) - \tfrac{i}{2} 0.089(7), \nonumber
\end{alignat}
with couplings

\vspace{0.2cm}
\begin{tabular}{r|cc}
						& $a_t\, c_{\pi K}$ 						& $a_t\, c_{\eta K}$ \\
						\hline\\[-1.4ex]
sheet $\mathsf{II}$\; 	& $0.191(18)\, e^{i\pi\, 0.015(23) }$\;		& $0.076(32)\, e^{i\pi\, 0.41(8) }$ \\[0.6ex]
sheet $\mathsf{III}$\;	& $0.164(11)\, e^{i\pi\, 0.064(14) }$\;		& $0.052(11)\, e^{i\pi\, 0.27(10) }$
\end{tabular}
\vspace{.5cm}

\noindent
which may admit an interpretation as a broad resonance with large coupling to $\pi K$ and small coupling to $\eta K$. 

The same $S$-wave amplitude, Eq.~\ref{global}, is found to have another set of poles fairly close to the physical scattering region -- on each of sheets $\mathsf{II}$ and $\mathsf{III}$ there is a pole on the real $s$-axis below $\pi K$ threshold, located at $a_t \sqrt{s_0} = 0.120(8)$ on sheet $\mathsf{II}$ and $a_t \sqrt{s_0} = 0.147(7)$ on sheet $\mathsf{III}$. The coupling to $\pi K$ on $\mathsf{II}$ is $a_t \,c_{\pi K} = 0.114(5)\, i$ while the coupling to the $\eta K$ channel is smaller and badly determined. Thus we find that this amplitude features a virtual bound-state as well as a broad resonance and this feature may account for the relatively rapid rise of the phase-shift at threshold. 

The poles presented above might be considered to be relatively far from the physical scattering region, and this leads us to question whether poles at those positions are truly required to describe the real-$s$ behavior of the scattering amplitudes. In the next section we will find corresponding poles in roughly the same locations when describing the data using a wider range of $K$-matrix parameterizations, which does suggest that the singularity structure is not merely a result of the particular parameterization form utilized.

We have so far not considered another important class of singularities in the $t$-matrix, the ``left-hand" cuts which occur in the simplest case on the real $s$ axis below all kinematic thresholds. These can be thought of as being related to the ``forces" between hadrons, or as the effects of crossed-channel processes. For example, exchange of a meson of mass $\mu$ in the $t$-channel, when projected into \mbox{$s$-channel} partial-waves gives a logarithm with a cut starting at $s = -\mu^2 + 2(m_1^2+ m_2^2)$. The kind of $K$-matrix parameterizations we have used do not feature any such cuts and thus can only be considered to provide a description of the scattering amplitude in a limited energy region. If the left-hand cuts of an amplitude are sufficiently close to the energy region being considered then they should not be neglected -- an example would be $\pi\pi$ $I=0$ scattering near threshold, where the left-hand cut, beginning at $s=0$, is as close to the physical scattering region as the nearest resonant pole, the $\sigma$. We can estimate the position of the onset of the left-hand cut in our case using $K^\star$ exchange in the $t$-channel using the $K^\star$ bound-state mass determined above. This leads to a cut starting at $s = \big(0.032(1) \,a_t^{-1}\big)^2$. We note that the virtual bound-state pole discussed above is much closer to the physical region than this cut. The possibility of an Adler zero in the amplitude has not been explored at this stage -- it is not clear whether such features of chiral symmetry breaking are relevant in a calculation with $m_\pi \sim 400 \,\mathrm{MeV}$.

We found a description of the $D$-wave coupled $\pi K, \eta K$ amplitude in Eq.~\ref{Dfit}. We remind the reader that this result is not as rigorous as the $S,P$-waves presented above owing to our lack of consideration of the $\pi\pi K$ channel which is kinematically open in the energy region we described. The resulting $t$-matrix has resonance poles at
\begin{alignat}{2}
&a_t \sqrt{s_0}\big|_{\mathsf{II}}  &&= 0.2784(12) - \tfrac{i}{2} 0.0110(21) \nonumber \\
&a_t \sqrt{s_0}\big|_{\mathsf{III}} &&= 0.2785(12) - \tfrac{i}{2} 0.0117(13),
\end{alignat}
with a coupling to $\eta K$ that is consistent with zero and a coupling to $\pi K$ of value $a_t c_{\pi K} = 0.0628(31) e^{-i \pi\, 0.030(10)}$ on sheet $\mathsf{II}$ and a statistically compatible value on sheet $\mathsf{III}$. This $J^P=2^+$ pole is much closer to the real axis than the $0^+$ pole presented earlier, corresponding to a narrower resonance that appears to be only coupled to $\pi K$.


\subsection{Varying the parameterizations}
\label{sec_fit_systematics}

\begin{table*}
\begin{tabular}{r|c|c|c}
name & equation & num. params & $\chi^2/N_\mathrm{dof}$ \\
\hline&&\\[-0.9ex]
$K$-matrix pole + const 		& $K_{ij} =\frac{g_i g_j}{m^2 -s } + \gamma_{ij}$		& 6 & 0.89 \\[0.5ex]
$K$-matrix pole + linear 		& $K_{ij} = \frac{g_i g_j}{m^2 -s } + \gamma_{ij} s$		& 6 & 0.93 \\[0.5ex]
$K^{-1}$ poly $\{1,0,1\}$		& $K^{-1} = \begin{bmatrix} c^{(0)}_{\pi K, \pi K} + c^{(1)}_{\pi K, \pi K} s & c^{(0)}_{\pi K, \eta  K} \\ c^{(0)}_{\pi K, \eta  K} &  c^{(0)}_{\eta K, \eta K} + c^{(1)}_{\eta K, \eta K}s  \end{bmatrix}$ & 5 & 0.93 \\[0.5ex]
$K^{-1}$ poly $\{2,0,1\}$		& $K^{-1} = \begin{bmatrix} c^{(0)}_{\pi K, \pi K} + c^{(1)}_{\pi K, \pi K} s + c^{(2)}_{\pi K, \pi K} s^2 & c^{(0)}_{\pi K, \eta  K} \\ c^{(0)}_{\pi K, \eta  K} &  c^{(0)}_{\eta K, \eta K} + c^{(1)}_{\eta K, \eta K}s  \end{bmatrix}$ & 6 & 0.90 \\[0.5ex]
$K^{-1}$ poly $\{1,1,1\}$		& $K^{-1} = \begin{bmatrix} c^{(0)}_{\pi K, \pi K} + c^{(1)}_{\pi K, \pi K} s  & c^{(0)}_{\pi K, \eta  K} + c^{(1)}_{\pi K, \eta K} s \\ c^{(0)}_{\pi K, \eta  K} +  c^{(1)}_{\pi K, \eta K} s &  c^{(0)}_{\eta K, \eta K} + c^{(1)}_{\eta K, \eta K}s  \end{bmatrix}$ & 6 & 0.95 \\[0.5ex]
$K^{-1}$ poly $\{1,0,0\}$		& $K^{-1} = \begin{bmatrix} c^{(0)}_{\pi K, \pi K} + c^{(1)}_{\pi K, \pi K} s  & c^{(0)}_{\pi K, \eta  K}  \\ c^{(0)}_{\pi K, \eta  K}  &  c^{(0)}_{\eta K, \eta K}   \end{bmatrix}$ & 4 & 0.93\\[0.5ex]
$K^{-1}$ poly $\{2,0,0\}$		& $K^{-1} = \begin{bmatrix} c^{(0)}_{\pi K, \pi K} + c^{(1)}_{\pi K, \pi K} s + c^{(2)}_{\pi K, \pi K} s^2  & c^{(0)}_{\pi K, \eta  K}  \\ c^{(0)}_{\pi K, \eta  K}  &  c^{(0)}_{\eta K, \eta K}   \end{bmatrix}$ & 5 & 0.93\\[0.5ex]
$K^{-1}$ poly $\{2,1,0\}$		& $K^{-1} = \begin{bmatrix} c^{(0)}_{\pi K, \pi K} + c^{(1)}_{\pi K, \pi K} s + c^{(2)}_{\pi K, \pi K} s^2  & c^{(0)}_{\pi K, \eta  K} + c^{(1)}_{\pi K, \eta K} s \\ c^{(0)}_{\pi K, \eta  K} + c^{(1)}_{\pi K, \eta K} s &  c^{(0)}_{\eta K, \eta K}   \end{bmatrix}$ & 6 & 0.87
\end{tabular}

\caption{Parameterizations of coupled-channel $J^P=0^+$ $t$-matrix.}
\label{param}
\end{table*}

Parameterizing the $t$-matrix using the $K$-matrix given in Eqs. \ref{eq_t_matrix_k} and \ref{k-matrix} is an arbitrary choice that was selected because it respects physically important properties of the $t$-matrix such as unitarity, but also has the flexibility to describe the physics present in resonant and non-resonant coupled-channel scattering. It is important to establish that the resonance properties presented above are generic properties of the scattering amplitude and not specific to the particular choice of parameterization we have made.

\begin{figure}
\includegraphics[width=0.5\textwidth]{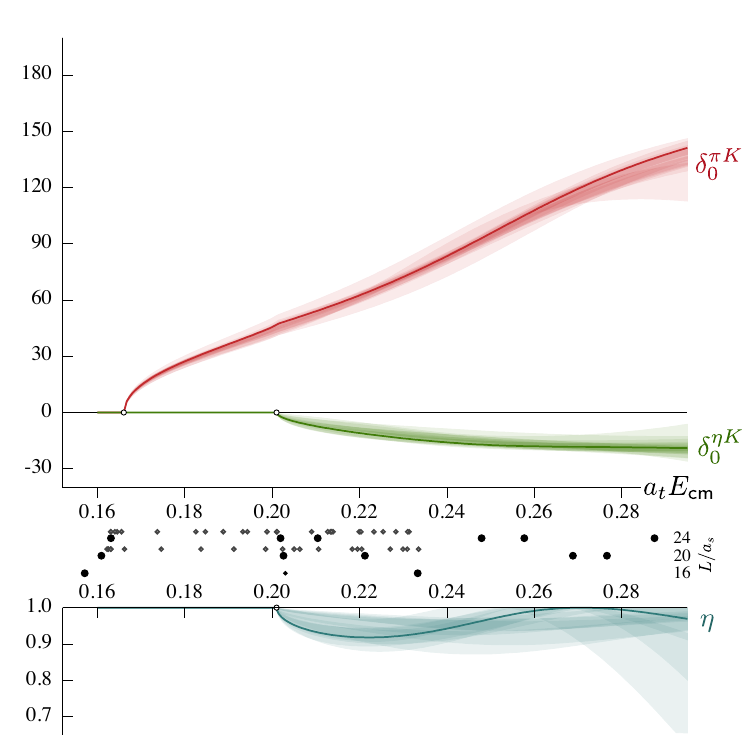}
\caption{Variation of the $J^P=0^+$ scattering amplitudes under changes in $K$-matrix parameterization -- described in Table \ref{param}. The solid line shows the result of the ``pole plus constant" form, Eq.~\ref{global}, previously presented. Each band shows the 1$\sigma$ statistical variation on the phase-shifts and inelasticity for the entries in Table~\ref{param}.
}
\label{param_var}
\end{figure}

\begin{figure*}
\includegraphics[width=1.0\textwidth]{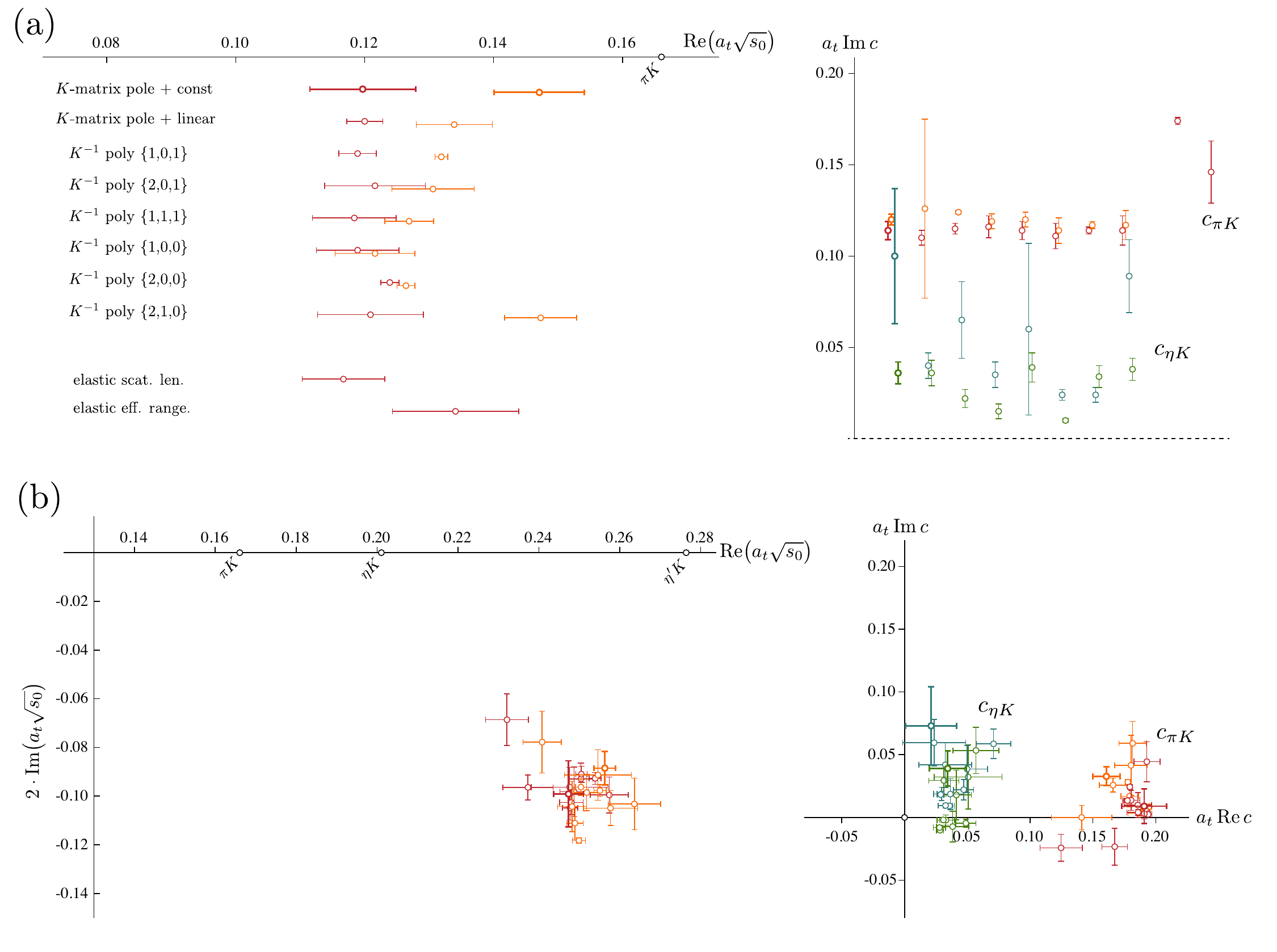}
\caption{ Complex-$s$ plane singularities of the $S$-wave amplitude.
(a) virtual bound-state position (left) and channel couplings (right) under parameterization variation. In the left plot the red points indicate the pole position on sheet $\mathsf{II}$ while the orange points indicate sheet  $\mathsf{III}$. The lowest two points correspond to fits to the elastic $\pi K$ scattering region using a scattering length or scattering length plus effective range parameterization. In the right plot red/orange points represent $c_{\pi K}$ on sheets $\mathsf{II}$/$\mathsf{III}$ and blue/green points represent $c_{\eta K}$ on sheets $\mathsf{II}$/$\mathsf{III}$. 
(b) resonance pole position (left) and channel couplings (right). Color scheme as above. 
}
\label{fig_poles_residues}
\end{figure*}

A range of possible parameterizations have been explored to describe the finite-volume spectra -- a subset are presented in Table \ref{param}. Many of them are based upon the form given in Eq.~\ref{k-poly-inv} and are labelled by the order of the polynomial in each entry of $K^{-1}$: $\{N_{\pi K, \pi K}, N_{\pi K, \eta K}, N_{\eta K, \eta K} \}$. The resulting amplitudes are presented in Figure~\ref{param_var} where we observe that they all show the same gross structure. The pole positions and residues in the corresponding $t$-matrices prove to vary rather little under the parameterization chnages, indicating that the particular form of the amplitude is not overly biasing the resonance determination. We plot the positions of the resulting poles and their associated residues in Fig.~\ref{fig_poles_residues}. 

In Figure~\ref{fig_poles_residues}, the only place that we see any significant variation between parameterizations is for the virtual bound-state in the $J^P=0^+$ channel. The variation occurs whenever we use a simple phase-space description (rather than the Chew-Mandelstam form) and the effect may be due to the form of the analytic continuation of the phase-space factors through the lowest threshold. For example, scattering length fits implicitly use a simple phase-space prescription, and this is not particularly well behaved far below threshold. Conversely, the Chew-Mandelstam form varies only slowly below threshold, which is one of the principal reasons for using it. Using the simple phase-space in a coupled $\pi K, \eta K$ $K$-matrix description can lead to spurious poles with large $\eta K$ coupling below $\pi K$ threshold which originate in the unrealistic behavior of the $\eta K$ phase-space far below the $\eta K $ threshold.



\subsection{Experimental and theoretical comparisons}

With a description of the resonant content of our amplitudes in hand, we proceed to compare our results to previous lattice QCD calculations and, recalling that the computation is performed with 391 MeV pions, to compare qualitatively to experimental observations. We present results in physical units using the scale-setting procedure outlined at the end of Section \ref{sec_calc_details}.

Beginning with the $J^P=1^-$ $\pi K$ amplitude, we may compare with the corresponding $\pi\pi$ $I=1$ amplitude that we computed on the same lattices in \cite{Dudek:2012xn}. There we found a narrow $\rho$ resonance, lying only slightly above the $\pi\pi$ threshold. In this case we find that the strange vector resonance, the $K^\star$, appears to be a bound-state only slightly below the $\pi K$ threshold. That the $K^\star$ does not appear as a resonance is almost certainly an accident of the quark masses used; a slightly larger quark mass would lead to a more deeply bound state and a slightly lighter quark mass to a resonant state whose width would increase with decreasing quark mass as the available phase-space increases. At this quark mass, we find the vector bound-state to lie at a pole position $m = 933(1)\,\mathrm{MeV}$ for any sensible parameterization. Using a relativistic Breit-Wigner form, Eq.~\ref{eq_bw}, even in this case of a bound-state, to describe energies straddling the $\pi K$ threshold gives $m_R = 933(1)$ MeV and $g_R= 5.93(26)$. This coupling can be compared to the coupling extracted from the physical mass and width \cite{Beringer:1900zz}, $g_R^{\mathrm{phys.}} = 5.52(16)$. There is reasonable agreement which may signal that the proposed approximate quark-mass independence of $g_R$ for vector mesons \cite{Kawarabayashi:1966kd,Riazuddin:1966sw, Nebreda:2010wv} may even extend to the case when the state goes below threshold.

In this calculation we are restricted from saying anything about higher vector resonances owing to our neglect of $\pi\pi K$ and other multi-hadron channels. Our large basis of $q\bar{q}$-like operators do show overlap onto high-lying levels that we might identify as corresponding to the presence of excited vector mesons~\cite{Dudek:2010wm}, but without including three-meson operators and considering an extension of Eq.~\ref{eq_luescher_t} to include three-body channels we cannot rigorously determine scattering amplitudes and their resonant content.

The $J^P=0^+$ $\pi K$, $\eta K$ partial-wave contains a broad scalar resonance with pole mass and width\footnote{in this section we expand our uncertainties to include a spread over reasonable parameterizations forms -- see the previous section.} of ${m = 1370(45)\,\mathrm{MeV}}$, ${\Gamma = 530(45)\,\mathrm{MeV}}$. The couplings $\big| c_{\pi K} \big| = 1050(110)\,\mathrm{MeV}$, $\big| c_{\eta K} \big| = 400(170)\,\mathrm{MeV}$ indicate that the resonance dominantly couples to $\pi K$. This state has a significantly larger width than the experimental $K^\star_0(1430)$, which it otherwise resembles.

Lang {\it et al}, Ref.~\cite{Lang:2012sv}, in a calculation without dynamical strange quarks in a 2 fm box with 266 MeV pions, compute the rest-frame $A_1^+$ spectrum, and extract a subthreshold energy level plus one other level below their $\pi\pi\pi K$ threshold. They did not attempt to describe the resonant content of the amplitude.

Considering $S$-wave scattering close to threshold, we may describe the amplitude in terms of a scattering length, $\lim_{k\to 0} \, k \cot \delta_{\ell=0} = 1/a_{\ell=0}$. The value we extract depends slightly upon whether we describe only the elastic scattering region with a scattering length parameterization, Eq.~\ref{elastic_SP_K}, or if we extract the threshold behavior of our more global fit, Eq.~\ref{global}, which also describes the scalar resonance discussed above. For these two descriptions we find $m_\pi \!\!\cdot\! a_{\ell=0} = 1.20(6), 1.00(6)$, or in physical units, $a_{\ell=0} = 0.60(3), 0.50(3) \,\mathrm{fm}$, respectively. This scattering length is consistent with values found in other lattice QCD computations~\cite{Lang:2012sv,Sasaki:2013vxa} at similar quark masses.

The physical $\pi K$ $J^P=0^+$ amplitude at low energy has long been suspected to be strongly influenced by the presence of a broad resonance called the $\kappa$, the strange analogue of the $\sigma$ in $\pi\pi$ scattering. The most precise estimate of the low-energy physical amplitude is obtained from the Roy-Steiner equations that incorporate analyticity, unitarity and crossing symmetry~\cite{Buettiker:2003pp,DescotesGenon:2006uk} together with the available low-energy scattering data, through dispersion equations, to show that the amplitude features a somewhat distant pole on the unphysical sheet identified as the $\kappa$. Nebreda and Pel\`aez \cite{Nebreda:2010wv} consider what happens to the $\kappa$ as the pion mass increases away from its physical value. Using the Inverse Amplitude Method to unitarize $SU(3)$ chiral perturbation theory at one loop level, they find that as the pion mass increases from its physical value, the distant $\kappa$ poles on the unphysical sheet of $\pi K$ scattering move toward the real axis, becoming a single pole on the real axis below threshold but still on the unphysical sheet, i.e. a virtual bound-state. As the pion mass is increased further, the pole separates into two which then leap onto the physical sheet becoming bound states. 

In the qualitative picture laid out by Nebreda and Pel\`aez, our calculation at $m_\pi = 391\,\mathrm{MeV}$ appears to be in the intermediate region in which the $\kappa$ appears as a virtual bound-state. In all successful descriptions of the finite-volume spectrum we found a virtual bound-state, although its precise pole position did depend upon the parameterization used.

The $J^P=2^+$ $\pi K, \eta K$ partial-wave was found to feature a narrow resonance, essentially decoupled from $\eta K$, with pole mass $m = 1576(7)\,\mathrm{MeV}$ and pole width $\Gamma = 62(12)\,\mathrm{MeV}$. This state closely resembles the experimental $K_2^\star(1430)$ in most regards apart from one: we extracted this state neglecting altogether the kinematically open $\pi\pi K$ channel, while the physical state has a 50\% branching fraction into $\pi\pi K$.

\section{$\pi K$ scattering with $I=3/2$}
\label{sec_three_half}

In addition to $\pi K,\eta K$ scattering with $I=\frac{1}{2}$, we have also obtained correlation functions for the $\pi K$ $I=\frac{3}{2}$ channel. In this flavor-exotic sector the calculation is somewhat simpler: Quark line annihilations do not feature and ``single-meson" operators with $q\bar{q}$-like structure cannot appear. Inelasticity can appear through $\pi\pi K$ in $P$-wave and higher and $\pi \pi \pi K$ in all waves. Experimentally~\cite{Estabrooks:1977xe} we know that the scattering is weak and repulsive in $S,P$ and $D$-waves with no sign of resonant behavior in the energy region up to 1.72 GeV.

\begin{figure}
\includegraphics[width=0.92\columnwidth]{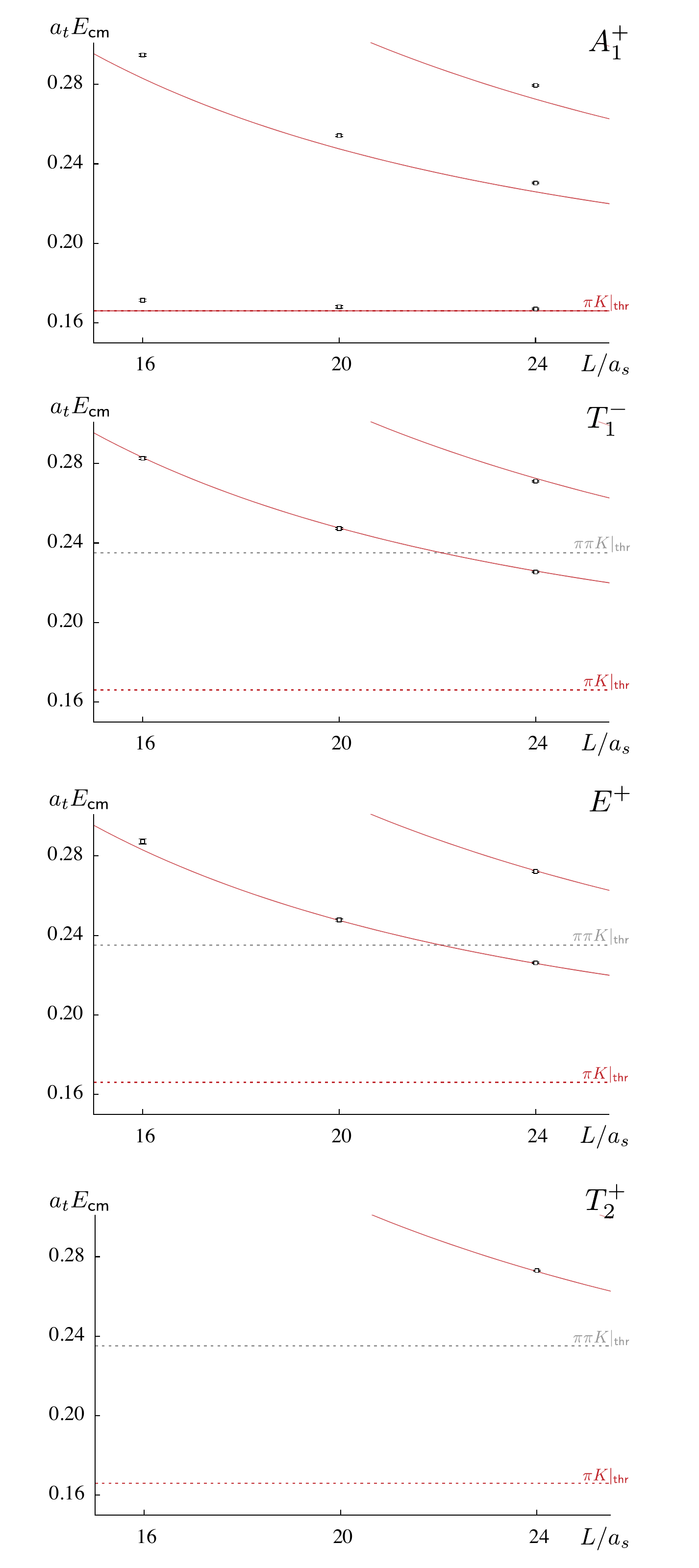}
\caption{$\pi K$ $I=3/2$ spectra with $\vec{P}=[000]$.}
\label{spectrum_I3o2_P000}
\end{figure}

\subsection{Finite-volume spectrum}

\begin{figure*}
\includegraphics[width=1.00\textwidth]{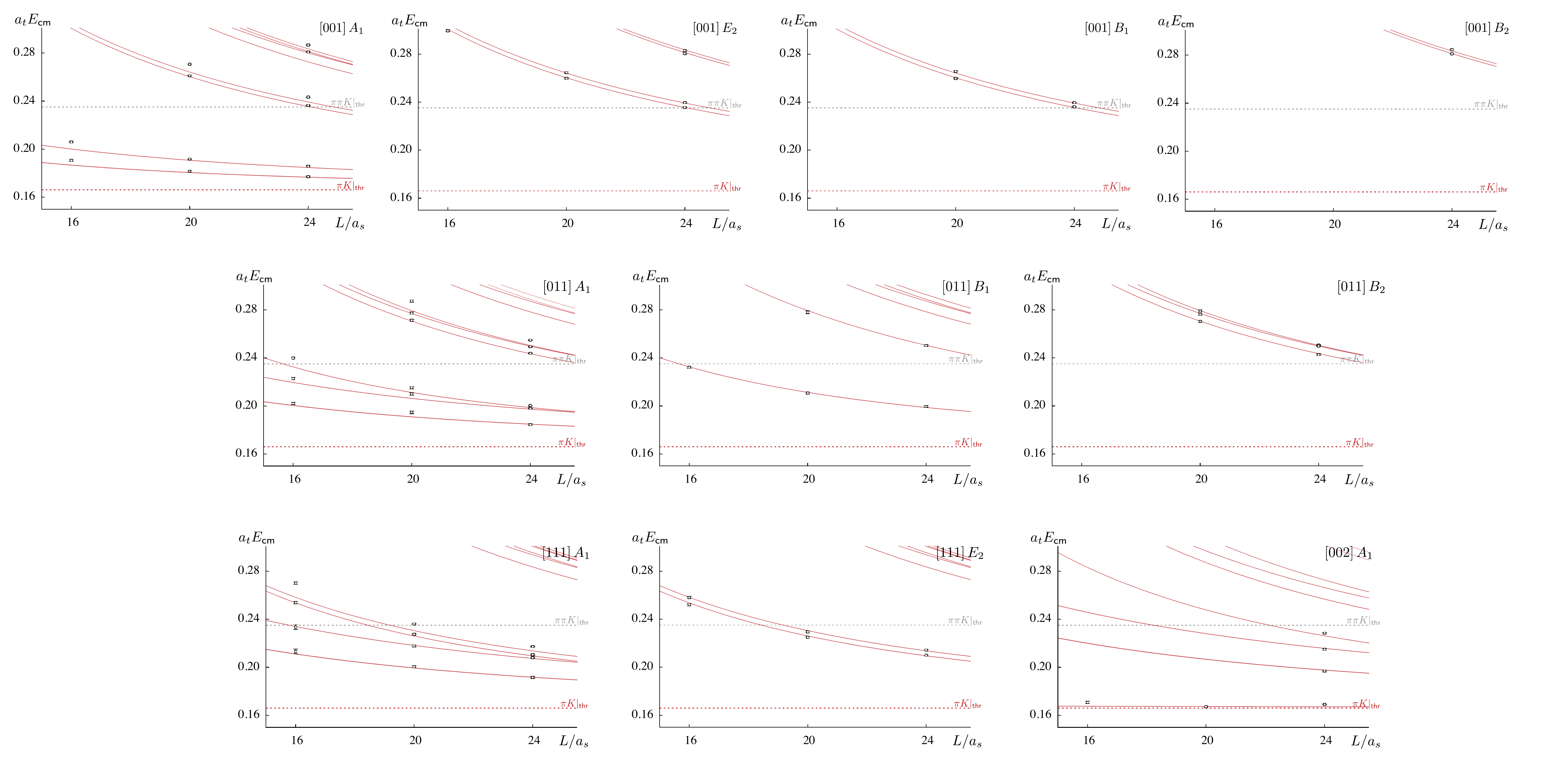}
\caption{$\pi K$ $I=3/2$ spectra with $\vec{P}=[001],\, [011], \, [111], \, [002]$. Note that in the $\vec{P}=[002]$, $A_1$ case there is a non-interacting level only slightly above threshold.}
\label{spectrum_I3o2_P100}
\end{figure*}

The spectra are obtained as described above for the $I=\frac{1}{2}$ case. The key difference is that there are no ``single-meson" operators so our basis is built entirely from $\pi K$ operators as described by Eq.~\ref{eq_op_meson_pair}. The contributions of each partial-wave in each lattice irrep are as given in Table~\ref{tab_pwa_irrep}. 

In Fig.~\ref{spectrum_I3o2_P000} we show the energies determined when the system is at rest with respect to the lattice. In $A_1^+$, which has overlap onto the $S$-wave, significant positive shifts with respect to the non-interacting energies are observed, which likely indicates some repulsion in the system. In $T_1^-$, which overlaps onto $P$-wave and higher, small negative shifts are observed which only become significant above $\pi\pi K$ threshold. In $E^+$ and $T_2^+$, the $D$-wave is the lowest contributing partial-wave, and we see no significant shifts from the non-interacting spectrum.

Similar patterns are visible in the data obtained when the $\pi K$ systems are considered in-flight, Figure \ref{spectrum_I3o2_P100}. The largest shifts are observed in the in-flight $A_1$ irreps, presumably due to the $S$-wave interaction.

The situation is very similar to that observed in the corresponding $\pi\pi$ $I=2$ calculations \cite{Dudek:2012gj}. Investigating the $\pi K$ operator overlaps for each extracted eigenstate we find that the basis is approximately orthogonal, corresponding to the eigenstates being relatively similar to the non-interacting states, as was presented in Figure 11 of \cite{Dudek:2012gj}.

\subsection{Scattering amplitudes}

\begin{figure}
\includegraphics[width=0.49\textwidth]{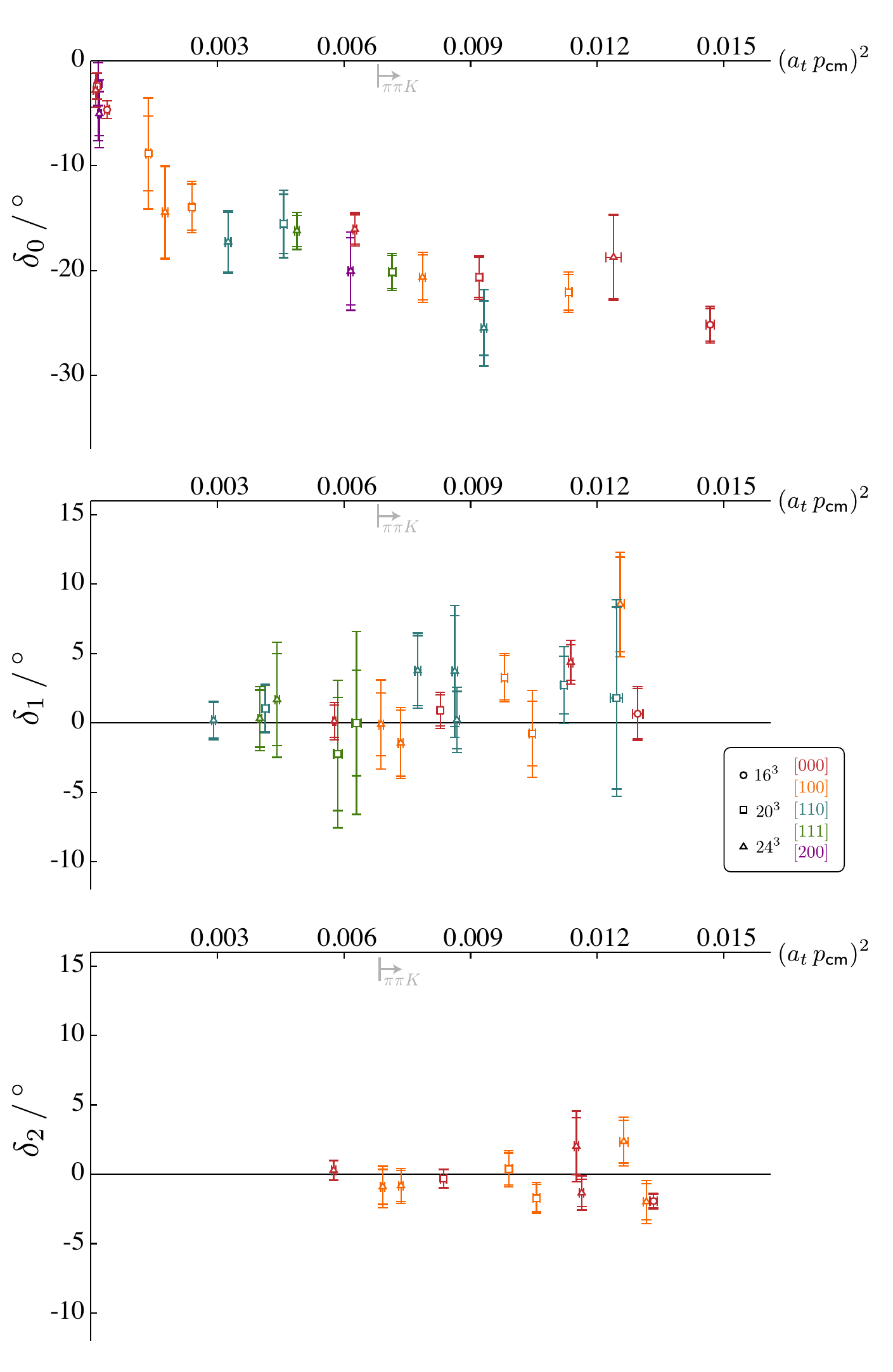}
\caption{The phase shift points obtained by applying Eq.~\ref{eq_luescher_t} directly to the energy levels shown in Figs.~\ref{spectrum_I3o2_P000}, \ref{spectrum_I3o2_P100}. In $\ell=1$ and $\ell=2$ effects due to $\pi\pi K$ inelasticities above $a_t E_\cm = 0.235$ have been neglected. The innermost errorbars follow from the statistical uncertainty on the energy levels, while the outer errorbars include variation of $m_K, m_\pi$ and $\xi$ within their uncertainties.}
\label{elastic_phase_shift_I3o2_data}
\end{figure}

We proceed as for the $I=\frac{1}{2}$ case, parameterizing the infinite-volume $t$-matrix using a simple model. In this case we only consider elastic amplitudes in the channel $\pi K \to \pi K$. We begin with the $A_1^+$ irrep, having overlap onto $\ell=0, 4$ and higher. We will neglect the role of an $\ell=4$ amplitude (and higher) over the energy region we consider on the grounds that it will be highly suppressed by the angular-momentum barrier. In the left-most plot in Fig.~\ref{spectrum_I3o2_P000}, seven levels are shown below $\pi\pi\pi K$ threshold, and we begin by fitting these using a scattering length parameterization, given by Eq.~\ref{eq_ere} with $r_{\ell=0}=0$, obtaining

\vspace{0.1cm}
\begin{tabular}{rll}
$a_0=$ & $( -3.81 \pm 0.14 \pm 0.14) \cdot a_t$\\
&\multicolumn{2}{l}{ $\chi^2/ N_\mathrm{dof} = \frac{2.03}{7 - 1} = 0.34 $} \,.
\end{tabular}
\vspace{0.2cm}

\noindent No improvement is obtained by allowing an effective range term in the fit, with the determined $a_{\ell=0}$ and $r_{\ell=0}$ being highly correlated. 

In addition to $A_1$ irreps at-rest and in-flight, which have $\ell=0$ as their lowest contributing partial wave, we may consider irreps which have $\ell=1$ and $\ell=2$ as the lowest partial wave. We first assume $\ell=3$ and higher are negligible. There are eight data points in the elastic region below the $\pi\pi K$ threshold at $a_t E_\cm=0.235$, and parameterizing the ${\ell=1,2}$ amplitudes by scattering lengths we obtain a fit

\vspace{0.1cm}
\begin{tabular}{rll}
$a_1=$ & $(-2.1 \pm 29.3 \pm 25.8) \cdot a_t^3$ & \multirow{2}{*}{ $\begin{bmatrix*}[r] 1 & -0.34 \\ & 1\end{bmatrix*}$ } \\
$a_2=$ & $(-2.8 \pm 1.8 \pm 2.1)\times 10^3 \cdot a_t^5$   & \\[1.3ex]
&\multicolumn{2}{l}{ $\chi^2/ N_\mathrm{dof} = \frac{2.33}{8 - 2} = 0.39\, , $} \\
\end{tabular}
\vspace{.2cm}

\noindent indicating no significant interaction in the elastic region for $P$ and $D$-waves. If we assume that there is negligible inelasticity into $\pi\pi K$ at low-energy, we may consider the other energy level values we have obtained up to $\pi\pi\pi K$ threshold. There are a total of 31 points relaxing this restriction, and a scattering length description gives,

\vspace{0.2cm}
\begin{tabular}{rll}
$a_1=$ & $(\,\,\,\; 42.4 \pm 4.7 \pm 13.9) \cdot a_t^3$ & \multirow{2}{*}{ $\begin{bmatrix*}[r] 1 &  0.04 \\ & 1\end{bmatrix*}$ } \\
$a_2=$ & $(-1.19 \pm 0.25 \pm  0.53)\times 10^3 \cdot a_t^5$   & \\[1.3ex]
&\multicolumn{2}{l}{ $\chi^2/ N_\mathrm{dof} = \frac{22.5}{31 - 2} = 0.77\, , $} \\
\end{tabular}
\vspace{.2cm}

\noindent which suggests there may be a slight attractive tendency in the $P$-wave at higher energy.

With the $\ell=1$ and $\ell=2$ partial-wave amplitudes determined above, we may now make use of the in-flight $A_1$ irreps to better constrain the $S$-wave scattering amplitude. The most conservative approach is to fix the contribution of $P$ and $D$ waves in the $A_1$ irreps according to the above fits and to then determine what the $S$-wave contribution must be. Doing so proves to give results essentially identical to performing a global fit where all of the $\ell=0,1,2$ amplitude parameters are allowed to float in a fit to all irreps. Considering a scattering length description of each partial-wave, the following fit describes the complete set of $A_1^+$ energy levels at rest and levels below the $\pi\pi K$ threshold in all other irreps: 

\vspace{0.2cm}
\begin{tabular}{rll}
$a_0=$ & $(-4.03 \pm 0.08 \pm 0.20) \cdot a_t$ & \multirow{3}{*}{ $\begin{bmatrix*}[r] 1 & -0.07 & -0.51 \\ & 1 &  -0.30 \\& & 1\end{bmatrix*}$ } \\
$a_1=$ & $(\,\,\,\; 50.1 \pm 17.1 \pm 24.7) \cdot a_t^3$ & \\
$a_2=$ & $(-1.08 \pm 2.80 \pm 1.62)\!\times\! 10^3 \cdot a_t^5$   & \\[1.3ex]
&\multicolumn{2}{l}{ $\chi^2/ N_\mathrm{dof} = \frac{24.9}{37 - 3} = 0.73\, . $}  \\
\end{tabular}
\vspace{0.2cm}

Again, extending the energy region up to $\pi\pi\pi K$ threshold by assuming that $\pi\pi K$ amplitudes are negligible we obtain,

\begin{tabular}{rll}
$a_0=$ & $(-4.04 \pm 0.05 \pm 0.15) \cdot a_t$ & \multirow{3}{*}{ $\begin{bmatrix*}[r] 1 & 0.01 &  0.04 \\ & 1 & 0.01 \\& & 1\end{bmatrix*}$ } \\
$a_1=$ & $(\,\,\,\;43.2 \pm 3.7 \pm 15.4) \cdot a_t^3$ & \\
$a_2=$ & $(-1.13 \pm 0.14 \pm 0.58)\!\times\! 10^3 \cdot a_t^5$   & \\[1.3ex]
&\multicolumn{2}{l}{ $\chi^2/ N_\mathrm{dof} = \frac{69.2}{75 - 3} = 0.96\, , $}  \\
\end{tabular}
\vspace{.2cm}

\noindent and as previously, including an effective range in the $S$-wave amplitude does not improve the description. Adding a scattering length amplitude for $\ell = 3$ and minimizing leads to a negligible change in the $\ell=0,1,2$ scattering lengths and a value of $a_{\ell=3}$ that is statistically compatible with zero, justifying our previous neglect of the $F$-wave.

\begin{figure}
\includegraphics[width=0.49\textwidth]{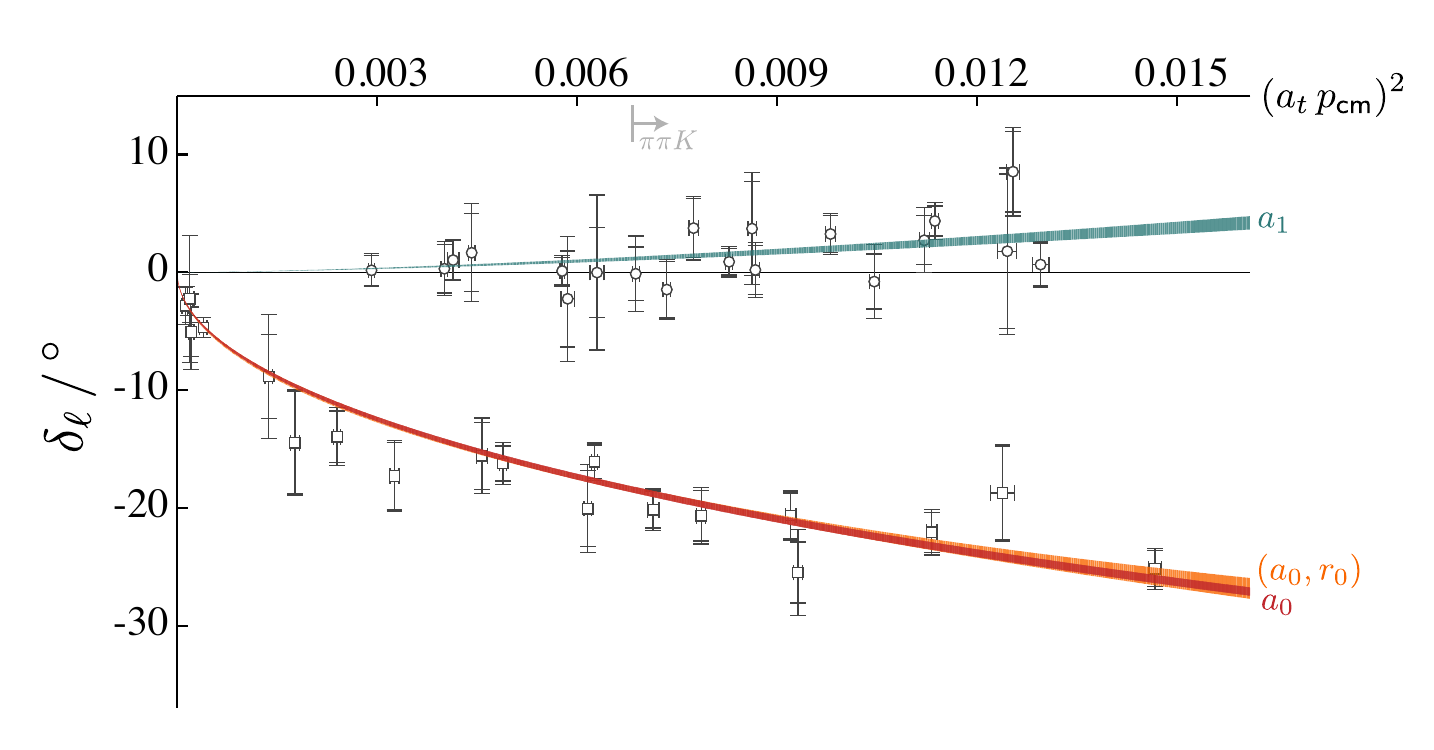}
\caption{The $\ell=0,1$ $I=3/2$ $\pi K$ phase-shifts and fits described in the text. Also shown, a fit including an $S$-wave effective range term which is observed to be negligibly different from the fit with only a scattering length.
}
\label{elastic_phase_shift_I3o2_S+P_fits}
\end{figure}

In Figure \ref{elastic_phase_shift_I3o2_data} we show phase-shift values extracted from Eq.~\ref{eq_luescher_t} assuming elastic scattering. In all cases, if more than one partial-wave appears in Eq.~\ref{eq_luescher_t}, the final parameterization given above is used to specify the higher partial-waves, with the remaining lowest partial-wave $\delta$ being extracted. We clearly see what was being described in the fits above, that the $S$-wave is significantly repulsive, while the $P$-wave may have some slight attraction at large energies and the $D$-wave is compatible with no interaction. Figure \ref{elastic_phase_shift_I3o2_S+P_fits} superimposes the phase-shift points over a plot of the parameterized solution presented above.

\begin{figure}
\includegraphics[width=0.49\textwidth]{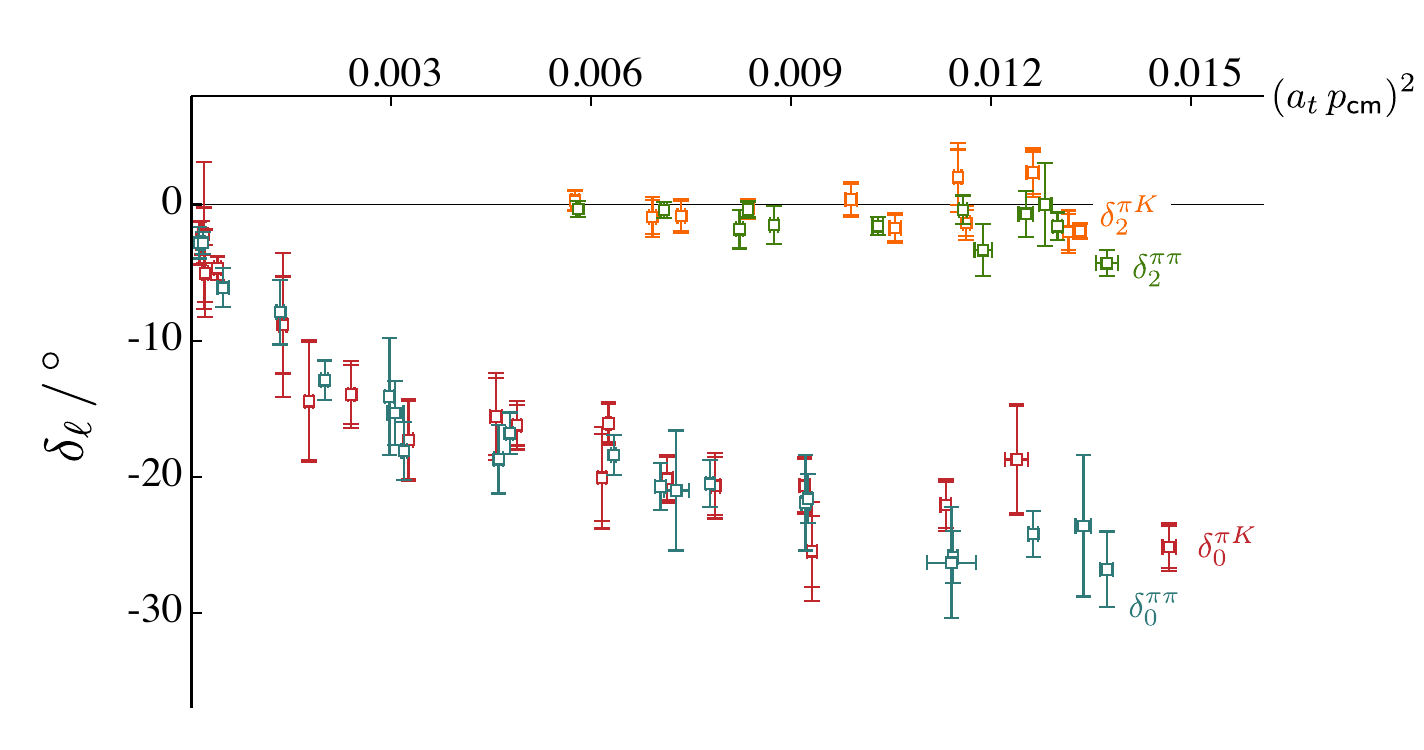}
\caption{$\ell=0,2$ scattering phase-shifts for $\pi K$ with ${I=3/2}$ (this paper) and $\pi\pi$ with $I=2$~\cite{Dudek:2012gj}. These two channels correspond to different rows of the $\mathbf{27}$-plet that appears in $\mathbf{8} \otimes \mathbf{8}$ scattering. 
}
\label{elastic_phase_shift_Kpi_I3o2+pipi_I2}
\end{figure}

Scattering with the exotic quantum numbers ${S=1,\, I=3/2}$ is similar to $\pi\pi$ $I=2$ scattering, and in the limit of equal light and strange quark masses they are identical. In our lattice calculation with $m_K/m_\pi = 1.4$, we are closer to having an $SU(3)$ flavor symmetry than in the physical limit where $m_K/m_\pi = 3.6$, and as such we might expect relatively small differences within multiplets of $SU(3)_F$. In $S$- and $D$-waves, the $I=3/2$ $\pi K$ scattering channel is part of a $\mathbf{27}$, which also contains $\pi\pi$ scattering with $I=2$. In \cite{Dudek:2012gj} we computed the corresponding $\pi\pi$ scattering amplitudes on the same gauge-field configurations using very similar techniques to those used in this paper. In Figure \ref{elastic_phase_shift_Kpi_I3o2+pipi_I2} we compare the $I=3/2,\, S=1$ and $I=2,\, S=0$ elements of the $\mathbf{27}$, observing that indeed there is very close agreement. On the other hand, the $\pi K$ $P$-wave amplitude lies in a $\mathbf{10}$ multiplet which does not contain $I=2,\, S=0$. The $\mathbf{10}$ can be constructed at the quark level if $qq\bar{q}\bar{q}$ configurations appear \cite{Chung:2003qy} -- the absence of any significant phase-shift behavior suggests that any putative $qq\bar{q}\bar{q}$ resonance must be at higher energy, although we remind the reader that we have not included explicit \emph{local} $qq\bar{q}\bar{q}$ operators in our basis.

\begin{figure}
\includegraphics[width=0.49\textwidth]{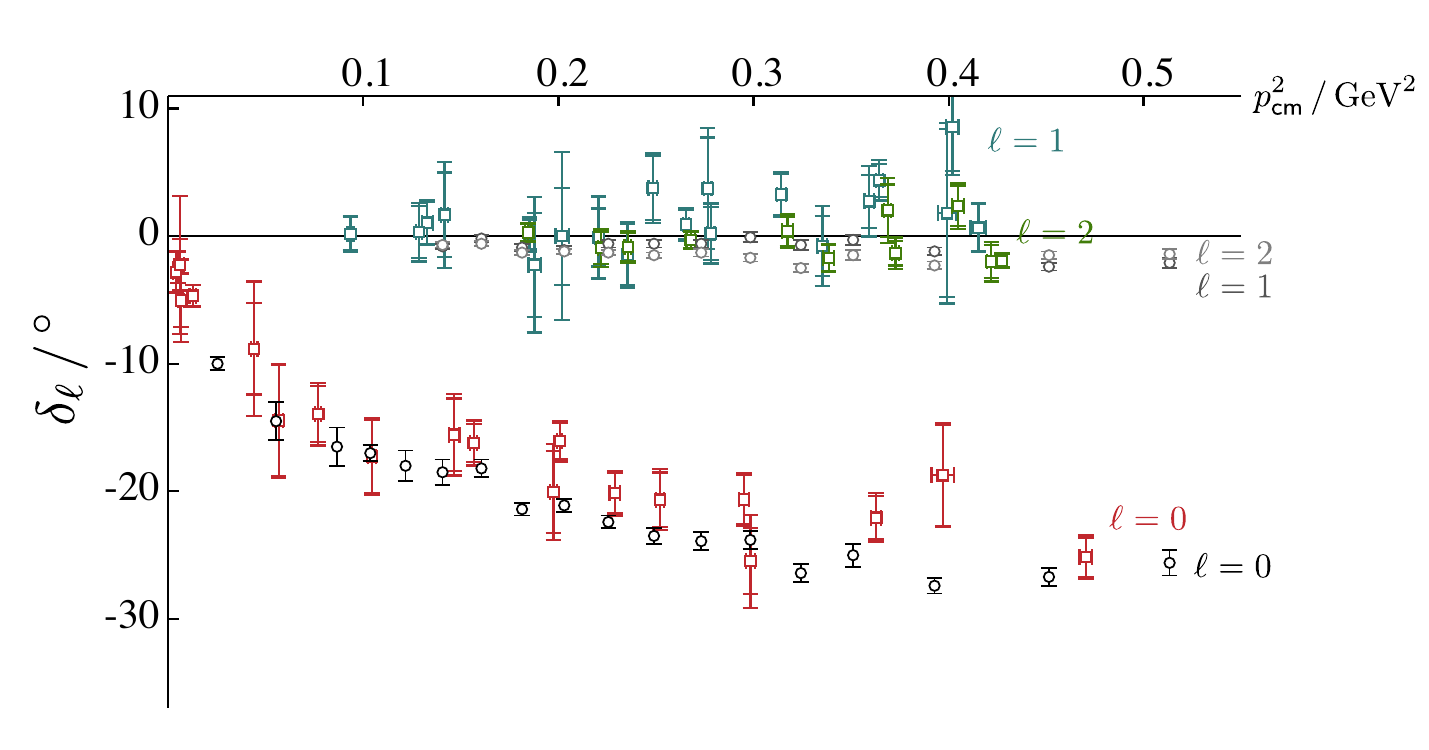}
\caption{$I=3/2$ $\pi K$ scattering. Colored points show the $\ell=0,1,2$ phase-shifts determined in this calculation with $m_\pi = 391\,\mathrm{MeV}$, with the energy scale set using the $\Omega$-baryon mass as described in Section~\ref{sec_calc_details}. Black and gray points show the Estabrooks \emph{et al} partial-wave analysis of experimental $\pi K$ scattering~\cite{Estabrooks:1977xe}. 
}
\label{elastic_phase_shift_I3o2_versus_Estabrooks}
\end{figure}

Within this calculation in which the $u,d$ quark masses are somewhat heavier than the true physical values, we are not justified in making a direct comparison of our determined phase-shifts with experimental data. Nevertheless we may superimpose the two and observe that we are replicating the qualitative features of the Estabrooks \emph{et al} partial-wave analysis \cite{Estabrooks:1977xe}, Figure \ref{elastic_phase_shift_I3o2_versus_Estabrooks}.

\subsection{Comparison to other studies}

Our best estimate of the $S$-wave scattering length when the pion mass is 391 MeV is ${m_\pi \!\cdot\! a_{\ell=0}^{I=3/2}= -0.278(15)}$, or expressed in physical units, ${a_{\ell=0}^{I=3/2}= -0.140(8)\,\mathrm{fm}}$.

A number of previous lattice QCD calculations have considered $\pi K$ scattering in isospin--3/2 at threshold~\cite{Beane:2006gj,Fu:2011wc,Lang:2012sv,Sasaki:2013vxa}. They typically extract a scattering length from the single energy level near threshold corresponding to a pion and a kaon each at rest. The scattering length has been determined for a range of quark masses. Our result for $a^{I=3/2}_{\ell=0}$ is in good agreement with Refs.~\cite{Beane:2006gj,Sasaki:2013vxa} who have obtained this quantity at similar values of $m_\pi$. 

Our result is based on a description of a much larger set of finite-volume energy levels compared to the studies above. Our aim was to obtain the energy dependence of the scattering amplitude and not just the threshold behavior. As it happens, we find that, at this pion mass, the next term in the effective range expansion, the range parameter, is consistent with zero and the extra data in our fit does not significantly improve the precision on the determination of the scattering length.

\section{Summary}
\label{sec_summary}

In this paper we have reported on the first application of the formalism relating coupled-channel scattering amplitudes to the discrete spectra of hadrons in a finite-volume. In order to overcome the underconstrained nature of the problem, where the position of each energy level is a volume-dependent function of multiple scattering amplitudes, we parameterized the energy dependence of the $t$-matrix and attempted to describe the entire spectrum globally. We found that relatively simple parameterizations, satisfying $S$-matrix unitarity, could be used successfully. 

In order to strongly constrain the energy dependence of the scattering amplitudes, we required detailed and precise excited-state spectra across a range of irreducible representations of the lattice symmetry in many moving frames. Variational analysis of correlation matrices computed using a large basis of operators, including some resembling $q\bar{q}$-like single-mesons and others resembling ``meson-meson" pairs with definite relative and total momentum, leads to such spectra. Distillation offers an efficient method of correlation construction in this case where quark-line annihilation features in a large number of required Wick contractions.

When, as in the case considered here, the scattering hadrons have unequal masses, the typical situation is for an irrep to receive contributions from a dense set of low-lying partial-waves. By computing a wide range of irreps, each featuring different combinations of partial-waves, we were able to decompose into a partial-wave basis even in this case where mixing is significant.

The parameterized scattering amplitudes obtained, constrained by over 100 real values of the energy variable, can be analytically continued into the complex energy plane, where their pole singularities correspond to resonances, bound-states etc. The residues of the amplitudes at the poles can be used to determine couplings of the states to their allowed decay channels. In our calculation at $m_\pi = 391 \,\mathrm{MeV}$ we examined the determined amplitudes for their singularity content, finding a set of states in isospin--1/2 which can be compared qualitatively with those observed in experiment.  

In the $J^P=0^+$ channel we found a broad resonance, coupled dominantly to $\pi K$ and not $\eta K$ with a pole mass of ${m = 1370(45)\,\mathrm{MeV}}$ and width of ${\Gamma = 530(45)\, \mathrm{MeV}}$. In this same $S$-wave amplitude, we found a second singularity -- a pole on the real axis below $\pi K$ threshold on unphysical sheets, a ``virtual" bound-state. These features appeared to be robust under changes in the parameterization form utilized.

We extracted a $J^P=1^-$ bound-state at $933(1)$ MeV, barely below our $\pi K$ threshold. By determining the position of the corresponding finite-volume state in many irreps, we were able to map-out the phase-shift across the threshold, giving a degree of energy dependence which allows us to extract a Breit-Wigner coupling of ${g_R = 5.93(26)}$.  

In the $J^P=2^+$ channel, if we assumed that $\pi\pi K$ was not significantly coupled to $\pi K$ or $\eta K$, we found that the spectrum obtained (without $\pi\pi K$-like operators) could be described consistently in terms of $\pi K, \eta K$ scattering with a narrow resonance coupled dominantly to $\pi K$. The resonance had a pole mass $m = 1576(7)\,\mathrm{MeV}$ and width $\Gamma = 62(12)\,\mathrm{MeV}$.

In addition to the isospin--1/2 channel, we also obtained 75 energy levels constraining scattering amplitudes in isospin--3/2 for the entire elastic scattering region for the lowest three partial-waves. This flavor-exotic process was found to have interactions that are rather weak and each partial-wave could be adequately described by a scattering length approximation. The $S$-wave scattering length was found to be ${a_{\ell=0}^{I=3/2}= -0.140(8)\,\mathrm{fm}}$ at $m_\pi = 391 \,\mathrm{MeV}$, in line with earlier lattice QCD calculations.

To compare quantitively with experimental observations we should perform calculations at the physical light quark mass, but even here at $m_\pi = 391$ MeV we may make some qualitative comparisons.

The broad scalar resonance we extract resembles somewhat the experimental $K^\star_0(1430)$, although we find a significantly larger width. The high-energy side of the projection of this resonance onto the real axis lies in the region above $\eta' K$ threshold, which in this first study we did not consider rigorously. Inclusion of $\eta' K$ operators into the variational basis may lead to an adjusted finite-volume spectrum and correspondingly altered resonance parameters. The tensor resonance we extract has some of the properties of the $K_2^\star(1430)$, notably the decay into $\pi K$ and not $\eta K$, but we lack a coupling to $\pi\pi K$ -- something that must be generated as the pion mass is reduced if the experimental state is to be described. The vector bound-state we extract is expected to become a resonance as the pion mass is decreased, the $\pi K$ threshold falls, and the phase-space for decay increases. There are theoretical expectations that the coupling does not change significantly with quark mass, and indeed we find a value that is in reasonable agreement with the value extracted from the PDG width. The property of the scattering amplitudes that we expect to change most drastically is the $\kappa$ pole; the virtual bound-state we found must, if our understanding of the experimental amplitude is correct, evolve into a resonant pole just above threshold, but far from the real axis, as the quark mass is reduced toward its physical value. We note that this is precisely the behavior suggested within unitarized chiral perturbation theory~\cite{Nebreda:2010zz}.


In the case of coupled $\pi K$, $\eta K$ scattering we have observed that there is relatively little coupling between the channels for even-$\ell$ partial-waves in the energy region considered, and as such the true diversity of possible behaviors in a coupled-channel system has not yet been explored. Further calculations are now warranted in such systems as $\pi \eta, K\overline{K}$ scattering, in which the $a_0(980)$ is expected to appear as a resonance coupled strongly to both channels, and $\pi\pi, K\overline{K}, \eta \eta$, where explanation of the scalar sector remains a phenomenological challenge.

A restriction was placed on the energy region we could consider in this calculation by the opening of the three-body $\pi\pi K$ channel. Such restrictions will only become more severe as the light quark mass is reduced towards its physical value. Including into the calculation operators resembling three-meson states presents no serious problem; a simple extension of the two-meson constructions used in this paper can be utilized. The difficulty lies in the formalism relating the finite-volume spectrum to scattering amplitudes featuring three-body states, which is not at this time completely mature, although significant progress is being made~\cite{Hansen:2013dla}. We have reason to believe that the fullest possible complexity of many-body final states may not be present within QCD -- experimentally it is observed that true high-multiplicity final states are not significantly directly populated in hadron resonance decays, rather that most decays proceed through intermediate two-body states featuring isobar resonances. Whether this simplification can be observed in amplitudes computed within QCD is a question for future computations.

The successful extraction of coupled-channel amplitudes in several partial-waves presented in this paper is an important milestone in progress towards a QCD description of the excited hadron spectrum.


 
\begin{acknowledgments}

We thank our colleagues within the Hadron Spectrum Collaboration. We also thank R. Briceno, M.R. Pennington,  C.J.Shultz and A.P. Szczepaniak for fruitful discussions. {\tt Chroma}~\cite{Edwards:2004sx} and {\tt QUDA}~\cite{Clark:2009wm,Babich:2010mu} were used to perform this work on clusters at Jefferson Laboratory under the USQCD Initiative and the LQCD ARRA project. Gauge configurations were generated using resources awarded from the U.S. Department of Energy INCITE program at Oak Ridge National Lab, the NSF Teragrid at the Texas Advanced Computer Center and the Pittsburgh Supercomputer Center, as well as at Jefferson Lab. RGE and JJD acknowledge support from U.S. Department of Energy contract DE-AC05-06OR23177, under which Jefferson Science Associates, LLC, manages and operates Jefferson Laboratory. JJD acknowledges support from the U.S. Department of Energy Early Career award contract DE-SC0006765. CET acknowledges partial support from the Science and Technology Facilities Council (U.K.) [grant number ST/L000385/1].

\end{acknowledgments}

\bibliography{kpi-refs}

\appendix


\section{$SU(3)$ flavor relations}
\label{app_su3f}

$\big(S,I,I_z\big)=\big(1,\tfrac{3}{2}, +\tfrac{3}{2}\big)$ states lie in two irreducible representations of $SU(3)_F$, the $\mathbf{27}$ and the $\mathbf{10}$. Two-meson states in the rest frame with definite angular momentum $(\ell,m)$ which transform irreducibly under $SU(3)$ can be obtained using the isoscalar factors tabulated in Ref.~\cite{deSwart:1963gc}:
\begin{align}
&\Big| \mathbf{27}; S=1, I=\tfrac{3}{2}, I_z = +\tfrac{3}{2}; \ell, m \Big\rangle \nonumber \\
&\quad=    \int \!\!d\hat{p}\; Y_\ell^m(\hat{p}) \Big[ \tfrac{1}{\sqrt{2}}\big| K^+_{\vec{p}} \pi^+_{-\vec{p}}\big\rangle + \tfrac{1}{\sqrt{2}} \big| \pi^+_{\vec{p}} K^+_{-\vec{p}}\big\rangle \Big] \nonumber \\
&\quad=  \big[\,  1 + (-1)^\ell\, \big]\, \tfrac{1}{\sqrt{2}}\! \int \!\!d\hat{p}\; Y_\ell^m(\hat{p}) \big| K^+_{\vec{p}} \pi^+_{-\vec{p}} \big\rangle,
\nonumber \\[1.2ex]
&\Big| \mathbf{10}; S=1, I=\tfrac{3}{2}, I_z = +\tfrac{3}{2}; \ell, m \Big\rangle \nonumber \\
&\quad=    \int \!\!d\hat{p}\; Y_\ell^m(\hat{p}) \Big[ \tfrac{1}{\sqrt{2}}\big| K^+_{\vec{p}} \pi^+_{-\vec{p}}\big\rangle - \tfrac{1}{\sqrt{2}}\big| \pi^+_{\vec{p}} K^+_{-\vec{p}}\big\rangle \Big] \nonumber \\
&\quad=   \big[\,  1 - (-1)^\ell\, \big]\, \tfrac{1}{\sqrt{2}}\!\int \!\!d\hat{p}\; Y_\ell^m(\hat{p}) \big| K^+_{\vec{p}} \pi^+_{-\vec{p}} \big\rangle.
\nonumber
\end{align}

\begin{widetext}

It thus follows that even-spin waves, $\ell = 0,2\ldots$, lie in the $\mathbf{27}$ representation, while odd-spin waves, $\ell=1,3\ldots$, lie in the $\mathbf{10}$ representation. Both representations are flavor exotic, in the sense that they cannot be constructed from $q\bar{q}$, but there need not be any simple relationship between them, unless specific dynamics causes there to be.

$\pi\pi$ scattering in $I=2$ is restricted to $\ell=$even and also lies in the $\mathbf{27}$ representation. In the computation presented in this paper, $m_\pi = 391$ MeV and $m_K=549$ MeV, such that $SU(3)$ flavor is an even better approximate symmetry than for physical quark masses, and as shown in Fig.~\ref{elastic_phase_shift_Kpi_I3o2+pipi_I2}, we observe rather good agreement between $\pi K$ and $\pi\pi$ in $S$- and $D$-wave scattering.

Turning to non-exotic scattering channels, we note that the Clebsch-Gordan series for ${\mathbf{8} \otimes \mathbf{8} = \mathbf{27} \oplus \mathbf{10} \oplus \overline{\mathbf{10}} \oplus \mathbf{8_1} \oplus \mathbf{8_2} \oplus \mathbf{1}}$ contains two octet representations. The $\mathbf{8_1}$ representation contains symmetric even-$\ell$ $\pi\pi$ scattering and thus couples to isoscalar states like $f_0, f_2 \ldots$ (the $\mathbf{1}$ representation also couples to these states), while the $\mathbf{8_2}$ representation contains odd-$\ell$ $\pi\pi$ scattering and hence the $\rho, \rho_3 \ldots$.

Two-meson states with $\big(S,I,I_z\big)=\big(1,\tfrac{1}{2}, +\tfrac{1}{2}\big)$ in the rest frame with definite angular momentum $(\ell,m)$ which transform irreducibly under $SU(3)$ in non-exotic multiplets are:

\begin{align}
&\Big| \mathbf{8_1}; S=1, I=\tfrac{1}{2}, I_z = +\tfrac{1}{2}; \ell, m \Big\rangle \nonumber \\
&=  \int \!\!d\hat{p}\; Y_\ell^m(\hat{p}) 
	 \Big[ \tfrac{3\sqrt{5}}{10} \Big( 
			-\!\sqrt{\tfrac{2}{3}} \big| K^0_{\vec{p}} \pi^+_{-\vec{p}} \big\rangle 
			+\sqrt{\tfrac{1}{3}} \big| K^+_{\vec{p}} \pi^0_{-\vec{p}} \big\rangle 
			-\sqrt{\tfrac{2}{3}} \big| \pi^+_{\vec{p}} K^0_{-\vec{p}} \big\rangle 
			+\sqrt{\tfrac{1}{3}} \big| \pi^0_{\vec{p}} K^+_{-\vec{p}} \big\rangle  \Big)
			-\tfrac{\sqrt{5}}{10} \Big( 
			\big| K^+_{\vec{p}} \eta_{-\vec{p}} \big\rangle + \big| \eta_{\vec{p}} K_{-\vec{p}} \big\rangle \Big)
			\Big] \nonumber \\
&=   \big[\,  1 + (-1)^\ell\, \big]\,	\int \!\!d\hat{p}\; Y_\ell^m(\hat{p}) 
	 \Big[ \tfrac{3\sqrt{5}}{10} \Big( 
			-\!\sqrt{\tfrac{2}{3}} \big| K^0_{\vec{p}} \pi^+_{-\vec{p}} \big\rangle 
			+\sqrt{\tfrac{1}{3}} \big| K^+_{\vec{p}} \pi^0_{-\vec{p}} \big\rangle 
			 \Big)
			-\tfrac{\sqrt{5}}{10} 
			\big| K^+_{\vec{p}} \eta_{-\vec{p}} \big\rangle
			\Big]	\nonumber
\\[1.2ex]
&\Big| \mathbf{8_2}; S=1, I=\tfrac{1}{2}, I_z = +\tfrac{1}{2}; \ell, m \Big\rangle \nonumber \\
&=   \int \!\!d\hat{p}\; Y_\ell^m(\hat{p}) 
	 \Big[ \tfrac{1}{2} \Big( 
			-\!\sqrt{\tfrac{2}{3}} \big| K^0_{\vec{p}} \pi^+_{-\vec{p}} \big\rangle 
			+\sqrt{\tfrac{1}{3}} \big| K^+_{\vec{p}} \pi^0_{-\vec{p}} \big\rangle 
			+\sqrt{\tfrac{2}{3}} \big| \pi^+_{\vec{p}} K^0_{-\vec{p}} \big\rangle 
			-\sqrt{\tfrac{1}{3}} \big| \pi^0_{\vec{p}} K^+_{-\vec{p}} \big\rangle  \Big)
			+\tfrac{1}{2} \Big( 
			\big| K^+_{\vec{p}} \eta_{-\vec{p}} \big\rangle - \big| \eta_{\vec{p}} K_{-\vec{p}} \big\rangle \Big)
			\Big] \nonumber \\
&= \big[\,  1 - (-1)^\ell\, \big]\,	\int \!\!d\hat{p}\; Y_\ell^m(\hat{p}) 
	 \Big[ \tfrac{1}{2} \Big( 
			-\!\sqrt{\tfrac{2}{3}} \big| K^0_{\vec{p}} \pi^+_{-\vec{p}} \big\rangle 
			+\sqrt{\tfrac{1}{3}} \big| K^+_{\vec{p}} \pi^0_{-\vec{p}} \big\rangle 
			 \Big)
			+\tfrac{1}{2} 
			\big| K^+_{\vec{p}} \eta_{-\vec{p}} \big\rangle
			\Big]	\nonumber
\end{align}
\end{widetext}

This indicates that, as in the non-strange case, the even-$\ell$ waves couple to $\mathbf{8_1}$, and the odd-$\ell$ waves couple to $\mathbf{8_2}$. The relative couplings to $\pi K $ and $\eta K$ differ significantly though -- in the even-$\ell$ case, the amplitude for $\pi K$ is three times larger than for $\eta K$, while in the odd-$\ell$ case the couplings are equal.

These $SU(3)$ flavor expectations appear to hold qualitatively in experiment; LASS~\cite{Aston:1987ey} observed the $K^\star_3$ as an enhancement in the $\eta K$ final state, but did not observe at any significant level the $K^\star_2$. The modern PDG averages~\cite{Beringer:1900zz} have $K^\star_3$ decaying to $\pi K$ and $\eta K$ with $19(1)\%$ and $30(13)\%$ branches respectively. The $K_2^\star$, on the other hand, has a $50\%$ branch to $\pi K$ and less than $1\%$ into $\eta K$.

\section{The Chew-Mandelstam phase-space}
\label{app_chew_man}

In Eq.~\ref{eq_t_matrix_k}, which relates the $K$-matrix to the $t$-matrix, there appears a matrix $I_{ij}(s)$ which is constrained by $S$-matrix unitarity to have a certain imaginary part above threshold $\mathrm{Im}\, I_{ij}(s) = -\rho_i(s)\,  \Theta( s- s^{(i)}_\mathrm{thr})\,  \delta_{ij}$. A convenient choice for the real part is supplied by the Chew-Mandelstam function, which relates the real part to the imaginary part through a dispersion integral and which provides a smooth transition across the kinematic threshold. The matrix is diagonal $I_{ij}(s) = \delta_{ij} I_i(s)$, and if in channel $i$ the two scattering particles have mass $m_1, m_2$, then the once subtracted dispersion integral is 
$$ I(s) = I(s_\mathrm{thr}) - \frac{s -  s_\mathrm{thr}}{\pi} \int_{s_\mathrm{thr}}^\infty \!\! ds' \, \frac{\rho(s')}{(s'-s)(s' - s_\mathrm{thr})} $$
where 
$$\rho(s) = \frac{2 k(s)}{\sqrt{s}} = \left(1 - \frac{(m_1+m_2)^2}{s}\right)^{\!\!1/2} \!\left(1 - \frac{(m_1-m_2)^2}{s}\right)^{\!\!1/2}$$ 
with a threshold at $s_\mathrm{thr} = (m_1+m_2)^2$.

The form of the integral is such that at $s + i \epsilon$, the real part is given by the principal value, and the imaginary part, $ \mathrm{Im}\, I(s) = - \rho(s)\,  \Theta( s- s_\mathrm{thr}) $, is as it should to satisfy unitarity. The integral can be performed~\cite{Pennington:2013pc} to give
\begin{align} I(s) &= I(s_\mathrm{thr})  \nonumber \\
&\quad+ \frac{\rho(s)}{\pi} \log \left[ \frac{\xi(s) + \rho(s)}{\xi(s) - \rho(s)} \right] - \frac{\xi(s)}{\pi} \frac{m_2-m_1}{m_1+m_2} \log \frac{m_2}{m_1} \nonumber
\end{align}
with $\xi(s) = 1-\frac{(m_1 + m_2)^2}{s}$. In this closed form the imaginary part resides in the $\log \left[ \frac{\xi(s) + \rho(s)}{\xi(s) - \rho(s)} \right]$ term when the argument is negative, which occurs for $s > (m_1+m_2)^2$.

We may choose $I(s_\mathrm{thr})$ as we see fit, a common choice is to have the function zero at threshold. Another convenient option arises when dealing with a resonance: as an example consider a single-channel in \mbox{$S$-wave} where we parameterize $K(s) = \frac{g^2}{m^2 - s}$ such that we have ${t(s) = \frac{g^2}{m^2 -s + g^2 I(s)}}$. If we choose $I(s_\mathrm{thr})$ such that $\mathrm{Re}\, I(s=m^2) = 0$, then in the region around $s=m^2$ the $t$-matrix resembles a Breit-Wigner pole with the mass $m$ being the Breit-Wigner mass, $m_R$.

\vspace{.5cm}

\section{Operator tables}
\label{app_op_tabs}

Table~\ref{table_ops} presents a shorthand of the momentum constructions used in our ``meson-meson" operators. In Tables~\ref{ops_24}, \ref{ops_20}, \ref{ops_16} we show the particular set of $\pi K$, $\eta K$ operators and the number of ``single-meson" operators used in our determination of the $I=1/2$ spectra. In Table~\ref{ops_I3o2} we show the set of $\pi K$ operators used to determine the $I=3/2$ spectra.

\begin{table}[b]
\begin{tabular}{c | l l l }
\hline \hline
&&&\\[-2ex]
 $\vec{P}$ & \quad $\vec{k}_1$ & \quad$\vec{k}_2$ & \multicolumn{1}{c}{$\Lambda^{(P)}$} \\
&&&\\[-2.2ex]
\hline \hline
&&&\\[-2ex]
\multirow{4}{*}{$\begin{matrix}[0,0,0] \\ \text{O}^{\text{D}}_h\end{matrix}$}
 & $[0,0,0]$ & $[0,0,0]$ 							& $A_1^+$ 						\\
 & $[0,0,1]$ & $[0,0,\text{-}1]$ 					& $A_1^+, T_1^-, E^+$			\\
 & $[0,1,1]$ & $[0,\text{-}1,\text{-}1]$ 			& $A_1^+, T_1^-,E^+,T_2^+$		\\
 & $[1,1,1]$ & $[\text{-}1,\text{-}1,\text{-}1]$ 	& $A_1^+, T_1^-, T_2^+$		\\
 &&&\\[-2ex]
\hline
&&&\\[-2ex]
\multirow{3}{*}{$\begin{matrix}[0,0,1] \\ \text{Dic}_4 \end{matrix}$}
 & $[0,0,0]$ & $[0,0,1]$ 							& $A_1$ 					\\
 & $[0,\text{-}1,0]$ & $[0,1,1]$ 					& $A_1, E_2, B_1$			  \\
 & $[\text{-}1,\text{-}1,0]$ & $[1,1,1]$ 			& $A_1, E_2, B_2$ 			\\
&&&\\[-2ex]
\hline
&&&\\[-2ex]
\multirow{4}{*}{$\begin{matrix}[0,1,1] \\ \text{Dic}_2 \end{matrix}$}
 & $[0,0,0]$ & $[0,1,1]$ 							& $A_1$ 					\\
 & $[0,1,0]$ & $[0,0,1]$ 							& $A_1, B_1$ 				\\
 & $[\text{-}1,0,0]$ & $[1,1,1]$ 					& $A_1, B_2$ 				\\
 & $[1,1,0]$ & $[\text{-}1,0,1]$ 					& $A_1, B_1, B_2$ 			\\
&&&\\[-2ex]
\hline
&&&\\[-2ex]
\multirow{2}{*}{$\begin{matrix}[1,1,1] \\ \text{Dic}_3 \end{matrix}$}
 & $[0,0,0]$ & $[1,1,1]$ 							& $A_1$ 					\\
 & $[1,0,0]$ & $[0,1,1]$ 							& $A_1, E_2$ 				\\
&&&\\[-2ex]
\hline
&&&\\[-2ex]
\multirow{3}{*}{$\begin{matrix}[0,0,2] \\ \text{Dic}_4 \end{matrix}$}
 & $[0,0,0]$ & $[0,0,2]$ 							& $A_1$ 					\\
 & $[0,0,1]$ & $[0,0,1]$ 							& $A_1$ 				\\
 & $[0,\text{-}1,1]$ & $[0,1,1]$ 					& $A_1$ 				\\
&&&\\[-2ex]
\hline \hline
\end{tabular}
\caption{``meson-meson" operator constructions presented for each $\vec{P}$; also shown is $\textrm{LG}(\vec{P})$.  Example momenta $\vec{k}_1$ and $\vec{k}_2$ are given -- all momenta in $\{\vec{k}_1\}^{\star}$ and $\{\vec{k}_2\}^{\star}$ are summed over in Eq.~\ref{eq_op_meson_pair}.  When ${|\vec{k}_1| \neq |\vec{k}_2|}$, the distinct operators with ${\vec{k}_1 \leftrightarrow \vec{k}_2}$, having the same distribution across irreps, are usually also included as an independent operator in the basis.}
\label{table_ops}
\end{table}

\begin{table*}
\begin{tabular}{c|c|c|c}
\multicolumn{4}{c}{$[000]$}\\
$A_1^+$ 	& $T_1^-$	&$E^+$		& $T_2^+$\\
\hline
$\pi_0 K_0$		& 					& 					&			 \\
$\pi_1 K_1$ 	& $\pi_1 K_1$ 		& $\pi_1 K_1$		&\\
$\pi_2 K_2$ 	& $\pi_2 K_2$ 		& $\pi_2 K_2$		& $\pi_2 K_2$\\
$\pi_3 K_3$ 	&  					&					& $\pi_3 K_3$ \\[0.9ex]
$\eta_0 K_0$ 	&					&\\
$\eta_1 K_1$ 	& $\eta_1 K_1$ 		& $\eta_1 K_1$\\[0.9ex]
8 				& 9 				& 13 				& 14 	
\end{tabular}
\quad\quad
\begin{tabular}{c|c|c|c}
\multicolumn{4}{c}{$[001]$}\\
$A_1$ 	& $E_2$		&$B_1$		& $B_2$\\
\hline
$\pi_0 K_1$		& 					&					&\\
$\pi_1 K_0$ 	&					&					&\\
$\pi_1 K_2$ 	& $\pi_1 K_2$		& $\pi_1 K_2$		& \\
$\pi_2 K_1$ 	& $\pi_2 K_1$		& $\pi_2 K_1$		& \\
$\pi_2 K_3$		& $\pi_2 K_3$ 		&					&$\pi_2 K_3$\\
$\pi_3 K_2$ 	& $\pi_3 K_2$		&					&$\pi_3 K_2$\\[0.9ex]
$\eta_0 K_1$ 	&					&					&\\	
$\eta_1 K_0$ 	&					&					&\\
$\eta_1 K_2$ 	&					& $\eta_1 K_2$		& \\	
$\eta_2 K_1$	& $\eta_2 K_1$		& $\eta_2 K_1$		&  \\
				&					&					& $\eta_2 K_3$\\
				&					&					& $\eta_3 K_2$\\[0.9ex]
17 				& 16 				& 11 				& 8 	
\end{tabular}
\quad\quad
\begin{tabular}{c|c|c}
\multicolumn{3}{c}{$[011]$}\\
$A_1$ 		& $B_1$		&$B_2$ \\	
\hline
$\pi_0 K_2$			&					&				\\
$\pi_2 K_0$			&					&				\\
$\pi_1 K_1$			& $\pi_1 K_1$		&				\\
$\pi_1 K_3$ 		&					& $\pi_1 K_3$	\\
$\pi_3 K_1$			&					& $\pi_3 K_1$	\\
$\pi_2 K_2$			& $\pi_2 K_2$		& $\pi_2 K_2$	\\[0.9ex]
$\eta_0 K_2$ 		&					&				\\
$\eta_2 K_0$ 		&					&				\\
$\eta_1 K_1$ 		& $\eta_1 K_1$		&				\\
$\eta_1 K_3$		&					&				\\
$\eta_3 K_1$		&					&				\\
$\eta_2 K_2$		& $\eta_2 K_2$		&				\\[0.9ex]
15 					& 18 				& 13 			\\
\end{tabular}
\quad\quad
\begin{tabular}{c|c}
\multicolumn{2}{c}{$[111]$}\\
$A_1$ 		& $E_2$		\\
\hline
$\pi_0 K_3$			&					\\
$\pi_3 K_0$			&					\\
$\pi_1 K_2$			& $\pi_1 K_2$		\\
$\pi_2 K_1$			& $\pi_2 K_1$		\\[0.9ex]
$\eta_0 K_3$		&					\\
$\eta_3 K_0$		&					\\
$\eta_1 K_2$		& $\eta_1 K_2$		\\
$\eta_2 K_1$		& $\eta_2 K_1$		\\[0.9ex]
16 					& 13 	
\end{tabular}
\quad\quad
\begin{tabular}{c}
$[002]$\\
$A_1$ \\		
\hline
$\pi_0 K_4$	\\	
$\pi_4 K_0$ \\
$\pi_1 K_1$ \\
$\pi_2 K_2$ \\[0.9ex]
$\eta_0 K_4$ \\		
$\eta_4 K_0$ \\
$\eta_1 K_1$ \\
$\eta_2 K_2$ \\[0.9ex]
17 	
\end{tabular}
\vspace{-.3cm}
\caption{Operator basis used to determine $I=1/2$ spectrum on $24^3$ lattice. Final row shows the number of ``single-meson" operators included.}
\label{ops_24}
\end{table*}

\begin{table*}
\begin{tabular}{c|c|c}
\multicolumn{3}{c}{$[000]$}\\
$A_1^+$ 	& $T_1^-$	&$E^+$		\\
\hline
$\pi_0 K_0$		& 					& 							 \\
$\pi_1 K_1$ 	& $\pi_1 K_1$ 		& $\pi_1 K_1$		\\
$\pi_2 K_2$ 	& $\pi_2 K_2$ 		& $\pi_2 K_2$		\\
$\pi_3 K_3$		& $\pi_3 K_3$  		&					\\[0.9ex]
$\eta_0 K_0$ 	&					&\\
$\eta_1 K_1$ 	& $\eta_1 K_1$ 		& $\eta_1 K_1$\\[0.9ex]
6 				& 9 				& 12 	
\end{tabular}
\quad\quad
\begin{tabular}{c|c|c|c}
\multicolumn{4}{c}{$[001]$}\\
$A_1$ 	& $E_2$		&$B_1$		& $B_2$\\
\hline
$\pi_0 K_1$		& 					&					&\\
$\pi_1 K_0$ 	&					&					&\\
$\pi_1 K_2$ 	& $\pi_1 K_2$		& $\pi_1 K_2$		& \\
$\pi_2 K_1$ 	& $\pi_2 K_1$		& $\pi_2 K_1$		& \\
$\pi_2 K_3$		& $\pi_2 K_3$ 		&					&$\pi_2 K_3$\\
$\pi_3 K_2$ 	& $\pi_3 K_2$		&					&$\pi_3 K_2$\\[0.9ex]
$\eta_0 K_1$ 	&					&					&\\	
$\eta_1 K_0$ 	&					&					&\\
				& $\eta_1 K_2$		& $\eta_1 K_2$		& \\	
				& $\eta_2 K_1$		& $\eta_2 K_1$		& \\[0.9ex]
13 				& 16 				& 8 				& 11 	
\end{tabular}
\quad\quad
\begin{tabular}{c|c|c}
\multicolumn{3}{c}{$[011]$}\\
$A_1$ 		& $B_1$		&$B_2$ \\	
\hline
$\pi_0 K_2$			&					&				\\
$\pi_2 K_0$			&					&				\\
$\pi_1 K_1$			& $\pi_1 K_1$		&				\\
$\pi_1 K_3$			&					& $\pi_1 K_3$	\\
$\pi_3 K_1$			&					& $\pi_3 K_1$	\\
$\pi_2 K_2$			& $\pi_2 K_2$		& $\pi_2 K_2$	\\[0.9ex]
$\eta_0 K_2$ 		&					&				\\
$\eta_2 K_0$ 		&					&				\\
$\eta_1 K_1$ 		& $\eta_1 K_1$		&				\\[0.9ex]
14				 	& 16 				& 18 	\\
\end{tabular}
\quad\quad
\begin{tabular}{c|c}
\multicolumn{2}{c}{$[111]$}\\
$A_1$ 		& $E_2$		\\
\hline
$\pi_0 K_3$			&					\\
$\pi_3 K_0$			&					\\
$\pi_1 K_2$			& $\pi_1 K_2$		\\
$\pi_2 K_1$			& $\pi_2 K_1$		\\[0.9ex]
$\eta_0 K_3$		&					\\
$\eta_3 K_0$		&					\\
$\eta_1 K_2$		& $\eta_1 K_2$		\\
$\eta_2 K_1$		& $\eta_2 K_1$		\\[0.9ex]
16 					& 12 	
\end{tabular}
\quad\quad
\begin{tabular}{c}
$[002]$\\
$A_1$ \\		
\hline
$\pi_0 K_4$	\\	
$\pi_4 K_0$ \\
$\pi_1 K_1$ \\
$\pi_2 K_2$ \\[0.9ex]
$\eta_0 K_4$ \\		
$\eta_4 K_0$ \\
$\eta_1 K_1$ \\
$\eta_2 K_2$ \\[0.9ex]
16
\end{tabular}

\caption{Operator basis used to determine $I=1/2$ spectrum on $20^3$ lattice. Final row shows the number of ``single-meson" operators included.}
\label{ops_20}
\end{table*}

\begin{table*}
\begin{tabular}{c|c|c}
\multicolumn{3}{c}{$[000]$}\\
$A_1^+$ 	& $T_1^-$	&$E^+$		\\
\hline
$\pi_0 K_0$		& 					& 							 \\
$\pi_1 K_1$ 	& $\pi_1 K_1$ 		& $\pi_1 K_1$		\\
$\pi_2 K_2$		& $\pi_2 K_2$ 		& $\pi_2 K_2$		\\
$\pi_3 K_3$		& $\pi_3 K_3$ 		&					\\[0.9ex]
$\eta_0 K_0$ 	&					&\\
$\eta_1 K_1$ 	& $\eta_1 K_1$ 		& $\eta_1 K_1$\\[0.9ex]
6 	& 11 	& 10 	
\end{tabular}
\quad\quad
\begin{tabular}{c|c}
\multicolumn{2}{c}{$[001]$}\\
$A_1$ 	& $E_2$	\\
\hline
$\pi_0 K_1$		& 			\\
$\pi_1 K_0$ 	&				\\
				& $\pi_1 K_2$		\\
				& $\pi_2 K_1$	\\[0.9ex]
$\eta_0 K_1$ 	&					\\	
$\eta_1 K_0$ 	&					\\[0.9ex]
8 & 16 	
\end{tabular}
\quad\quad
\begin{tabular}{c|c|c}
\multicolumn{3}{c}{$[011]$}\\
$A_1$ 		& $B_1$		&$B_2$ \\	
\hline
$\pi_0 K_2$			&					&				\\
$\pi_2 K_0$			&					&				\\
$\pi_1 K_1$			& $\pi_1 K_1$		&				\\
					& $\pi_2 K_2$		& $\pi_2 K_2$	\\
					&					& $\pi_1 K_3$	\\
					&					& $\pi_3 K_1$	\\[0.9ex]
$\eta_0 K_2$ 		&					&				\\
$\eta_2 K_0$ 		&					&				\\
$\eta_1 K_1$ 		& $\eta_1 K_1$		&				\\[0.9ex]
12 	& 15 	& 20 	\\
\end{tabular}
\quad\quad
\begin{tabular}{c|c}
\multicolumn{2}{c}{$[111]$}\\
$A_1$ 		& $E_2$		\\
\hline
$\pi_0 K_3$			&					\\
$\pi_3 K_0$			&					\\
$\pi_1 K_2$			& $\pi_1 K_2$		\\
$\pi_2 K_1$			& $\pi_2 K_1$		\\[0.9ex]
$\eta_1 K_2$		& $\eta_1 K_2$		\\
$\eta_2 K_1$		& $\eta_2 K_1$		\\[0.9ex]
15 	& 12 	
\end{tabular}
\quad\quad
\begin{tabular}{c}
$[002]$\\
$A_1$ \\		
\hline
$\pi_0 K_4$	\\	
$\pi_4 K_0$ \\
$\pi_1 K_1$ \\[0.9ex]
$\eta_1 K_1$ \\[0.9ex]
10 
\end{tabular}

\caption{Operator basis used to determine $I=1/2$ spectrum on $16^3$ lattice. Final row shows the number of ``single-meson" operators included.}
\label{ops_16}
\end{table*}

\begin{table*}
\begin{tabular}{c|c|c|c}
\multicolumn{4}{c}{$[000]$}\\
$A_1^+$ 	& $T_1^-$	&$E^+$		& $T_2^+$\\
\hline
$\pi_0 K_0$		& 					& 					&			 \\
$\pi_1 K_1$ 	& $\pi_1 K_1$ 		& $\pi_1 K_1$		&\\
$\pi_2 K_2$ 	& $\pi_2 K_2$ 		& $\pi_2 K_2$		& $\pi_2 K_2$\\
$\pi_3 K_3$ 	& $\pi_3 K_3$ 		&					& $\pi_3 K_3$
\end{tabular}
\quad\quad
\begin{tabular}{c|c|c|c}
\multicolumn{4}{c}{$[001]$}\\
$A_1$ 	& $E_2$		&$B_1$		& $B_2$\\
\hline
$\pi_0 K_1$		& 					&					&\\
$\pi_1 K_0$ 	&					&					&\\
$\pi_1 K_2$ 	& $\pi_1 K_2$		& $\pi_1 K_2$		& \\
$\pi_2 K_1$ 	& $\pi_2 K_1$		& $\pi_2 K_1$		& \\
$\pi_2 K_3$		& $\pi_2 K_3$ 		&					& $\pi_2 K_3$\\
$\pi_3 K_2$ 	& $\pi_3 K_2$		&					& $\pi_3 K_2$
\end{tabular}
\quad\quad
\begin{tabular}{c|c|c}
\multicolumn{3}{c}{$[011]$}\\
$A_1$ 		& $B_1$		&$B_2$ \\	
\hline
$\pi_0 K_2$			&					&				\\
$\pi_2 K_0$			&					&				\\
$\pi_1 K_1$			& $\pi_1 K_1$		&				\\
$\pi_1 K_3$ 		&					& $\pi_1 K_3$	\\
$\pi_3 K_1$			&					& $\pi_3 K_1$	\\
$\pi_2 K_2$			& $\pi_2 K_2$		& $\pi_2 K_2$
\end{tabular}
\quad\quad
\begin{tabular}{c|c}
\multicolumn{2}{c}{$[111]$}\\
$A_1$ 		& $E_2$		\\
\hline
$\pi_0 K_3$			&					\\
$\pi_3 K_0$			&					\\
$\pi_1 K_2$			& $\pi_1 K_2$		\\
$\pi_2 K_1$			& $\pi_2 K_1$		
\end{tabular}
\quad\quad
\begin{tabular}{c}
$[002]$\\
$A_1$ \\		
\hline
$\pi_0 K_4$*	\\	
$\pi_4 K_0$* \\
$\pi_1 K_1$ \\
$\pi_2 K_2$* 
\end{tabular}

\caption{Operator basis used to determine $I=3/2$ spectrum on all three lattice volumes. * indicates the operator was only used on the $24^3$ lattice.}
\label{ops_I3o2}
\end{table*}

\end{document}